\documentclass{emulateapj}

\def\sun{\hbox{$\odot$}}
\def\cm3{cm$^{-3}$}
\def\kms{km~s$^{-1}$}
\def\msunyr{M$_{\odot}$\,yr$^{-1}$\,}
\def\lsun{L$_{\odot}$}

\def\mdot{$\dot{\rm M}$}
\def\mo{M$_{\odot}$\,}

\def\beq{\begin{equation}}
\def\eeq{\end{equation}}

\def\sles{\lower2pt\hbox{$\buildrel {\scriptstyle <}
   \over {\scriptstyle\sim}$}}

\begin{document}

\title{Using Quantitative Spectroscopic Analysis to Determine the Properties 
and Distances of Type II-Plateau Supernovae: SN\lowercase{e} 2005\lowercase{cs}  and 2006\lowercase{bp}}

\author{Luc Dessart\altaffilmark{1,2},
        St\'{e}phane Blondin\altaffilmark{3},
        Peter J.~Brown\altaffilmark{4},
        Malcolm Hicken \altaffilmark{3},
        D. John Hillier \altaffilmark{5},
        Stephen~T.~Holland\altaffilmark{6,7},
        Stefan Immler\altaffilmark{6,7},
        Robert P.~Kirshner\altaffilmark{3}, 
        Peter Milne\altaffilmark{1},
        Maryam Modjaz\altaffilmark{3}, \& 
        Peter W.~A.~Roming\altaffilmark{4}
        }

\altaffiltext{1}{Department of Astronomy and Steward Observatory,
                University of Arizona,
                Tucson, AZ 85721, USA}

\altaffiltext{2}{luc@as.arizona.edu}

\altaffiltext{3}{Harvard-Smithsonian Center for Astrophysics,
                Cambridge, MA 01238, USA}

\altaffiltext{4}{Pennsylvania State University,
                 Department of Astronomy \& Astrophysics,
                 University Park, PA 16802, USA}

\altaffiltext{5}{Department of Physics and Astronomy,
                 University of Pittsburgh,
                 Pittsburgh, PA 15260}

\altaffiltext{6}{Astrophysics Science Division,
                X-Ray Astrophysics Branch, Code 662,
                 NASA Goddard Space Flight Center,
                 Greenbelt, MD 20771, USA}

\altaffiltext{7}{Universities Space Research Association,
                10211 Wincopin Circle, Columbia MD 21044, USA}

\begin{abstract}

We analyze the Type II Plateau supernovae (SN II-P) 2005cs and 2006bp with the
non-LTE model atmosphere code CMFGEN.  We fit 13 spectra in the first month
for SN 2005cs and 18 for SN 2006bp.  {\sl Swift} ultraviolet photometry and
ground-based optical photometry calibrate each spectrum. 
Our analysis shows both objects were
discovered less than 3 days after they exploded, making these the earliest SN
II-P spectra ever studied. They reveal broad and very weak lines from
highly-ionized fast ejecta with an extremely steep density profile.
We identify He{\,\sc ii}\,4686\AA\ emission in the SN 2006bp ejecta. Days later,
the spectra resemble the prototypical Type II-P SN 1999em, which had a
supergiant-like photospheric composition. Despite the association of SN 2005cs
with possible X-ray emission, the emergent UV and optical light comes from the
photosphere, not from circumstellar emission.

We surmise that the very steep density fall-off we infer at early times may be
a fossil of the combined actions of the shock wave passage and radiation driving at
shock breakout. Based on tailored CMFGEN models, the direct-fitting
technique and the Expanding Photosphere Method both yield distances and
explosion times that agree within a few percent. We derive a distance to NGC 5194, the host of
SN 2005cs, of 8.9$\pm$0.5\,Mpc and 17.5$\pm$0.8\,Mpc for SN 2006bp in NGC
3953. The luminosity of SN 2006bp is 1.5 times that of SN 1999em, and 6 times
that of SN 2005cs.  Reliable distances to Type II-P supernovae that do not
depend on a small range in luminosity provide an independent route to the
Hubble Constant and improved constraints on other cosmological parameters.

\end{abstract}


\section{Introduction}

  Although radiation represents a negligible fraction of the energy in core-collapse 
supernova (SN) explosions, most of what we know about these
events is inferred from the analysis of spectra and light curves.
The gravitational binding-energy of the newly-formed protoneutron star is on the order of 
10$^{53}$\,erg. Within milliseconds after core bounce, a few $\times$ 10$^{52}$\,erg 
of this energy is used to photodissociate the infalling outer iron core. Starting with an
electron-neutrino burst when the core reaches nuclear densities, the radiation of neutrinos of all 
flavors operates over a few tens of seconds after the bounce as the protoneutron star cools, 
and carries away a few $\times$ 10$^{52}$\,erg. These neutrinos 
are believed to deposit $\sim$10$^{51}$\,erg (1\% of the total) into the infalling progenitor mantle,
reversing the accretion, and to provide the internal and kinetic energy for the SN ejecta 
(Woosley \& Janka 2005). Only $\sim$10$^{49}$\,erg (0.01\% of the total) is eventually processed into light, that gets
radiated with a rate equivalent to a few $\times$ 10$^8$\lsun\, sustained for three months.
 
Depending on the progenitor radius, the shock emerges at the surface between a few  
hours and a day after core bounce, with a radiative precursor that is expected to heat and accelerate the surface
layers, producing a soft X-ray flash (Chevalier 1982; Ensman \& Burrows 1992; 
Matzner \& McKee 1999; Blinnikov et al. 2000). Cooling, due to adiabatic expansion and 
radiative losses, is moderated as time progresses by recombination of hydrogen during the
photospheric phase of Type II SN and by non-thermal excitation by unstable isotopes.  Although UV emission is significant for the first few days after shock breakout, the spectral energy distribution (SED) subsequently peaks further and further into
the red, first in the visual and then in the near-IR, a few months after explosion 
(Kirshner et al. 1973,1975; Mitchell et al. 2002; Leonard et al. 2002a; Brown et al. 2007).

  Core-collapse SN explosions constitute an excellent laboratory for inferring the properties
of massive stars at the end of their lives and, indirectly, their pre-SN evolution. 
As the material expands, the photosphere recedes to deeper and deeper  
layers of the star's former envelope, allowing the observer to probe  
all the way from the surface at shock breakout to the innermost  
layers just outside the compact remnant. The nebular phase, the epoch when
the ejecta become optically thin throughout ($\sim$3-4 months after explosion for Type II SNe), 
permits the inspection of the regions where
the blast originated, offering a means to constrain the mechanism and the morphology of the 
explosion as well as the nucleosynthetic yields.

  Detailed quantitative analyzes focus on the early, 
photospheric, evolution (Ho\"{e}flich 1988; Eastman \& Kirshner 1989; Schmutz et al. 1990;
Baron et al 1996, 2000, 2003, 2004, 2007; Mitchell et al. 2002; Dessart \& Hillier 2005a,2006ab), because the ejecta are best suited for 
``standard'' model atmosphere computations: There is an optically-thick base where the radiation thermalizes;
there is no contribution from non-thermal excitation by isotopes like $^{56}$Ni and its daughter
product $^{56}$Co; electron scattering dominates the opacity; and
line and continuum formation is spatially confined. Advances in computer technology permit 
better handling of non-Local-Thermodynamic-Equilibrium (non-LTE) 
effects and the critical role played by metals is now systematically
accounted for (Baron et al. 2004; Dessart \& Hillier 2005a). Non-LTE effects are important even at
early times, for example, to reproduce the He{\,\sc i} lines observed in optical spectra.
% Studies done in the nebular phase are an observational challenge since the
% SN are faint at these epochs. The radiative transfer problem is also
% significantly different: Non-thermal excitation is paramount, heating, exciting, and  
% ionizing
% the ejecta in a non-local fashion. Continuum absorption, scattering, and emission are less % important than line
% transfer. Further, densities are very low so that forbidden-line transitions play a major % % role in
% cooling the ejecta, so time-dependent effects are important (Fransson \& Kozma 1993; 
% Kozma \& Fransson 1998ab).
% Much is to be learnt from studies of both the photospheric and the nebular phases, in
% particular
% the transition between the two when the photosphere resides in the mass shells just above the
% progenitor iron core, but no such comprehensive study exists to this date. 
% CMFGEN is currently under 
% development to allow a modeling in a time-dependent
% fashion from one week after shock breakout until the nebular phase. 
% One step on the way to a fully-blown time-dependent model has been made, accounting for time 
% dependence in the statistical and radiative equilibrium equations, providing consequently a natural 
% explanation for the observed strength of hydrogen Balmer lines during the recombination epoch of 
% Type II SN (Utrobin \& Chugai 2005; Dessart \& Hillier 2006b). In this work, we employ 
% the time-dependent version of CMFGEN only for illustrative purposes, to document the important
% effects associated with time dependence in Type II SNe spectra a few weeks after explosion.

\begin{figure*}
\epsscale{1.1}
\plottwo{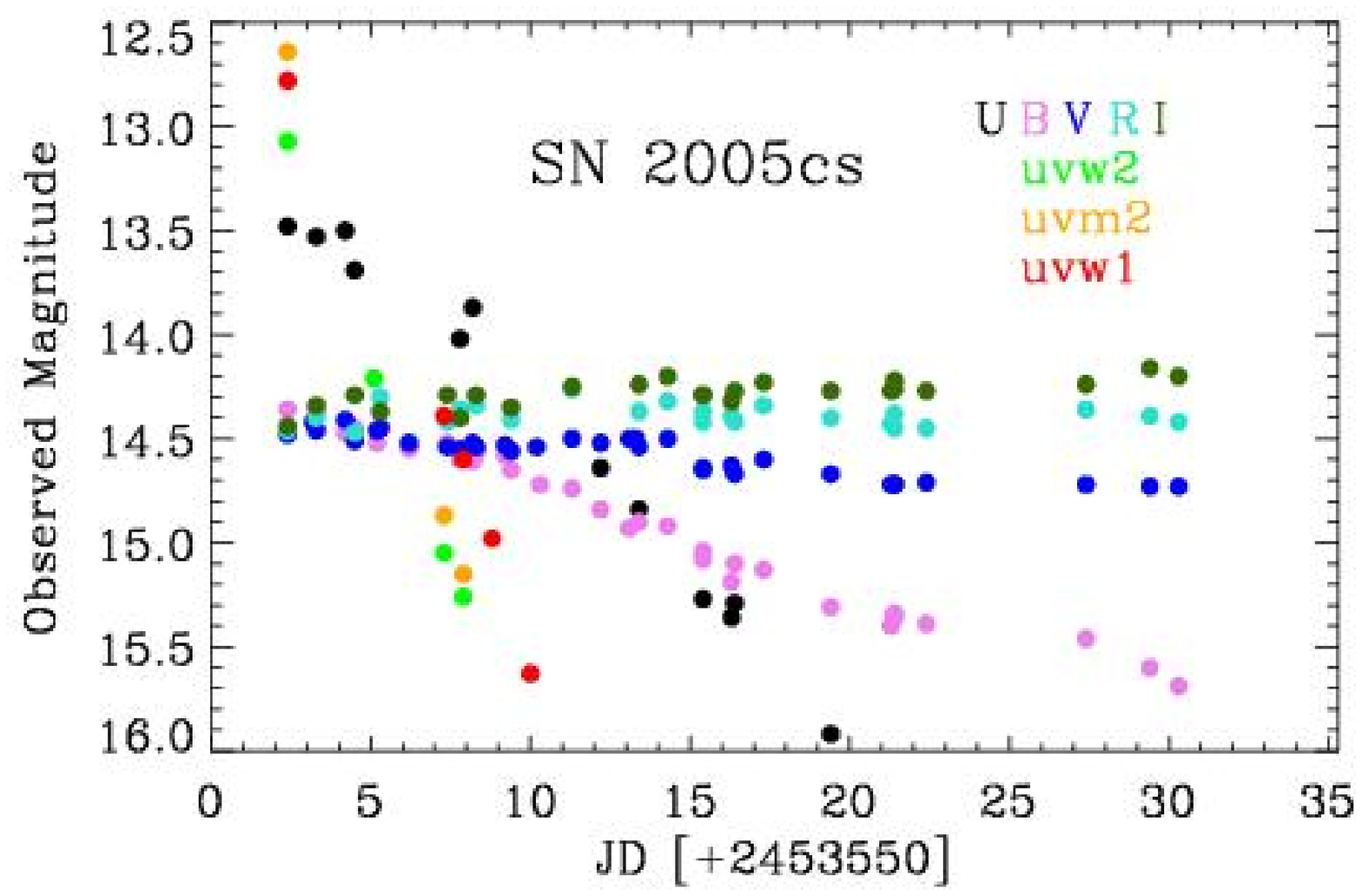}{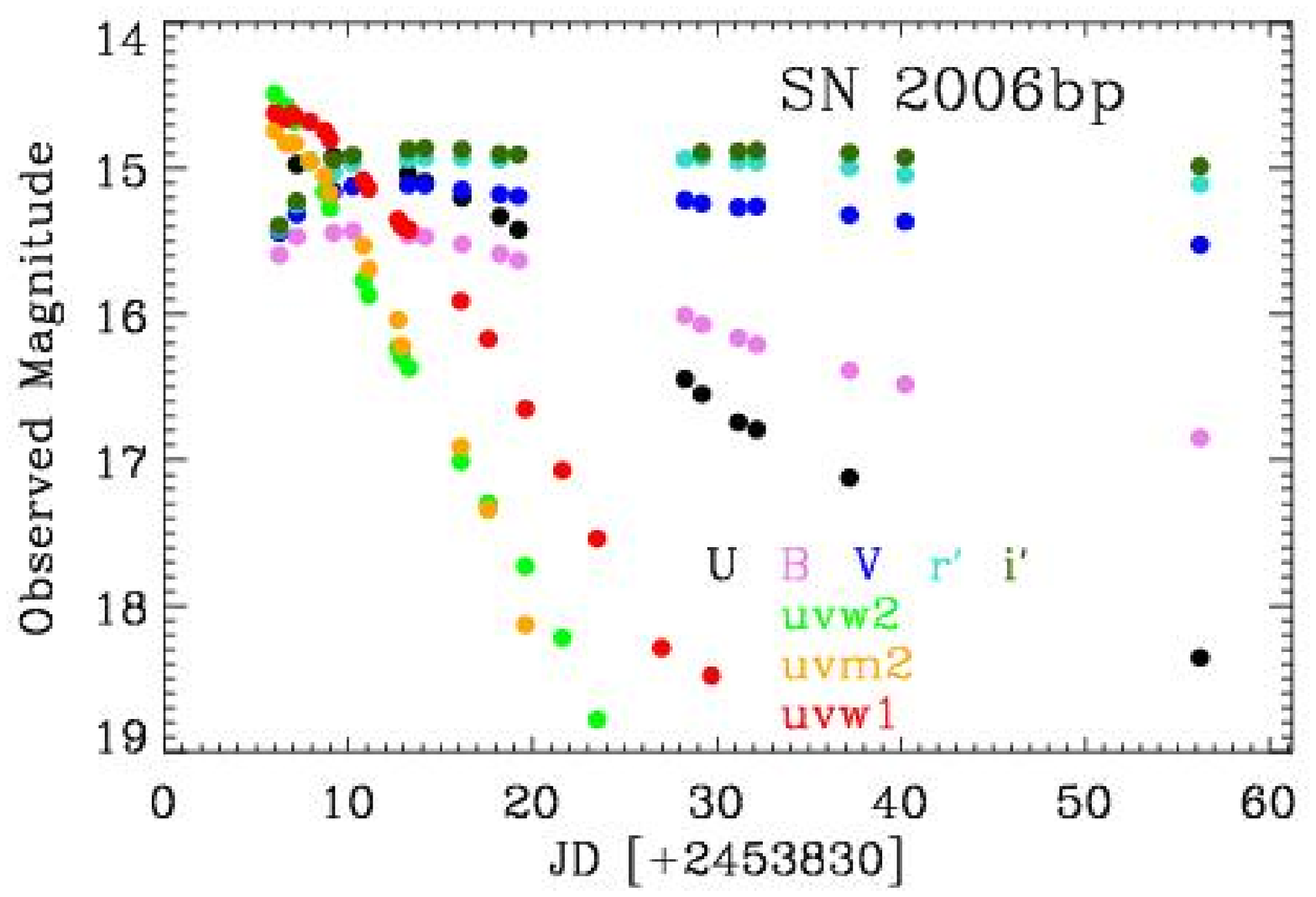}
% \plottwo{obs_lc_05cs}{obs_lc_06bp}
\caption{
Observed light curves for SN 2005cs (left; optical data include the observations of  
T06, P06, and those of the CfA) and for SN 2006bp (right).
Besides optical magnitudes in $UBVRI$ ($i'$ is shown instead of $I$ for SN 2006bp),
we also include the {\it Swift} UVOT magnitudes in the filters $uvw2$, $uwm2$, and $uvw1$
(corresponding values and photometric errors are given in 
Table~\ref{tab_phot_05cs} \& \ref{table_swift_05cs}
for SN 2005cs and Table~\ref{tab_phot_06bp} \& \ref{table_swift_06bp} 
for SN 2006bp, for each filter and date).
[See the electronic edition of the Journal for a color version of this figure, 
and see \S\ref{sect_obs_phot} and \S\ref{swift_obs} for discussion.\label{fig_lc_obs}]
}
%%%\label{05cs_early}
\end{figure*}

  Here, we present a quantitative spectroscopic analysis of the early photospheric-phase evolution of
the Type II SNe 2005cs in NGC 5194 and SN 2006bp in NGC 3953, both with plateau-type light curves as  
shown in Fig.~\ref{fig_lc_obs}.
We employ the non-LTE model atmosphere code CMFGEN (Hillier \& Miller 1998; Dessart \& Hillier 2005a),
in its steady-state configuration, following a similar approach to Dessart \& Hillier 
(2006a; hereafter DH06) and their analysis of SN 1999em. 
A novel component of our study is the
use of {\sl Swift} (Gehrels et al. 2004) Ultra-Violet/Optical Telescope (UVOT; Roming et al. 2005) ultraviolet photometry
that places important constraints on the SED in the UV, the early cooling of the photosphere, and 
the reddening. We infer from our models that SN 2005cs and SN 2006bp, 
were discovered only $\sim$2 days after explosion, the earliest detections
after explosion since SN 1987A (Type II peculiar) and SN 1993J (Type IIb; see, e.g., Matheson et al. 2000). 
As a result of this prompt discovery, we have the opportunity to observe a bright UV spectrum and to 
see high-excitation lines that have not previously been observed in SN II-P.

Finally, the fine time sampling of our spectra during the first month,
along with dense photometric monitoring allows us to make an accurate distance determination. This is a good demonstration of the potential of Type II-P SNe, when coupled with detailed models, to provide accurate extragalactic distances.  Unlike the standard approach that depend on the Cepheid period-luminosity relation 
in the LMC and other nearby galaxies, our method is based on physical models fitted directly to
observations, and the distances it provides are independent of all the lower rungs of the distance ladder.  We aim to assess the accuracy of distance 
determinations based on Type II-P SNe, and to refine the value 
of the Hubble constant $H_0$, in the spirit of the earlier efforts of Schmidt et al. (1994).
Measurements of the total matter and baryonic content of the Universe via CMB experiments or
large-scale structure correlations depend on the square of $H_0$ (Spergel et al. 2007; 
Eisenstein et al. 2005). In concordance solutions, these uncertainties affect measurements of the Dark Energy density 
and its associated equation-of-state parameter $w$ (Garnavich et al. 1998; Riess et al 2005;
see Macri et al. 2006 for an illustration of the effect an improved accuracy on $H_0$ has on $w$). A small improvement in our knowledge of $H_0$ will lead to a significant improvement in the understanding the energy content of the Universe.

% \clearpage
% \clearpage
This paper is structured as follows. In the next section, we present the photometric and spectroscopic
datasets.
In \S\ref{sect_mod_pres}, we present the model atmosphere code CMFGEN and describe our
methods.
In \S\ref{sect_05cs}, we present a quantitative spectroscopic analysis at 13 photospheric-phase 
epochs for SN 2005cs, and in \S\ref{sect_06bp}, analyze 18 epochs for SN 2006bp. 
In \S\ref{sect_dist}, we apply the Expanding Photosphere Method (EPM) as well as a direct-fitting 
method (in the spirit of the Spectral-fitting Expanding Atmosphere Method, SEAM, of Baron et al. 2000) 
to infer the distance and the explosion date for both SN events.
In \S\ref{sect_discussion}, we discuss the implications of our results, and in \S\ref{sect_conclusion}, 
we conclude.
% \clearpage
\begin{deluxetable*}{lccccccccc}
% \rotate
\tablewidth{0pt}
\tabletypesize{\scriptsize}
\tablecaption{Journal of photometric observations of SN 2005\lowercase{cs}\label{tab_phot_05cs}}
\tablehead{
\colhead{UT Date} & 
\colhead{JD} &
\colhead{U} & 
\colhead{B} & 
\colhead{V} & 
\colhead{R} & 
\colhead{I} & 
\colhead{r'} & 
\colhead{i'} & 
\colhead{Source} \\
\colhead{} & 
\colhead{(+2,453,000)} &
\colhead{} & 
\colhead{} & 
\colhead{} & 
\colhead{} & 
\colhead{} & 
\colhead{} & 
\colhead{} & 
\colhead{} 
}   
\startdata
%# JD-2400000    U       Uerr           B           Berr    V           Verr    R            Rerr    I            Ierr    r'      r'err        i'      i'err   Reference
%#----------------------------------------------------------------------------------------------------------------------------------------
2005-06-30  &552.4 &13.48 $\pm$  0.05  &14.36 $\pm$  0.05  &14.48 $\pm$  0.02  &14.46 $\pm$  0.04 &14.44  $\pm$ 0.04  &..               &..                 &  P06 \\
2005-07-01  &553.2 &..                 &14.42 $\pm$  0.02  &14.42 $\pm$  0.02  &..                &..                 &14.68 $\pm$ 0.03 &14.34  $\pm$ 0.03  &  CfA \\
2005-07-01  &553.3 &13.53 $\pm$  0.06  &14.36 $\pm$  0.05  &14.46 $\pm$  0.05  &14.40 $\pm$  0.02 &14.34  $\pm$ 0.06  &..               &..                 &  P06 \\
2005-07-02  &554.2 &13.50 $\pm$  0.02  &14.47 $\pm$  0.01  &14.41 $\pm$  0.01  &..                &..                 &14.64 $\pm$ 0.02 &14.32  $\pm$ 0.02  &  CfA \\
2005-07-02  &554.5 &13.69 $\pm$  0.10  &14.45 $\pm$  0.04  &14.51 $\pm$  0.03  &14.47 $\pm$  0.03 &14.29  $\pm$ 0.10  &..               &..                 &  P06 \\
2005-07-03  &555.2 &..                 &14.52 $\pm$  0.02  &14.46 $\pm$  0.02  &..                &..                 &..               &14.35  $\pm$ 0.02  &  CfA \\
2005-07-03  &555.3 &..                 &14.45 $\pm$  0.02  &14.45 $\pm$  0.02  &14.30 $\pm$  0.02 &14.37  $\pm$ 0.05  &..               &..                 &  T06 \\
2005-07-04  &556.2 &..                 &14.55 $\pm$  0.01  &14.52 $\pm$  0.02  &..                &..                 &14.66 $\pm$ 0.02 &14.36  $\pm$ 0.02  &  CfA \\
2005-07-05  &557.4 &..                 &14.52 $\pm$  0.04  &14.54 $\pm$  0.04  &14.42 $\pm$  0.05 &14.29  $\pm$ 0.08  &..               &..                 &  P06 \\
2005-07-06  &557.8 &14.02 $\pm$  0.06  &14.60 $\pm$  0.05  &14.55 $\pm$  0.03  &14.36 $\pm$  0.03 &14.40  $\pm$ 0.03  &..               &..                 &  P06 \\
2005-07-06  &558.2 &13.87 $\pm$  0.02  &14.61 $\pm$  0.02  &14.52 $\pm$  0.02  &..                &..                 &14.60 $\pm$ 0.03  &14.33  $\pm$ 0.03  &  CfA \\
2005-07-06  &558.3 &..                 &14.57 $\pm$  0.02  &14.54 $\pm$  0.02  &14.34 $\pm$  0.02 &14.29  $\pm$ 0.03  &..               &..                 &  T06 \\
2005-07-07  &559.2 &..                 &14.58 $\pm$  0.03  &14.53 $\pm$  0.01  &..                &..                 &14.59 $\pm$ 0.02 &14.33  $\pm$ 0.01  &  CfA \\
2005-07-07  &559.4 &..                 &14.65 $\pm$  0.04  &14.56 $\pm$  0.09  &14.41 $\pm$  0.09 &14.35  $\pm$ 0.11  &..               &..                 &  P06 \\
2005-07-08  &560.2 &..                 &..                 &14.54 $\pm$  0.01  &..                &..                 &..               &14.23  $\pm$ 0.01  &  CfA \\
2005-07-08  &560.3 &..                 &14.72 $\pm$  0.01  &..                 &..                &..                 &..               &..                 &  CfA \\
2005-07-09  &561.3 &..                 &14.74 $\pm$  0.02  &14.50 $\pm$  0.02  &14.26 $\pm$  0.02 &14.25  $\pm$ 0.03  &..               &..                 &  T06 \\
2005-07-10  &562.2 &14.64 $\pm$  0.03  &14.84 $\pm$  0.02  &14.52 $\pm$  0.02  &..                &..                 &..               &..                 &  CfA \\
2005-07-11  &563.1 &..                 &14.93 $\pm$  0.01  &14.50 $\pm$  0.02  &..                &..                 &..               &..                 &  CfA \\
2005-07-11  &563.3 &..                 &..                 &14.50 $\pm$  0.02  &..                &..                 &..               &..                 &  T06 \\
2005-07-11  &563.4 &14.84 $\pm$  0.05  &14.90 $\pm$  0.04  &14.54 $\pm$  0.04  &14.37 $\pm$  0.02 &14.24  $\pm$ 0.05  &..               &..                 &  P06 \\
2005-07-12  &564.3 &..                 &14.92 $\pm$  0.02  &14.50 $\pm$  0.02  &14.32 $\pm$  0.02 &14.20  $\pm$ 0.03  &..               &..                 &  T06 \\
2005-07-13  &565.4 &..                 &15.08 $\pm$  0.05  &14.65 $\pm$  0.04  &14.42 $\pm$  0.04 &14.29  $\pm$ 0.06  &..               &..                 &  T06 \\
2005-07-13  &565.4 &15.27 $\pm$  0.06  &15.04 $\pm$  0.05  &14.64 $\pm$  0.02  &14.37 $\pm$  0.03 &14.29  $\pm$ 0.05  &..               &..                 &  P06 \\
2005-07-14  &566.3 &15.36 $\pm$  0.06  &15.19 $\pm$  0.02  &14.63 $\pm$  0.02  &14.39 $\pm$  0.02 &14.32  $\pm$ 0.03  &..               &..                 &  T06 \\
2005-07-14  &566.4 &15.29 $\pm$  0.05  &15.10 $\pm$  0.05  &14.67 $\pm$  0.02  &14.42 $\pm$  0.04 &14.27  $\pm$ 0.02  &..               &..                 &  P06 \\
2005-07-15  &567.3 &..                 &15.13 $\pm$  0.03  &14.60 $\pm$  0.02  &14.34 $\pm$  0.02 &14.23  $\pm$ 0.03  &..               &..                 &  T06 \\
2005-07-17  &569.4 &15.92 $\pm$  0.07  &15.31 $\pm$  0.05  &14.67 $\pm$  0.03  &14.40 $\pm$  0.03 &14.27  $\pm$ 0.04  &..               &..                 &  P06 \\
2005-07-19  &571.3 &..                 &15.40 $\pm$  0.02  &14.72 $\pm$  0.02  &14.43 $\pm$  0.02 &14.27  $\pm$ 0.03  &..               &..                 &  T06 \\
2005-07-19  &571.4 &..                 &15.34 $\pm$  0.02  &14.72 $\pm$  0.02  &14.38 $\pm$  0.04 &14.22  $\pm$ 0.04  &..               &..                 &  T06 \\
2005-07-19  &571.4 &..                 &15.36 $\pm$  0.12  &14.72 $\pm$  0.05  &14.45 $\pm$  0.09 &14.26  $\pm$ 0.07  &..               &..                 &  P06 \\
2005-07-20  &572.4 &..                 &15.39 $\pm$  0.07  &14.71 $\pm$  0.03  &14.45 $\pm$  0.03 &14.27  $\pm$ 0.03  &..               &..                 &  P06 \\
2005-07-25  &577.4 &..                 &15.46 $\pm$  0.09  &14.72 $\pm$  0.04  &14.36 $\pm$  0.06 &14.24  $\pm$ 0.06  &..               &..                 &  P06 \\
2005-07-27  &579.4 &..                 &15.60 $\pm$  0.07  &14.73 $\pm$  0.04  &14.39 $\pm$  0.04 &14.16  $\pm$ 0.04  &..               &..                 &  P06 \\
2005-07-28  &580.3 &..                 &15.69 $\pm$  0.05  &14.73 $\pm$  0.03  &14.42 $\pm$  0.03 &14.20  $\pm$ 0.04  &..               &..                 &  T06 \\
\enddata 
\tablecomments{Sources for photometric observations are CfA, P06 and Tsvetkov et al. (2006; T06).
For the distance determinations performed in \S\ref{sect_dist_05cs}, we use the photometry corresponding 
to the dates on which spectroscopic observations were obtained. 
If there is no photometry on the date of a spectrum, we linearly interpolate the slowly varying magnitude obtained at bracketing dates.}
% \tablenotetext{}{}
\end{deluxetable*}

\begin{deluxetable*}{lcccccc}
% \rotate
\tablewidth{18cm} 
\tabletypesize{\scriptsize}
\tablecaption{Journal of photometric observations of SN 2006\lowercase{bp}\label{tab_phot_06bp}}
\tablehead{
\colhead{UT Date} & 
\colhead{JD} &
\colhead{U} & 
\colhead{B} & 
\colhead{V} & 
\colhead{r'} & 
\colhead{i'} \\
\colhead{} & 
\colhead{(+2,453,000)} &
\colhead{} & 
\colhead{} & 
\colhead{} & 
\colhead{} & 
\colhead{} 
}   
\startdata
2006-04-10 &836.27  &   ..                 & 15.59 $\pm$  0.02    &   15.44 $\pm$ 0.02  &   15.42 $\pm$ 0.02  &     15.39 $\pm$ 0.02 \\
2006-04-11 &837.20  &   14.97 $\pm$ 0.03   & 15.47 $\pm$  0.02    &   15.31 $\pm$ 0.02  &   15.24 $\pm$ 0.02  &     15.23 $\pm$ 0.02 \\
2006-04-13 &839.20  &   14.92 $\pm$ 0.03   & 15.44 $\pm$  0.02    &   15.16 $\pm$ 0.02  &   15.03 $\pm$ 0.02  &     14.94 $\pm$ 0.02 \\
2006-04-14 &840.23  &   ..                 & 15.43 $\pm$  0.02    &   15.12 $\pm$ 0.02  &   14.97 $\pm$ 0.02  &     14.91 $\pm$ 0.01 \\
2006-04-17 &843.24  &   15.04 $\pm$ 0.03   & 15.46 $\pm$  0.02    &   15.11 $\pm$ 0.02  &   14.94 $\pm$ 0.02  &     14.87 $\pm$ 0.02 \\
2006-04-18 &844.16  &   15.09 $\pm$ 0.03   & 15.47 $\pm$  0.02    &   15.12 $\pm$ 0.02  &   14.93 $\pm$ 0.02  &     14.86 $\pm$ 0.02 \\
2006-04-20 &846.17  &   15.20 $\pm$ 0.02   & 15.52 $\pm$  0.02    &   15.15 $\pm$ 0.02  &   14.93 $\pm$ 0.02  &     14.87 $\pm$ 0.02 \\
2006-04-22 &848.22  &   15.33 $\pm$ 0.03   & 15.59 $\pm$  0.02    &   15.18 $\pm$ 0.02  &   14.94 $\pm$ 0.02  &     14.90 $\pm$ 0.02 \\
2006-04-23 &849.23  &   15.42 $\pm$ 0.03   & 15.63 $\pm$  0.02    &   15.19 $\pm$ 0.02  &    ..               &     14.91 $\pm$ 0.02 \\
2006-05-02 &858.27  &   16.44 $\pm$ 0.03   & 16.01 $\pm$  0.02    &   15.22 $\pm$ 0.02  &   14.93 $\pm$ 0.02  &      ..              \\
2006-05-03 &859.19  &   16.55 $\pm$ 0.04   & 16.07 $\pm$  0.02    &   15.24 $\pm$ 0.02  &   14.93 $\pm$ 0.02  &     14.89 $\pm$ 0.02 \\
2006-05-05 &861.17  &   16.74 $\pm$ 0.04   & 16.16 $\pm$  0.02    &   15.27 $\pm$ 0.02  &   14.95 $\pm$ 0.02  &     14.88 $\pm$ 0.02 \\
2006-05-06 &862.16  &   16.79 $\pm$ 0.05   & 16.21 $\pm$  0.02    &   15.26 $\pm$ 0.02  &   14.96 $\pm$ 0.02  &     14.88 $\pm$ 0.02 \\
2006-05-11 &867.19  &   17.12 $\pm$ 0.06   & 16.39 $\pm$  0.03    &   15.32 $\pm$ 0.02  &   15.00 $\pm$ 0.02  &     14.89 $\pm$ 0.02 \\
2006-05-14 &870.20  &   ..                 & 16.48 $\pm$  0.04    &   15.37 $\pm$ 0.02  &   15.05 $\pm$ 0.02  &     14.92 $\pm$ 0.02 \\
2006-05-30 &886.23  &   18.36 $\pm$ 0.07   & 16.85 $\pm$  0.02    &   15.52 $\pm$ 0.02  &   15.11 $\pm$ 0.02  &     14.98 $\pm$ 0.02 \\
\enddata 
\tablecomments{ Log of CfA photometric observations. For the distance determination done in \S\ref{sect_dist_06bp}, 
we only use the $B$, $V$, and $i'$ magnitudes.
We use  photometric data on the dates of the spectra, or, if there is no photometry on the day of a spectrum, linearly interpolate between magnitudes obtained at bracketing dates.}
% \tablenotetext{}{}
\end{deluxetable*}

\begin{deluxetable*}{ccccccccc}
% \rotate
\tablewidth{16cm} 
\tabletypesize{\scriptsize}
\tablecaption{Journal of spectroscopic observations of SNe 2005\lowercase{cs} and 2006\lowercase{bp} 
with FLWO 1.5\,m+FAST\label{Table:speclog}}
\tablehead{
\colhead{UT Date{\tablenotemark{a}}} & 
\colhead{JD{\tablenotemark{a}}} &
\colhead{Epoch{\tablenotemark{b}}} &
\colhead{$\Delta {\rm PA}${\tablenotemark{c}}} &
\colhead{Airmass} &
\colhead{Flux stds.{\tablenotemark{d}}} &
\colhead{Seeing{\tablenotemark{e}}} &
\colhead{Exp. time} &
\colhead{Observer{\tablenotemark{f}}} \\
\colhead{} & 
\colhead{(+2,453,000)} &
\colhead{(d)} &
\colhead{($^\circ$)} &
\colhead{} &
\colhead{} &
\colhead{($\arcsec$)} &
\colhead{(s)} &
\colhead{}
}   
\startdata
\multicolumn{9}{c}{SN 2005cs} \\     
\hline
2005-06-30.23 & 551.73 &   4 &  1.1 & 1.31 & BD28, BD17 & \nodata & 1200 & RH \\
2005-07-01.27 & 552.77 &   5 &  5.0 & 1.53 & BD28, BD17 & 4   &  900 & RH \\
2005-07-02.23 & 553.73 &   6 &  3.1 & 1.34 & BD28, BD26 & 2   & 1200 & RH \\ 
2005-07-03.15 & 554.65 &   7 & 32.7 & 1.09 & F34, BD33 & 2   &  720 & RH \\ 
2005-07-04.16{\tablenotemark{g}} & 555.66 & 8 & 85.4 & 1.11 & BD28,BD17 & 1 & 1200 & JG \\ 
2005-07-04.18{\tablenotemark{g}} & 555.68 & 8 & 78.4 & 1.15 & BD28,BD17 & 1 & 1200 & JG \\ 
2005-07-05.18 & 556.68 &   9 &  8.9 & 1.16 & F34, BD17 & 1   & 1200 & JG \\ 
2005-07-06.17 & 557.67 &  10 &  5.2 & 1.15 & BD28, BD17 & 1   & 1200 & JG \\ 
2005-07-07.15 & 558.65 &  12 &  1.8 & 1.11 & BD28, BD17 & 1   & 1200 & EF \\ 
2005-07-09.19 & 560.69 &  13 &  0.7 & 1.24 & BD28, BD17 & 1-2 & 1200 & PB \\ 
2005-07-10.18 & 561.68 &  14 &  1.1 & 1.22 & BD28, BD17 & 1-2 & 1200 & PB \\
2005-07-11.18 & 562.68 &  15 &  2.9 & 1.23 & BD28, BD17 & 1-2 & 1200 & MC \\
2005-07-12.16 & 563.66 &  16 &  2.8 & 1.17 & BD28, BD17 & 1-2 & 1200 & MC \\
2005-07-28.17 & 579.67 &  32 &  1.0 & 1.38 & BD28, BD17 & 1-2 & 1200 & PB \\
2005-07-29.15 & 580.65 &  33 &  2.1 & 1.30 & BD28, BD17 & 2-3 &  960 & MC \\
2005-12-25.56 & 730.04 & 182 &  5.0 & 1.11 & BD28, BD17 & 1   & 1800 & MC \\
\hline
\multicolumn{9}{c}{SN 2006bp} \\     
\hline
%%% Luc, you'll have to determine the epoch based on your estimate for t_exp
2006-04-20.30 & 845.80 & 12 & 12.3 & 1.13 & F34, HD84 & 1.5   & 1200 & WP \\
2006-04-21.21 & 846.71 & 13 &  3.5 & 1.07 & F34, HD84 & 1.5   & 1200 & WP \\
2006-04-24.19 & 849.69 & 16 & 67.9 & 1.08 & F34, BD26 & 1.5   & 1500 & JH \\
2006-05-01.35 & 856.85 & 23 &  0.8 & 1.41 & F34, HD84 & 1.5   &  780 & WB \\
2006-05-02.14 & 857.64 & 24 &  3.8 & 1.10 & F34, HD84 & 1.5-2 & 1200 & WB \\
2006-05-03.22 & 858.72 & 25 & 64.4 & 1.08 & F34, HD84 & 1.5-2 & 1200 & TG \\
2006-05-04.15 & 859.65 & 26 &  2.8 & 1.09 & F34, HD84 & 1.5-2 & 1200 & TG \\
2006-05-07.19 & 862.69 & 29 & 12.2 & 1.07 & F34, BD17 & 1.5-2 & 1200 & WP \\
\enddata 

\tablecomments{All the above spectra were obtained with the 1.5\,m
Tillinghast telescope using the FAST spectrograph
(Fabricant et al. 1998). The grating used yields a resolution of
$\sim 7$\AA. The observed wavelength range roughly spans
3500-7400\,\AA. (See \S\ref{sect_obs_spec} for discussion).}

\tablenotetext{a}{UT Date and Julian Date at midpoint of observations.}
\tablenotetext{b}{Epoch of spectrum relative to the estimated explosion date.}
\tablenotetext{c}{Absolute difference between the observed position
angle and average parallactic angle over the course of the observation.} 
\tablenotetext{d}{Standard star pairs used for flux calibration: BD17
$=$ BD+17$^\circ$4708; BD26 $=$ BD+26$^\circ$2606; BD28 $=$
BD+28$^\circ$4211; BD33 $=$ BD+33$^\circ$2642; F34 $=$ Feige 34; HD84
$=$ HD 84937.} 
\tablenotetext{e}{Seeing is based upon estimates by the observers.} 
\tablenotetext{f}{Observers: EF $=$ E.~Falco; JG $=$ J.~Gallagher; JH
$=$ J.~Hernandez; MC $=$ M.~Calkins; PB $=$ P.~Berlind; RH $=$
R.~Hutchins; TG $=$ T.~Groner; WB $=$ W.~Brown; WP $=$ W.~Peters.}  
\tablenotetext{g}{Both spectra of SN~2005cs on UT 2005-07-04 were
combined into a single spectrum for plotting and modeling purposes.} 

\end{deluxetable*}
% \clearpage
% LocalWords:  Pastorello Tillinghast FLWO Fabricant IRAF CCD HeNeAr IDL SNe bp
% LocalWords:  telluric Matheson parallactic Filippenko Cheimets al stds BD EF
% LocalWords:  Exptime HD JH WB TG Feige Falco Calkins Berlind Groner

%%%%%%%%%%%%%%%%%%%%%%%%%%%%%%%
%
%
%

\section{Summary of observations and data reduction}

\label{sect_obs}

To sample the early photospheric phase with as many optical observations as possible, we use 
multiple data sources. % for SN 2005cs. For SN 2006bp, we use only CfA observations.
For both objects, {\it Swift} UVOT photometry is used. Spectra are less frequently observed than the photometry, typically once every few days, and this sets the limit to our analysis.
In the next sections, we review the ground-based optical photometry, the optical spectroscopy, the 
{\it Swift} UVOT data.

\subsection{Optical photometry}
\label{sect_obs_phot}

UBVr'i' photometry was obtained for SN 2005cs and SN 2006bp at the FLWO
1.2m telescope on Mt. Hopkins.  The Minicam instrument was used for SN
2005cs while the Keplercam was used for SN 2006bp (for details on cameras
and photometric reduction see Hicken, 2007 in preparation).  All data were
bias subtracted and flat fielded using standard procedures. PSF-fitting
photometry was performed using Dophot (Schechter et al. 1993).  
Host-galaxy subtraction was
performed for SN 2006bp.  The host-subtracted photometry was approximately
0.02 mag fainter than the photometry from the unsubtracted images, across
all bands.  A host-galaxy subtraction template was not available for SN
2005cs but the supernova was on a relatively faint and smooth background and
observations were taken while SN 2005cs was bright so the absence of host
subtraction is negligible.  Comparison stars for
each SN were calibrated on only one night.  However, our confidence in
their accuracy is high as other SNe that had been
observed on other photometric nights were observed on these nights, and the results are consistent.  The
comparison stars of these other SNe showed good agreement across all
nights (the standard deviation of a typical comparison star across these
nights being about 0.015 mag in V).
We present the photometry obtained for SN 2005cs in Table~\ref{tab_phot_05cs} and in
Fig.~\ref{fig_lc_obs} (left panel), and for SN 2006bp in Table~\ref{tab_phot_06bp} 
and in Fig.~\ref{fig_lc_obs} (right panel), where, for completeness, we 
have also included, where appropriate, the observations obtained by 
Pastorello et al. (2006; hereafter, P06) and Tsvetkov et al. (2006),
and the {\it Swift} observations for the three bluest filters.

% We also employ Super-LOTIS (SLOTIS) photometric data for SN 2006bp, recorded in Table~\ref{tab_phot_06bp}.
% The 0.6m Super-LOTIS telescope is located on Kitt Peak, Arizona, USA. 
% The system utilizes a Spectral Instruments CCD camera with a E2V 2048x2048 
% CCD. The system images in the $BVRI$ filters with a $17'\times 17'$ field of view. 
% All magnitudes were obtained with a $1.8''$ radius aperture and corrected to a 
% $5.4''$ radius aperture utilizing the PSF of nearby stars in the field.   
% All data were analyzed using IRAF DAOPHOT routines. 
% In \S\ref{sect_dist_06bp}, we use the $I$-band magnitude to constrain the
% distance to SN 2006bp, in combination with the CfA $B$ and $V$ magnitudes. 

\subsection{Optical spectroscopy}
\label{sect_obs_spec}

% NOTE: due to IRAF problems, I haven't been able to run the
% deconvolution method on the SN 2005cs spectra that suffer from
% galaxy contamination. This *will* be done, and the paragraphs below
% can easily be adjusted to reflect this.

Some of the spectra for SN~2005cs presented and analyzed in this paper have been
previously published by P06, while a 
subset for SN 2006bp have been presented by Quimby et al. (2007).
We refer the reader to those papers for more information on those spectra. In 
the analysis section of SN 2005cs, we indicate this source as P06, 
wherever appropriate, while in the analysis section of SN 2006bp, we
indicate this source as Quimby. 

All the other spectra were obtained with the 1.5\,m Tillinghast
telescope at the Fred Lawrence Whipple Observatory (FLWO) using the
low-dispersion FAST spectrograph
(Fabricant et al. 1998). Observational details of the spectra are
listed in Table~\ref{Table:speclog}. The FAST data were all reduced in
the same manner, using standard IRAF\footnote{IRAF is
distributed by the National Optical Astronomy Observatories,
operated by the Association of Universities for Research in
Astronomy, Inc., under contract to the National Science Foundation
of the United States.} routines for CCD processing and spectrum
extraction. All the spectra were extracted using the optimal algorithm
of Horne (1986). The spectra are then wavelength calibrated
using calibration-lamp spectra; the calibration is
later checked against night sky lines and adjusted accordingly. We used
our own IDL procedures for flux calibration. These include the removal
of telluric lines using the well-exposed continua of the
spectrophotometric standard stars (Wade \& Horne 1988; Matheson et al. 2000).

The relative spectrophotometry is good to about 5\%, when
done at the parallactic angle (Matheson et al. 2007, in prep.),
but no attempt was made to determine an absolute flux calibration. To
avoid second-order light contamination, we calibrated each
spectrum using two standard stars of different colors (see
Table~\ref{Table:speclog}). In general, the spectra were observed at the parallactic angle to minimize the effects of
atmospheric differential refraction (Filippenko 1982). When this
is not the case, the spectra can suffer from a relative flux depletion
at bluer wavelengths, especially when observed at high airmass ($\sec z
> 1.1$, where $z$ is the zenith angle). In Table~\ref{Table:speclog}
we list the angle in degrees, $\Delta {\rm PA}$, between the
observed position angle and the parallactic angle for each
spectrum, along with the airmass. When $\Delta {\rm PA} > 10^{\circ}$,
the relative spectrophotometry is less accurate.

Around maximum light, both SNe~2005cs and 2006bp were bright with
respect to their host galaxy. Nevertheless, the earliest (June 30$^{\rm
th}$\,UT) and latest (Dec 25$^{\rm th}$ UT) FAST spectra of SN~2005cs
%%% Luc, I'm not even sure whether you're using the 25 Dec spectrum..
have substantial galaxy background, and the relative flux calibration is
affected by this contamination.
%%% This may change once I run the deconvolution tool.

Finally, all SN 2005cs spectra presented here have been de-redshifted, 
adopting our own measurement of 466$\pm$35\,\kms of the recession velocity 
of NGC\,5194, while for SN 2006bp and NGC 3953, we adopt 1280\,\kms 
(Quimby et al. 2007; Verheijen \& Sancisi 2001;
a comparable value of 1053\,\kms is given by Ho et al. 1997).

\subsection{{\sl Swift} UVOT photometry}
\label{swift_obs}

{\it Swift} UVOT photometry allows us to track the ultraviolet color evolution
of SNe from a few days to a few weeks after shock breakout.
   Observations of both SN 2005cs and SN 2006bp were obtained with {\it Swift} UVOT
(Gehrels et al. 2004; Roming et al. 2005; Poole et al. 2007)
at multiple epochs covering from just hours after discovery until the UV light had faded to
the background emission level of the SN environment. 
The Swift UVOT data for SNe 2005cs and 2006bp, originally presented in Brown
et al. (2007) and Immler et al. (2007), have been reanalyzed using the improved
calibration of Poole et al. (2007). The magnitudes in Table~\ref{table_swift_05cs}
and in Table~\ref{table_swift_06bp} are in the UVOT
system, distinguished by lower case letters.  The UVOT $u$, $b$, and $v$ are close to 
the Johnson $UBV$; however, the UVOT $u$ band is bluer than the Johnson $U$ so there
can be significant differences at early times when the SNe are blue. The $b$
and $v$ agree well with their Johnson counterparts.  
The magnitudes and flux densities were measured
using a $5''$ aperture source region, corresponding to that used to calibrate the
zeropoints and coincidence loss corrections, using $uvotmaghist$. Here, we subtract
the background light by means of late-time template images, selecting a
background region with the same average count rate per pixel as in the $5''$
source region.  When the background subtracted count rate drops below that 
of the background (as measured in the template images)  we report the
magnitude of the background as an upper limit.  The errors reported include
the systematic errors in the zeropoints.
% , with an additional systematic error
% of 10\% in the flux conversion added in quadrature with the statistical error
% due to the spectral shape being different (and varying) from that used to
% compute the conversion factors.  
% These observations were presented and described qualitatively in Brown et al. (2007) 
% for SN 2005cs, and Immler et al. (2007) for SN 2006bp, and we refer the reader to these
% papers for tables and figures depicting these data.

   We use the flux equivalent of the {\it Swift} UVOT magnitudes given 
in Table~\ref{table_swift_05cs} \& \ref{table_swift_06bp}, accounting for the different (and varying) energy distribution of the supernova by adding an additional systematic
error of 10\% in quadrature with the statistical error.  
We plot these fluxes at the effective wavelengths of the UVOT filters along 
with appropriately reddened synthetic flux distributions computed with
the non-LTE model atmosphere code CMFGEN (see \S\,\ref{sect_mod_pres}, \S\,\ref{sect_05cs},
and \S\,\ref{sect_06bp}).
{\it Swift} UVOT filters $uvw2$, $uvm2$, $uvw1$, $u$, $b$, and $v$
have their effective wavelength at 
2030\AA, 2231\AA, 2634\AA, 3501\AA, 4329\AA, and 5402\AA\ (Poole et al. 2007).

% Note that for SN 2005cs, a UV flux contribution from the background H{\,\sc ii} region 
% and measured at the SN location three months after explosion at a 
% $\sim$10$^{-15}$erg\,cm$^{-2}$\,s$^{-1}$\,\AA$^{-1}$ level is subtracted from all {\it Swift} 
% UVOT magnitudes (Brown et al. 2007).
% \clearpage
\begin{deluxetable*}{cccccccc}
\tablewidth{16cm} 
\tabletypesize{\scriptsize}
\tablecaption{{\it Swift} UVOT Photometry of SN 2005\lowercase{cs}\label{table_swift_05cs}}
\tablehead{\colhead{UT Date } & \colhead{JD} & \colhead{$uvw2$} & \colhead{$uvm2$}
& \colhead{$uvw1$} & \colhead{$u$} & \colhead{$b$} & \colhead{$v$} \\ 
\colhead{} & \colhead{} & \colhead{} & \colhead{} & \colhead{} & \colhead{} &
\colhead{} & \colhead{} } 
\startdata
2005-06-30 & 2453552.4 & 13.07 $\pm$ 0.03 & 12.64 $\pm$ 0.03 & 12.78 $\pm$
0.03 & \nodata & \nodata & 14.63 $\pm$ 0.03 \\
2005-07-03 & 2453555.1 & 14.21 $\pm$ 0.03 & \nodata & \nodata & \nodata &
\nodata & 14.63 $\pm$ 0.06 \\
2005-07-05 & 2453557.3 & 15.05 $\pm$ 0.04 & 14.87 $\pm$ 0.04 & 14.39 $\pm$
0.04 & 13.64 $\pm$ 0.03 & 14.75 $\pm$ 0.03 & 14.68 $\pm$ 0.03 \\
2005-07-06 & 2453557.9 & 15.26 $\pm$ 0.04 & 15.15 $\pm$ 0.04 & 14.60 $\pm$
0.04 & 13.75 $\pm$ 0.03 & 14.76 $\pm$ 0.03 & 14.66 $\pm$ 0.03 \\
2005-07-07 & 2453558.8 & $>$15.7 & $>$15.7 & 14.98 $\pm$ 0.04 & 13.85 $\pm$
0.03 & 14.76 $\pm$ 0.02 & 14.64 $\pm$ 0.03 \\
2005-07-08 & 2453560.0 & $>$15.7 & $>$15.7 & 15.63 $\pm$ 0.05 & 14.14 $\pm$
0.03 & 14.78 $\pm$ 0.03 & 14.59 $\pm$ 0.03 \\
2005-07-09 & 2453560.6 & $>$15.7 & $>$15.7 & $>$15.7 & 14.32 $\pm$ 0.04 &
14.78 $\pm$ 0.04 & 14.58 $\pm$ 0.04 \\
2005-07-10 & 2453562.0 & $>$15.7 & $>$15.7 & $>$15.7 & 14.67 $\pm$ 0.04 &
14.87 $\pm$ 0.05 & 14.63 $\pm$ 0.05 \\
2005-07-11 & 2453562.8 & $>$15.7 & $>$15.7 & $>$15.7 & 14.92 $\pm$ 0.05 &
14.98 $\pm$ 0.05 & 14.70 $\pm$ 0.05 \\
2005-07-13 & 2453564.5 & $>$15.7 & $>$15.7 & $>$15.7 & \nodata & \nodata &
14.67 $\pm$ 0.05 \\
2005-07-14 & 2453565.5 & $>$15.7 & $>$15.7 & $>$15.7 & 15.43 $\pm$ 0.05 &
15.21 $\pm$ 0.04 & 14.74 $\pm$ 0.04 \\
2005-07-17 & 2453568.5 & $>$15.7 & $>$15.7 & $>$15.7 & $>$15.8 & 15.48 $\pm$
0.05 & 14.76 $\pm$ 0.05 \\
2005-07-22 & 2453573.5 & $>$15.7 & $>$15.7 & $>$15.7 & $>$15.8 & 15.62 $\pm$
0.06 & 14.90 $\pm$ 0.05 \\
2005-09-14 & 2453628.5 & $>$15.7 & $>$15.7 & $>$15.7 & $>$15.8 & 16.33 $\pm$
0.06 & 14.95 $\pm$ 0.04 \\
\enddata
\end{deluxetable*}

\begin{deluxetable*}{cccccccc}
\tablewidth{16cm} 
\tabletypesize{\scriptsize}
\tablecaption{{\it Swift} UVOT Photometry of SN 2006\lowercase{bp}\label{table_swift_06bp}}
\tablehead{\colhead{UT Date} & \colhead{JD} & \colhead{$uvw2$} & \colhead{$uvm2$}
& \colhead{$uvw1$} & \colhead{$u$} & \colhead{$b$} & \colhead{$v$} \\ 
\colhead{} & \colhead{} & \colhead{} & \colhead{} & \colhead{} & \colhead{} &
\colhead{} & \colhead{} } 
\startdata
2006-04-10 & 2453836.0 & 14.49 $\pm$ 0.03 & 14.74 $\pm$ 0.03 & 14.62 $\pm$
0.03 & 14.61 $\pm$ 0.02 & 15.75 $\pm$ 0.02 & 15.63 $\pm$ 0.03 \\
2006-04-11 & 2453836.6 & 14.57 $\pm$ 0.04 & 14.83 $\pm$ 0.05 & 14.66 $\pm$
0.04 & 14.60 $\pm$ 0.04 & 15.60 $\pm$ 0.04 & 15.52 $\pm$ 0.05 \\
2006-04-11 & 2453836.9 & 14.62 $\pm$ 0.03 & 14.83 $\pm$ 0.04 & 14.63 $\pm$
0.03 & 14.55 $\pm$ 0.03 & 15.59 $\pm$ 0.03 & 15.52 $\pm$ 0.03 \\
2006-04-11 & 2453837.1 & 14.68 $\pm$ 0.03 & 14.83 $\pm$ 0.04 & 14.64 $\pm$
0.04 & 14.58 $\pm$ 0.03 & 15.53 $\pm$ 0.03 & 15.44 $\pm$ 0.04 \\
2006-04-12 & 2453837.9 & 14.95 $\pm$ 0.03 & 14.95 $\pm$ 0.04 & 14.68 $\pm$
0.03 & 14.49 $\pm$ 0.03 & 15.52 $\pm$ 0.02 & 15.31 $\pm$ 0.03 \\
2006-04-13 & 2453838.7 & 15.16 $\pm$ 0.03 & 15.05 $\pm$ 0.03 & 14.74 $\pm$
0.03 & 14.51 $\pm$ 0.02 & 15.47 $\pm$ 0.02 & 15.25 $\pm$ 0.03 \\
2006-04-13 & 2453839.0 & 15.27 $\pm$ 0.03 & 15.17 $\pm$ 0.04 & 14.80 $\pm$
0.03 & 14.52 $\pm$ 0.03 & 15.43 $\pm$ 0.02 & 15.18 $\pm$ 0.03 \\
2006-04-15 & 2453840.8 & 15.77 $\pm$ 0.03 & 15.53 $\pm$ 0.04 & 15.08 $\pm$
0.03 & 14.58 $\pm$ 0.03 & 15.41 $\pm$ 0.02 & 15.20 $\pm$ 0.03 \\
2006-04-15 & 2453841.1 & 15.87 $\pm$ 0.04 & 15.69 $\pm$ 0.05 & 15.14 $\pm$
0.04 & 14.60 $\pm$ 0.03 & 15.47 $\pm$ 0.03 & 15.13 $\pm$ 0.03 \\
2006-04-17 & 2453842.7 & 16.23 $\pm$ 0.04 & 16.04 $\pm$ 0.04 & 15.35 $\pm$
0.04 & 14.64 $\pm$ 0.03 & 15.42 $\pm$ 0.02 & 15.14 $\pm$ 0.03 \\
2006-04-17 & 2453842.9 & 16.30 $\pm$ 0.05 & 16.22 $\pm$ 0.05 & 15.40 $\pm$
0.05 & 14.66 $\pm$ 0.04 & 15.41 $\pm$ 0.04 & 15.19 $\pm$ 0.05 \\
2006-04-17 & 2453843.3 & 16.37 $\pm$ 0.04 & \nodata & 15.42 $\pm$ 0.04 & 14.67
$\pm$ 0.03 & 15.41 $\pm$ 0.03 & 15.19 $\pm$ 0.04 \\
2006-04-20 & 2453846.1 & 17.01 $\pm$ 0.04 & 16.91 $\pm$ 0.05 & 15.91 $\pm$
0.04 & 14.94 $\pm$ 0.03 & 15.50 $\pm$ 0.03 & 15.16 $\pm$ 0.03 \\
2006-04-22 & 2453847.6 & 17.30 $\pm$ 0.05 & 17.34 $\pm$ 0.07 & 16.17 $\pm$
0.04 & 15.00 $\pm$ 0.03 & 15.56 $\pm$ 0.02 & 15.19 $\pm$ 0.03 \\
2006-04-24 & 2453849.6 & 17.73 $\pm$ 0.08 & 18.13 $\pm$ 0.10 & 16.65 $\pm$
0.05 & 15.18 $\pm$ 0.03 & 15.60 $\pm$ 0.03 & 15.27 $\pm$ 0.03 \\
2006-04-26 & 2453851.6 & 18.22 $\pm$ 0.09 & $>$18.8 & 17.07 $\pm$ 0.07 & 15.46
$\pm$ 0.03 & 15.65 $\pm$ 0.03 & 15.23 $\pm$ 0.03 \\
2006-04-28 & 2453853.5 & 18.78 $\pm$ 0.10 & $>$18.8 & 17.54 $\pm$ 0.07 & 15.83
$\pm$ 0.03 & 15.76 $\pm$ 0.02 & 15.23 $\pm$ 0.03 \\
2006-05-01 & 2453857.0 & $>$18.8 & $>$18.8 & 18.29 $\pm$ 0.07 & 16.31 $\pm$
0.03 & 15.91 $\pm$ 0.02 & 15.22 $\pm$ 0.02 \\
2006-05-04 & 2453859.7 & \nodata & \nodata & 18.48 $\pm$ 0.11 & 16.64 $\pm$
0.04 & 15.98 $\pm$ 0.02 & 15.22 $\pm$ 0.03 \\
2006-05-16 & 2453871.8 & \nodata & \nodata & $>$18.6 & 17.43 $\pm$ 0.07 &
\nodata & 15.38 $\pm$ 0.02 \\
2006-05-26 & 2453882.0 & \nodata & \nodata & $>$18.6 & 17.91 $\pm$ 0.09 &
16.64 $\pm$ 0.04 & 15.50 $\pm$ 0.03 \\
2006-05-30 & 2453885.9 & \nodata & \nodata & $>$18.6 & $>$18.2 & 16.72 $\pm$
0.02 & 15.56 $\pm$ 0.02 \\
\enddata
\end{deluxetable*}
% \clearpage

\section{Theoretical Methodology}
\label{sect_mod_pres}

    The analysis of SNe 2005cs and SN 2006bp is analogous to that presented in DH06, which analyzed SN 1999em at 8 epochs in the first 
45 days after explosion. That paper determined its
distance by means of various EPM-based approaches (Kirshner \& Kwan 1974; Eastman \& Kirshner 1989; 
Schmidt et al. 1994; Eastman et al. 1996; Hamuy et al. 2001; Leonard et al. 2002a; 
Elmhamdi et al. 2003; Dessart \& Hillier 2005b) and another method akin to the SEAM 
(Baron et al. 2000). 
Here, we have two advantages: {\sl Swift} observations monitor the UV light and we have more densely sampled data. {\sl Swift} provides additional constraints on the reddening, the temperature, and the ionization state of the ejecta at early epochs. Our analysis
covers 13 epochs in ($\sim$30 days) for SN 2005cs and 18 epochs in ($\sim$45 days) for SN 2006bp, providing better tracking of the spectral and photometric evolution of each.

Time-dependent terms in the statistical- and 
radiative-equilibrium equations become important a few weeks after explosion 
(Utrobin \& Chugai 2005; Dessart \& Hillier 2007a,b).  We avoid that complication 
by focusing on the early-time photospheric phase. We know that assuming a steady state in the statistical
and radiative equilibrium equations produces a prediction of the 
ejecta ionization state that is somewhat below its true level,
that He{\,\sc i} lines persist over a longer period than we predict, and that strong Balmer lines at late times are a clear sign of the epoch when this assumption is no longer valid.

CMFGEN solves the radiative transfer equation in the comoving frame, subject
to the constraints of radiative and statistical equilibrium (Hillier \& Miller 1998).
The ejecta are assumed to be spherically symmetric and chemically homogeneous, a good approximation 
given the very spatially confined regions of continuum and line formation (Dessart \& Hillier 2005a), 
and low polarization observed in the ejecta of photospheric phase Type II SN ejecta (Leonard \& Filippenko 2001).
The expansion becomes truly homologous about a week after the explosion, but, for simplicity, is assumed homologous at
all times. We adopt a density distribution characterized by a base density $\rho_0$ (at the base radius $R_0$) 
and a density exponent $n$, with $\rho(R) = \rho_0 (R_0/R)^n$.
The value of $\rho_0$ is chosen so that the Rosseland-mean optical depth $\tau_{\rm Rosseland}$ 
at $R_0$ is $\sim$50. The hydrogen-rich ejecta are always ionized at least once at the base and electron scattering is the primary source of continuum
opacity, while the large H{\,\sc i} bound-free cross sections ensure the thermalization of the radiation 
below the photosphere (see \S4.2 and Figs\,7-8-9 in Dessart \& Hillier 2005b).
The value of $n$ is usually quite large, i.e., $\sim$10, but could be much larger at early times, and can be constrained from line profile shapes, 
through the magnitude of the blueshift of peak emission of optical P-Cygni 
profiles (Dessart \& Hillier 2005a; Blondin et al. 2006; Aldering et al. 2005), or through the strength
of P-Cygni profiles with respect to the continuum.
For example, lower values of n lead to P-Cygni profiles having stronger and broader 
absorption and emission components, and with a peak emission that is closer to line center. 
The early-time spectra of SNe 2005cs and 2006bp, show 
weak P-Cygni profiles with a strongly blueshifted peak emission, and point towards 
high values of n.
We quote properties of models at the photosphere, which corresponds to the location
where the continuum optical depth, integrated inwards from the outer grid radius, is 2/3.

The luminosity at depth is not
known in advance unless one uses a consistent hydrodynamical input of the ejecta properties,
following simultaneously the time-dependent evolution of the radiation field, internal energy, and 
level populations.
Instead, we adjust the base (comoving-frame) luminosity so that the synthetic 
emergent (observer-frame) luminosity matches approximately the observed luminosity at each epoch,
for an adopted distance and reddening. 
Usually, with our inferred distance and reddening, the emergent reddened synthetic flux is
within a factor of 3 of the observed flux.
We find that radial/luminosity scaling of that magnitude, preserving the ratio $L/R^2$ produce the same color and relative line strengths, line to continuum
flux ratios, and other essential features of the spectrum. (see Dessart \& Hillier 2005a for a discussion).

Reddening is constrained using the Cardelli et al. (1988) law for both SNe, assuming $R=3.1$.
We adopt a single value of $E(B-V)$ to obtain
satisfactory fits to the observed spectra at all epochs. This works best when
the relative flux calibration is accurate. For SN 2005cs, we find that the
reddening is very small, i.e., $E(B-V) \sim 0.04$ (Baron et al. 2007 infer a value of 0.035 from
a quantitative spectroscopic analysis with PHOENIX; Hauschildt \& Baron 1999), 
while for SN 2006bp, it is quite large, i.e.,
$E(B-V) \sim 0.4$ (Ho et al. 1997 infer a value of 0.4 to the galaxy host NGC\,3953
from the H$\alpha$/H$\beta$ ratio and assuming Case B recombination).
{\sl Swift} UVOT observations allow more leverage on the estimate of reddening than is
permitted by optical observations alone by lengthening blueward the observed wavelength 
range.
% \clearpage
\begin{deluxetable*}{cccccrcccc}
%\rotate
\tablewidth{16cm}
\tabletypesize{\scriptsize}
\tablecaption{Model Characteristics for SN2005{\lowercase{cs}}.
\label{tab_model_05cs}}
\tablehead{
\colhead{JD}&
\colhead{Day}&
\colhead{$\Delta t ^a$}&
\colhead{$L_{{\rm CMF},R_0}$}& 
\colhead{$L_{{\rm OBS},R_{\rm Max}}$}& 
\colhead{$T_{\rm phot}$}& 
\colhead{$R_{\rm phot}$}&
\colhead{$v_{\rm phot}$}&
\colhead{$\rho_{\rm phot}$}&
\colhead{$n$}
\\
\colhead{}&      
\colhead{}&      
\colhead{days}&      
\multicolumn{2}{c}{(10$^8$ $L_{\odot}$)}& 
\colhead{(K)}&    
\colhead{(10$^{14}$\,cm)}&  
\colhead{(km\,s$^{-1}$)}&    
\colhead{(10$^{-14}$\,g\,cm$^{-3}$)}& 
\colhead{}
}
\startdata
%                    t - tdiscovery  L_CMF_BASE   L_OBS_RMAX      T_phot        R_phot     V_phot     rho_phot N_RHO
2453552.25   &  2005-06-30  &     1.85 &  3.0 &   2.69  &    15750   &    2.00    &  6880  &    24.6 &  20   \\ % cs0_1_l1b1_rho_v1_B
2453553.25   &  2005-07-01  &     2.85 &  2.3 &   1.99  &    13350   &    2.55    &  6950  &    10.8 &  12   \\ % cs0_1_l1b1_rho3_B_v1  
2453554.50   &  2005-07-02  &     4.10 &  2.3 &   2.01  &    13420   &    2.54    &  6370  &    10.6 &  12   \\ % cs0_1_l1b1_rho3_B
2453555.25   &  2005-07-03  &     4.85 &  2.0 &   1.74  &    10850   &    3.26    &  6080  &    7.6  &  10   \\ % cs0_1_l1b_B1_v1
2453556.00   &  2005-07-04  &     5.60 &  1.8 &   1.64  &     9300   &    3.91    &  5450  &    6.8  &  10   \\ % cs0_1_l1b_B1_v1_2
2453557.75   &  2005-07-05  &     7.35 &  1.8 &   1.56  &     8620   &    4.93    &  4780  &    5.6  &  10   \\ % cs0_1_l1b_B1_v1_5
2453558.25   &  2005-07-06  &     7.85 &  1.5 &   0.97  &     8250   &    4.18    &  5230  &    6.0  &  10   \\ % cs1_7
2453561.25   &  2005-07-09  &    10.85 &  1.5 &   1.21  &     7770   &    5.97    &  4710  &    4.7  &  10   \\ % cs3_3
2453562.50   &  2005-07-10  &    12.10 &  1.5 &   1.19  &     7120   &    6.38    &  4580  &    6.0  &  10   \\ % cs3_5
2453563.50   &  2005-07-11  &    13.10 &  1.5 &   1.10  &     7300   &    5.88    &  4440  &    8.9  &  10   \\ % r3_e_abund
2453564.25   &  2005-07-12  &    13.85 &  1.5 &   1.08  &     6980   &    6.01    &  4320  &   10.9  &  10   \\ % r3_abund
2453566.00   &  2005-07-14  &    15.60 &  1.1 &   0.79  &     6700   &    5.47    &  3930  &   28.2  &  10   \\ % r3_g_abund
2453580.25   &  2005-07-28  &    29.85 &  1.0 &   1.17  &     6140   &    7.84    &  2240  &  125.6  &  10   \\ % r4_b                  
\enddata
\tablecomments{
For each date in our sample of observations, we provide the following 
CMFGEN model parameters: Base comoving-frame luminosity $L_{{\rm CMF},R_0}$ and 
emergent observer-frame luminosity $L_{{\rm OBS},R_{\rm Max}}$ (in \lsun; 
 note that for the model on the last date, the observer-frame luminosity has not fully relaxed to the base condition,
a problem that stems from the presence of a steep and hard-to-model recombination front at the photosphere), 
photospheric conditions describing
the electron temperature $T_{\rm phot}$ (in K), the radius $R_{\rm phot}$ (in 10$^{14}$\,cm), 
the velocity $v_{\rm phot}$ (in \kms), and the density (in 10$^{-14}$\,g\,cm$^{-3}$), 
together with the density exponent $n$ characterizing the density law 
$\rho(r) = \rho_{\rm phot} (R_{\rm phot}/r)^n$.   
(See \S\ref{sect_mod_pres} for a presentation of the modeling approach).
 $^a$: $\Delta t$ is the time elapsed since discovery.
}
\end{deluxetable*} 
% \clearpage
%
% Chemistry for all models unless otherwise specified (all metals are at solar metallicity
% unless otherwise specified):
% H/He = 5, C/He=0.0004, N/He=0.0013, O/He = 0.0016 
% MODELS r3_e_abund, r3__abund, r3_g_abund, and r4_b have even higher CO abundances. Models pending.. 
% HIGH CO abundance models have: C/He=0.0017, N/He=0.0068, O/He = 0.01
 
% SN 2005cs
% date    tphot     rphot/rcore n   rho/r3rho  rcore    vcore       modelname
% 0630    15750.    1.249    20.    86.0    1.6000D4    5500.0D0    cs0_1_l1b1_rho_v1_B
% 0701    13420.    1.420    12.    41.6    1.8000D4    4500.0D0    cs0_1_l1b1_rho3_B
% 0702    13420.    1.420    12.    41.6    1.8000D4    4500.0D0    cs0_1_l1b1_rho3_B
% 0703    11180.    1.530    10.    41.6    2.0000D4    4500.0D0    cs0_1_l1b_B
% 0704     9300     1.557    10    89.8     2.5100D4    3500.0D0    cs0_1_l1b_B1_v1_2
% 0705     8620     1.493    10    110.6    3.3000D4    3200.0D0    cs0_1_l1b_B1_v1_5
% 0706    8250.     1.743    10.   230.0    2.4000D4    3000.0D0    cs1_7
% 0709    7770.     1.568    10.0  230.     3.8000D4    3000.0D0    cs3_3
% 0710    7100.     1.530    10.   4.2D-12  4.1760D4    3000.0D0    cs3_5
% 0711    7000.     1.492    10.0  4.2D-12  4.1760D4    3000.0D0    r3
% 0712    7000.     1.492    10.0  4.2D-12  4.1760D4    3000.0D0    r3
 
% \clearpage
\begin{figure}
\plotone{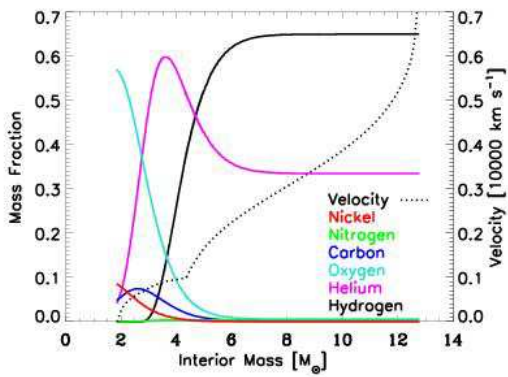}
% \plotone{s15_WH07.ps}
\caption{Illustration of the elemental mass fraction distribution versus interior mass for the most abundant species
in a ``reference'' 15\,\mo model from Woosley \& Heger (2007; a 1.83\,\mo neutron star is left 
behind in this explosion). The color coding differentiates species: Hydrogen (black),
helium (violet), carbon (blue), oxygen (turquoise), nitrogen (green), and nickel (red).
We also overplot the velocity (dotted black line), in units of 10,000\,\kms, to permit the inference of the
elemental stratification in {\it velocity} space.
The surface mass fractions are $X_{\rm H}$=0.65, $X_{\rm He}$=0.34, $X_{\rm C}$=0.0014, $X_{\rm N}$=0.003, 
$X_{\rm O}$=0.005, and with the adopted mixing, $X_{\rm ^{56}Ni}$ = 1.1 $\times$ 10$^{-9}$.
[See the electronic edition of the Journal for a color version of this figure, 
and see \S\ref{sect_mod_pres} for discussion.]
}
\label{fig_s15}
\end{figure}
% \clearpage

To guide our reasoning when interpreting Type II SN spectra, in Fig.~\ref{fig_s15},
we illustrate the chemical stratification and velocity distribution 
versus interior mass for a stellar model that is characteristic of the ejecta of Type II-P SNe, resulting from the explosion of a red supergiant (RSG) progenitor.  The figure shows the ejecta of a 15\,\mo (on the main sequence) progenitor star model 
(ignoring rotation), exploded with a kinetic energy of $\sim$1\,B ($\equiv$10$^{51}$erg) 
and with modest (artificial) nickel mixing 
(Woosley \& Heger 2007)\footnote{For the CMFGEN spectroscopic modeling that we present here, 
we do not use this ejecta structure at any time.}.

%\footnote{Note that we do 
%not claim here that such an ejecta structure matches that of SN 2005cs or SN 2006bp.
%We merely wish to guide our modeling with what represents a conservative prediction for the
%structure of a Type II-P SN, diversity in Nature stemming from different progenitor masses,
%explosion energy, rotation, multi-dimensional effects, etc.}.
At core collapse, this star has a mass of $\sim$13\,\mo ($\sim$2\,\mo have been lost through
a wind during the pre-SN evolution), 1.83\,\mo is left behind in
a protoneutron star and the rest escapes to form the SN ejecta.
Starting from the surface, there is $\sim$7\,\mo of essentially homogeneous
material, made up primarily of hydrogen ($X_{\rm H} = 0.65$) and helium ($X_{\rm He} = 0.34$) -- just slightly 
evolved from the primordial, solar metallicity, mixture (Grevesse \& Sauval 1998).
The velocity at the base of this hydrogen shell is about 2000\,\kms.
Just below is a $\sim$3\,\mo shell dominated by helium ($X_{\rm He} \sim 0.6$) and oxygen ($X_{\rm O} \sim 0.15$),
between velocities 1000 to 2000\,\kms.
Finally, just above the inner mass cut at 1.83\,\mo 
resides an oxygen-rich shell ($X_{\rm O} \sim 0.6$), with a $\sim$0.1 mass fraction of helium, carbon, and
nickel. The radioactive nickel is deep inside, at velocities below 2000\,\kms, and
separated by $\sim$10\,\mo of material from the photosphere at breakout. 

In such (1D) radiation-hydrodynamics simulations, explosion parameters are adjusted to 
obtain the right amount of nickel production and mixing, which influence the light curve shape or 
the potential appearance of lines from radioactive isotopes at the photosphere (from $^{56}$Co, for example). 
For the peculiar Type II SN 1987A, this required allowing mixing out to $\sim$4000\,\kms 
(Pinto \& Woosley 1988; Kumagai et al. 1989; Blinnikov et al. 2000), while
for the Type II-P SN 1999em, nickel mixing out to $\sim$650\,\kms seems optimal (Utrobin 2007).  More detailed 2D hydrodynamical
simulations of core-collapse supernovae explosions, assuming the energy demands are fulfilled,
do not naturally predict nickel at large velocities, placing it at $\sim$2500\,\kms 
(Hachisu et al. 1990, 1991; Kifonidis et al. 2000, 2003).
Observationally, there is a range of inferred values and 
a conservative assumption for Type II-P SNe
may be $\sim$1000-2000\,\kms, with 4000\,\kms representing an extreme circumstance. In the above 15\,\mo model, $^{56}$Ni 
is buried under about 10\,\mo of material, mostly composed of hydrogen and helium,
so that any non-thermal excitation and ionization of the photosphere by radioactive isotopes is delayed until 
the end of the Plateau phase, when the photosphere has receded to these deep and 
slow-moving ejecta layers. In our analysis of the early photospheric phase of Type II-P SN, we neglect
any contribution from radioactive decay.
We expect slow and small variations in chemistry during the plateau phase, and, thus, 
adopt chemical homogeneity at a given epoch.  This is consistent with our finding 
that C and O are only modestly enhanced at the photosphere one month after explosion.
Finally, in such explosions, the density profile follows closely a power law distribution with exponent 
$\sim$10 at depth in the hydrogen envelope, but is characterized by a much steeper drop-off near 
the surface layers, that may result from the strong radiation pressure in these layers 
at shock breakout (Blinnikov et al. 2000, Shigeyama \& Nomoto 1990). Such a steepening is not well-modeled by Lagrangean simulations, in particular for the shock-breakout phase, but our interpretation of our earliest observations supports this trend.

We adopt the surface composition of blue supergiants analyzed
by Crowther et al. (2006), which is in very close agreement with the 
outer composition of the 15\,\mo model described above.
By number, we adopt H/He = 5, C/He=0.0004, N/He=0.0013, O/He = 0.0016, with all metals taken
at their solar value (Grevesse \& Sauval 1998). Given as mass fractions, these abundances are:
$X_{\rm H}$=0.55, $X_{\rm He}$=0.44, $X_{\rm C}$=0.0005, $X_{\rm N}$=0.002, $X_{\rm O}$=0.003,
and $X_{\rm Fe}$=0.0013.
At later times, when in particular O{\,\sc i}\,7770\AA\ gains in strength, we need to enhance both the
carbon and the oxygen abundances, with C/He=0.001, O/He = 0.01, and to reduce the nitrogen 
abundance, N/He=0.001 to match the data.

A key asset of CMFGEN is the explicit treatment of line blanketing using a detailed
description of the atomic structure for a large number of species. The iterative march to
solving many non-LTE level populations together with the radiation field comes at great CPU expense. We optimize the problem by including only the relevant ions of the 
most important species, although the ejecta ionization stratification still requires
that at least three ionization stages of iron, for example, be included to capture
the sources of bound-bound and bound-free opacity at all depths.
Typically, any model crafted for very early conditions ( to match the first few, nearly featureless spectra of 
SNe 2005cs and 2006bp, with weak He{\,\sc i}\,5875\AA) will contain the following species and 
levels (the suffixed parenthesis 
contains the number of full- and super-levels; see Hillier \& Miller 1998 for details):
H{\,\sc i} (30,20), He{\,\sc i}(51,40), He{\,\sc ii} (30,13), C{\,\sc ii} (26,14), C{\,\sc iii} (112, 62),
C{\,\sc iv} (64,59), N{\,\sc ii} (3,3), N{\,\sc iii} (26,26), N{\,\sc iv} (60,34), N{\,\sc v} (67,45), 
O{\,\sc ii} (3,3), O{\,\sc iii} (115,79), O{\,\sc iv} (72,53), O{\,\sc v} (152,73), Si{\,\sc iv} (48,37), 
Si{\,\sc v} (71,33), Fe{\,\sc ii} (309,116), Fe{\,\sc iii} (563, 191), Fe{\,\sc iv} (787,120), 
Fe{\,\sc v} (191,47), Fe{\,\sc vi} (433,44), Ni{\,\sc iii} (67,15), and Ni{\,\sc iv} (242,36).
At one week after explosion, when He{\,\sc i}\,5875\AA\ is strong, we take out 
N{\,\sc iv}, O{\,\sc v}, Si{\,\sc iv}, Si{\,\sc v}, Fe{\,\sc vi}, and Ni{\,\sc iv}, and add 
N{\,\sc i} (104,54), O{\,\sc i} (75,23), Mg{\,\sc ii} (65,22), Ca{\,\sc ii} (77,21), 
Si{\,\sc ii} (59, 31), Ti{\,\sc ii} (152,37), Ti{\,\sc iii} (206,33), 
Ni{\,\sc ii} (93,19), and Ni{\,\sc iii} (67,15).
When hydrogen starts recombining, we put more emphasis on low ionization states and adopt 
the following model atoms:
H{\,\sc i} (30,20), He{\,\sc i}(51,40),  C{\,\sc i} (63,33),  C{\,\sc ii} (59,32), N{\,\sc i} (104,54), 
Na{\,\sc i} (71,22), Mg{\,\sc ii} (65,22), Si{\,\sc ii} (59, 31), O{\,\sc i} (75,23), 
Ca{\,\sc ii} (77,21), Fe{\,\sc i} (136,44), Fe{\,\sc ii} (309,116), Fe{\,\sc iii} (477,61), 
Ni{\,\sc ii} (93,19), and Ti{\,\sc ii} (152,37).
References to the corresponding atomic data sources are given in Dessart \& Hillier (2005a).

To summarize, inputs to a CMFGEN model are global characteristics of the ejecta such as base 
(comoving-frame) luminosity, radius, chemical composition, velocity and density. 
These parameters are iterated (see Dessart \& Hillier 2005a 
for details) until the synthetic spectrum matches the observations. Assessment of suitability is done by eye. 
Spectroscopic observations used here are accurate in relative flux,
while only accurate to within 10--20\% in absolute flux. For spectral fitting, 
relative flux calibration is essential
while absolute flux calibration is not.
When inferring distances, we use the photometry to set the absolute flux scale.
One important result of our study
is that the CMFGEN model parameters needed to fit the optical data of SNe 2005cs and 2006bp during the 
first 2-3 weeks after explosion also naturally explain the {\sl Swift} UVOT data {\it simultaneously}. There appears to be a common, photospheric, origin for the UV and optical light for both objects: circumstellar interaction contributes very little to the observed light in these spectral
regions (see also Brown et al. 2007). Note that SN 2006bp was, however, detected in the X-rays 
for up to 12 days after explosion, suggesting that interaction between the ejecta and the 
circumstellar medium (CSM) does occur (Immler et al. 2007), but without any sizable effect
on the UV and the optical flux.

The first month after explosion, when the photosphere resides
in the ionized hydrogen-rich layers of the ejecta, is not the best epoch to probe the innermost details of the explosion mechanism. However, it is ideal for distance determinations based on variants of the 
Expanding Photosphere Method:
A true continuum exists and the effects of line-blanketing are weak in the optical;
the properties of the ejecta are well determined by the model, using both the continuum
energy distribution and lines from the most abundant ions.
The homogeneous hydrogen-rich envelope ensures that a well defined photosphere exists and 
that radiation thermalizes at depth, allowing the adequate use of the diffusion
approximation at the model base. There, our non-LTE approach naturally predicts LTE 
conditions, the departure coefficients of all levels tending to unity, and at the thermalization
depth, the mean intensity is well represented by a blackbody (LTE is, in other words, never enforced). 
The analysis presented here allows the inference of distances to Type II-P SNe
(see \S\ref{sect_dist_05cs} and \S\ref{sect_dist_06bp}).

To be more concise, we present all model parameters in Table~\ref{tab_model_05cs} and \ref{tab_model_06bp},
but mention in the text only the parameters that are most directly related to the spectroscopic evolution, i.e.,
$v_{\rm phot}$, $T_{\rm phot}$, and $n$. 
% In the discussion section 
% (\S\ref{sect_discussion}), we summarize the temporal evolution of the optical absolute magnitudes, 
% which reveal the bolometric luminosity evolution of each SN, and of the photospheric radius, temperature,
% and velocity.

\section{SN 2005\lowercase{cs}}
\label{sect_05cs}

\begin{figure}
\epsscale{1}
\plotone{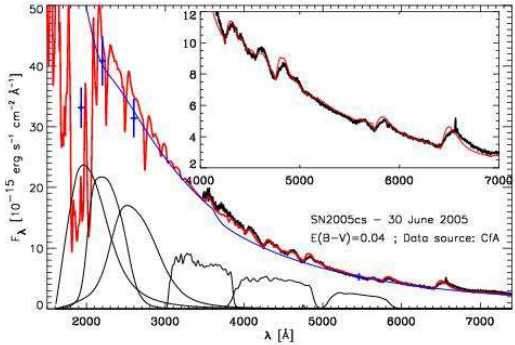}
% \plotone{sn2005cs_spectra/sn2005cs-20050630cs0_1_l1b1_rho_v1_B_uvot.eps}
\caption{Comparison between the reddened (E(B-V)=0.04) synthetic spectrum (red:full; 
blue: continuum only; $T_{\rm phot}=15750$\,K) and 
the observations of SN 2005cs on the 30th of June, 2005, including CfA observations (black) and 
{\it Swift} photometry (blue crosses).
We insert a panel that zooms on the optical range to allow a better assessment of the fit quality.
For completeness, we overlay the {\it Swift} UVOT filter passbands in black, (scaled for visibility, roughly in proportion to the observed flux), 
with, from short to long wavelengths, $uvw2$ ($\lambda_{\rm c}=$ 2030\AA), 
$uvm2$ ($\lambda_{\rm c}=$ 2231\AA), $uvw1$ ($\lambda_{\rm c}=$ 2634\AA), $u$ ($\lambda_{\rm c}=$ 3501\AA), 
$b$ ($\lambda_{\rm c}=$ 4329\AA), $v$ ($\lambda_{\rm c}=$ 5402\AA), 
where $\lambda_{\rm c}$ is their effective wavelength assuming 
a Vega-like spectrum (Poole et al. 2007). 
[See the electronic edition of the Journal for a color version of this figure, 
and see \S\ref{early_05cs} for discussion.]
}
\label{05cs_0630}
\end{figure}

\begin{figure*}
\plottwo{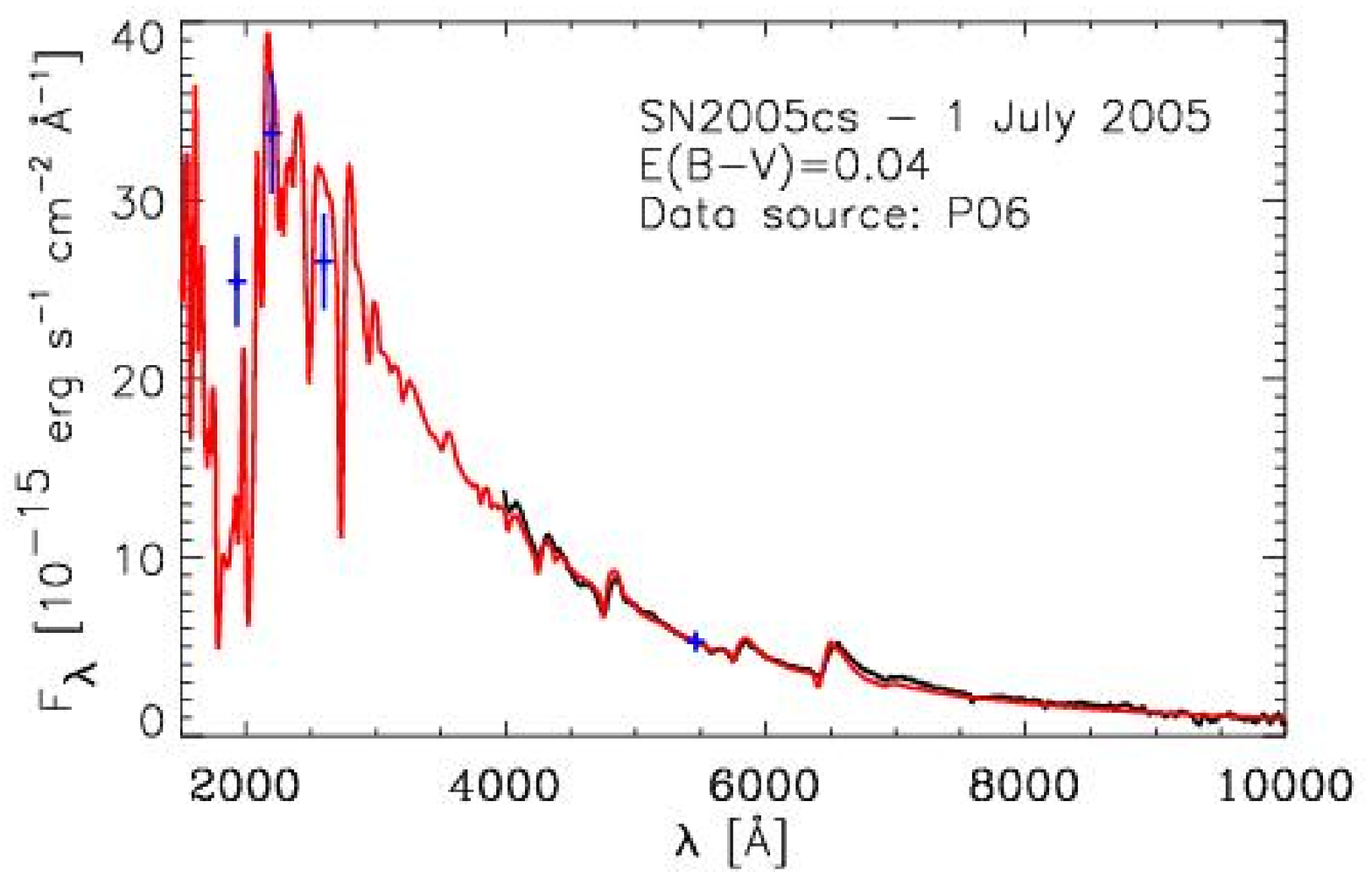}{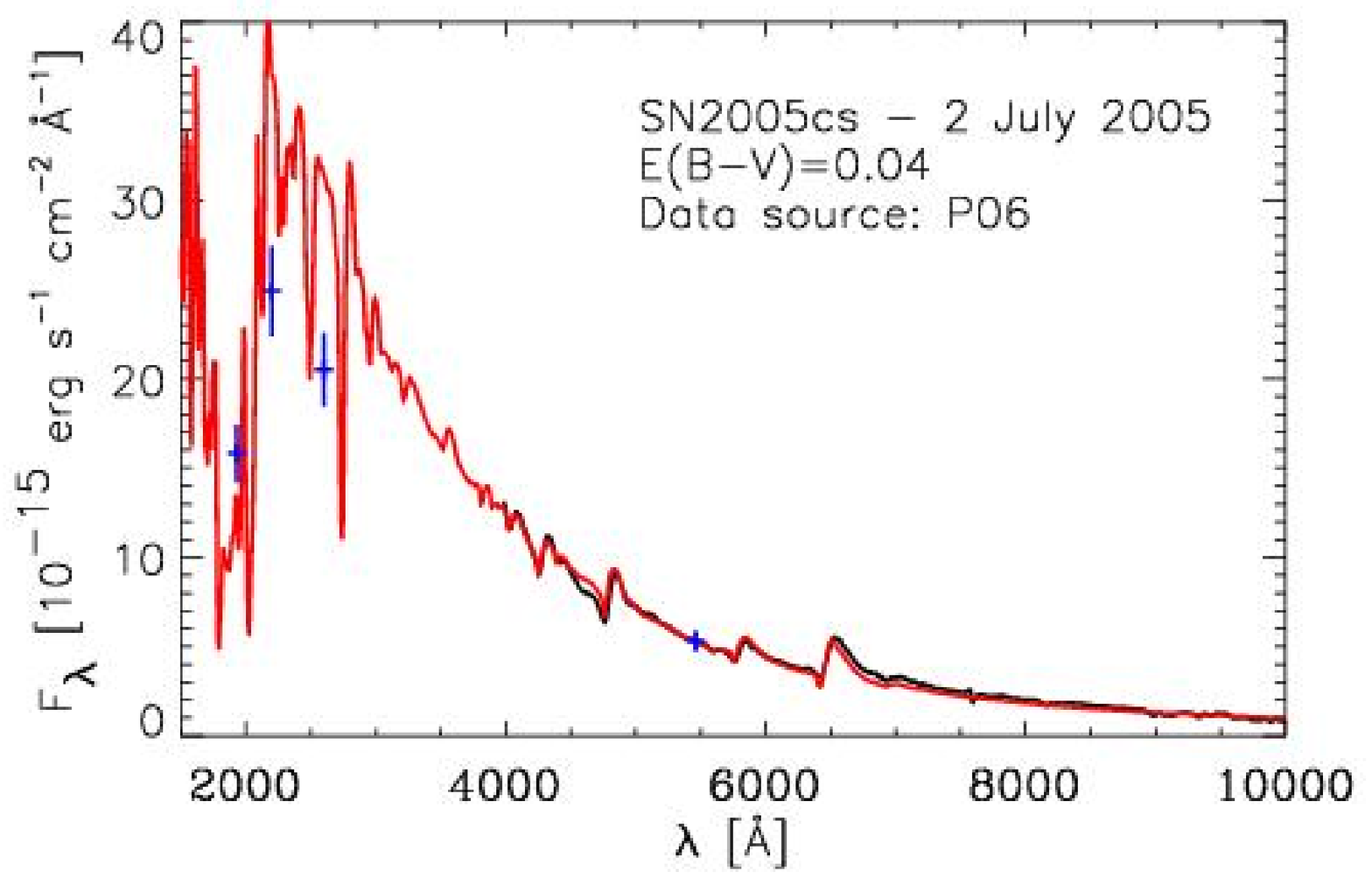}
\plottwo{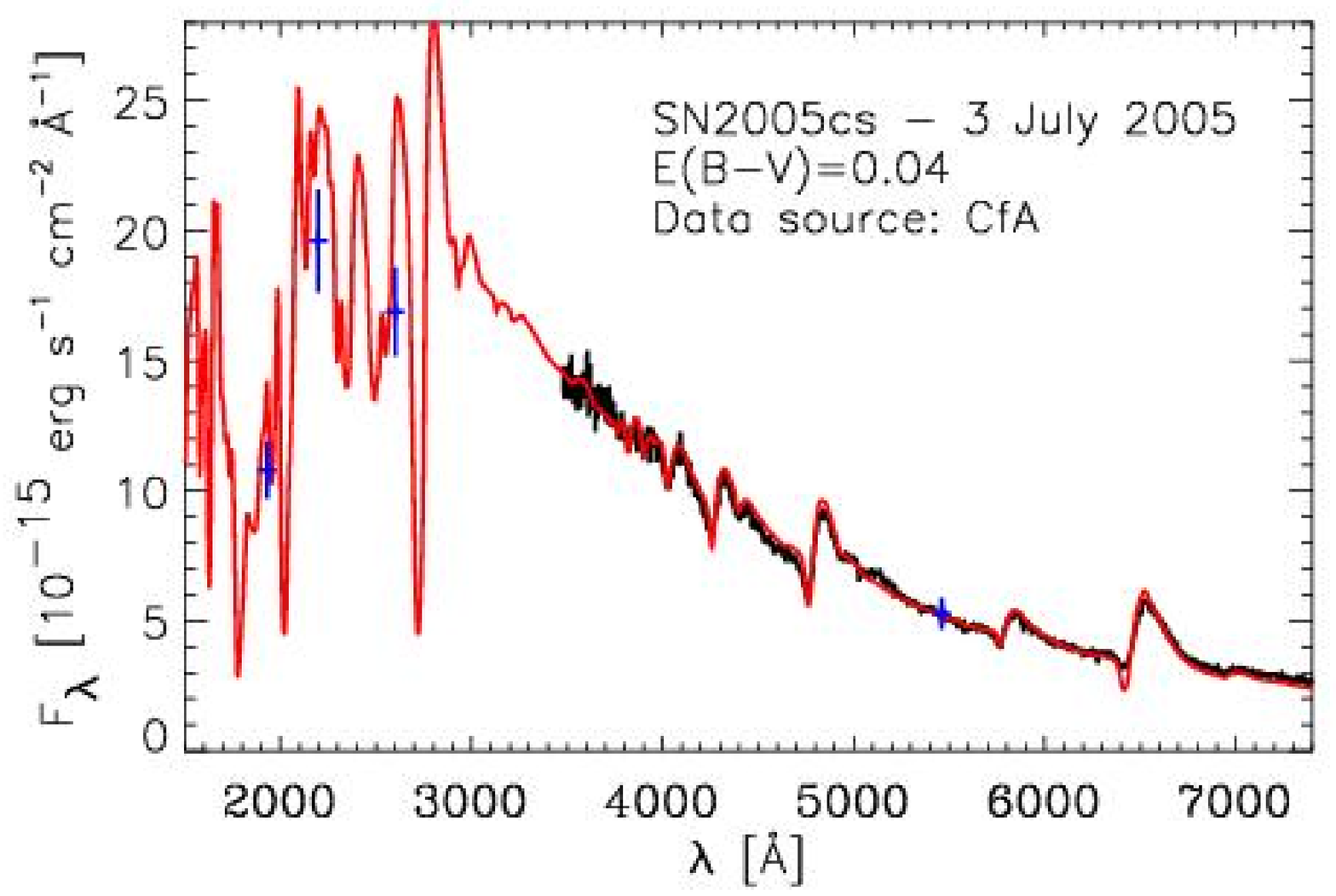}{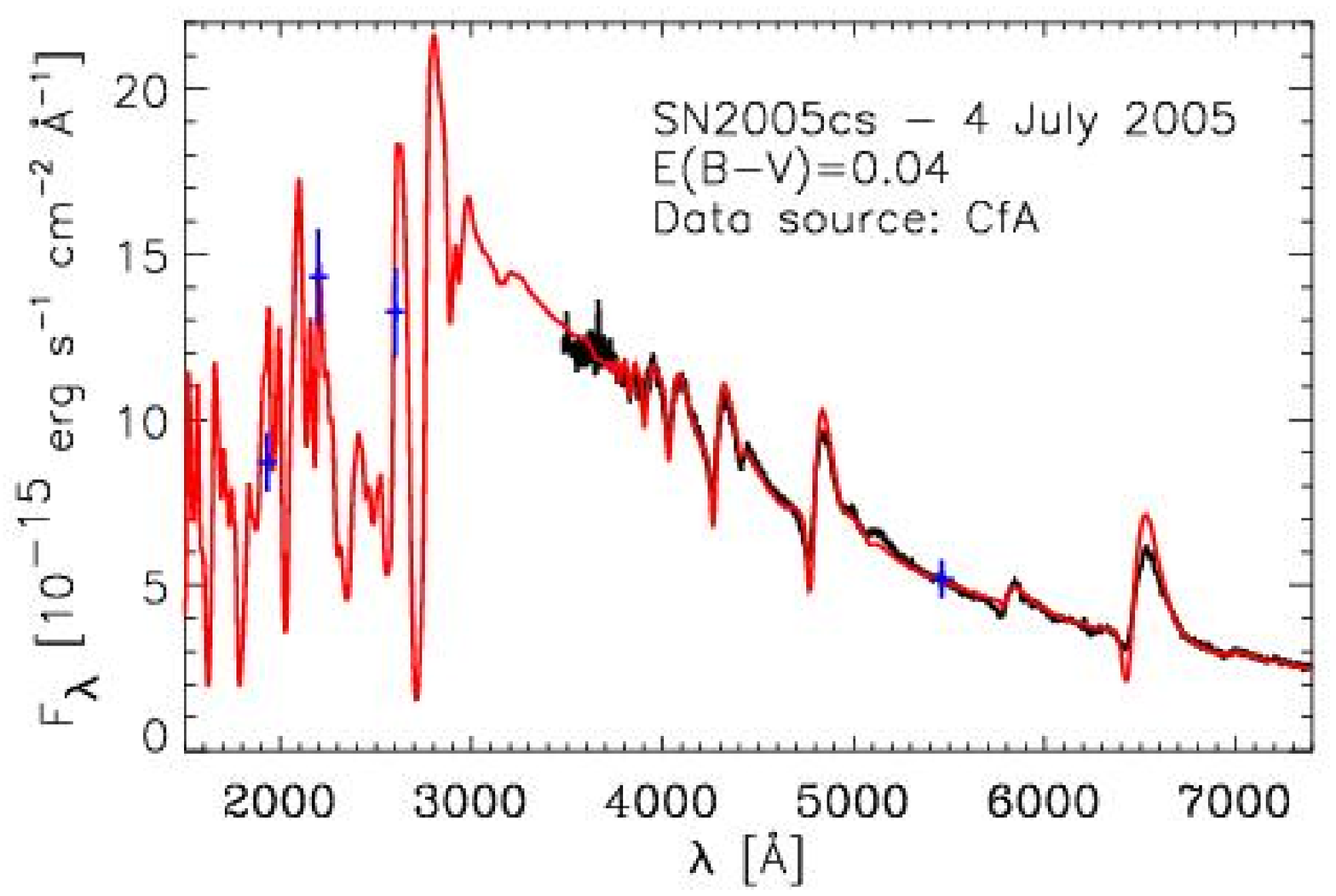}
% \plottwo{sn2005cs_spectra/sn2005cs-20050701cs0_1_l1b1_rho3_B_v1_uvot.eps}{sn2005cs_spectra/sn2005cs-20050702cs0_1_l1b1_rho3_B_uvot.eps}
% \plottwo{sn2005cs_spectra/sn2005cs-20050703cs0_1_l1b_B1_v1_uvot.eps}{sn2005cs_spectra/sn2005cs-20050704cs0_1_l1b_B1_v1_2_uvot.eps}
\caption{
Same as Fig.~\ref{05cs_0630}, but for the SN 2005cs observations of July 1st (top left; $T_{\rm phot}=13350$\,K), 
2nd (top right; $T_{\rm phot}=13420$\,K), 3rd (bottom left; $T_{\rm phot}=10850$\,K), 
and 4th (bottom right; $T_{\rm phot}=9300$\,K), 2005. A full set of model parameters is given 
in Table~\ref{tab_model_05cs}. 
[See the electronic edition of the Journal for a color version of this figure, 
and see \S\ref{early_05cs} for discussion.]
}
\label{05cs_early}
\end{figure*}

SN 2005cs was discovered in the Whirlpool galaxy (M\,51a; NGC\,5194) by Kloehr et al. (2005) on 28.905 June 2005 
(JD 2453550.4). It was not detected on 26.89 June below the limits 
$B\ge$ 17.3, $V\ge$ 17.7, and $R\ge$17.6 (P06), which places a narrow window of two days on the 
explosion date.
In this work, we employ observations from various sources at 13 epochs, starting 
on 30 June 2005, which could be within 3 days of the explosion.
Note that M\,51a, an SA(s)bc galaxy, also hosted the Type Ic SN 1994I (Wheeler et al. 1994;
Filippenko et al. 1995; Baron et al. 1996). Importantly, Maund et al. (2005) and Li et al. (2006) 
associate the SN 2005cs event with a 9$^{+3}_{-2}$\,\mo K0-M4 RSG progenitor star, based on the analysis 
of pre-explosion HST images, compatible with the observed plateau evolution of the light curve
in the optical bands. SN 2005cs was not detected in the X-rays and the upper limit to the 
X-ray luminosity yields a limit to the progenitor mass loss rate of 
\mdot\,$\sles\, 1 \times 10^{-5}$\,\msunyr\,$(v_{\rm w}/10)$\,\kms, where $v_{\rm w}$ is the asymptotic
velocity of the progenitor wind (Brown et al. 2007).
Model properties are given in Table~\ref{tab_model_05cs} and corresponding synthetic spectral
fits to observations are presented in Figs.~\ref{05cs_0630}--\ref{05cs_0728}.

% date    Lstar       TPHOT         RPHOT (10^14cm)   VPHOT              RHO_PHOT      NRHO
% 0630 &  3.0    &    15750   &          2.00    &     6880      &        24.6   &       20  % cs0_1_l1b1_rho_v1_B
% \clearpage

%
% Group dates from 1st to 4th of July together into one figure, 4 panels.
% 
% date    Lstar       TPHOT         RPHOT (10^14cm)   VPHOT              RHO_PHOT      NRHO
% 0701 &  2.3    &    13350   &          2.55    &     6950      &        10.8   &       12  & & & & & &  \\ % cs0_1_l1b1_rho3_B_v1  

% \clearpage

\subsection{The early times: 30th of June until 4th of July 2005}
\label{early_05cs}

   We show the first observations (blues crosses: {\sl Swift} UVOT photometry; 
black curve: optical spectrum of 30 June 2005 from CfA) of SN 2005cs in Fig.~\ref{05cs_0630}, 
together with synthetic spectra (red: full; blue: continuum only),
reddened with $E(B-V)=0.04$ and scaled by a factor 1.17. Observations
reveal a very hot spectrum, with a steep decline of the observed
flux longward of $\sim$2300\AA. We identify optical lines of the Balmer series 
(H$\alpha$, H$\beta$, and H$\gamma$), He{\,\sc i}\,5875\AA, the N{\,\sc ii} multiplets at
$\sim$4600\AA\ and $\sim$5400\AA, and, as suggested by Baron et al. (2007), an O{\,\sc ii} multiplet at
$\sim$4600\AA. Note that the lower ionization inferred for the earliest spectra of SN 1999em
suggested that  N{\,\sc ii} alone was sufficient to reproduce adequately the observed features
at both 4600\AA\ and 5400\AA. It is the higher ionization in the first spectrum of SN 2005cs
that calls for this novel O{\,\sc ii} identification, more consistent with solar CNO 
chemistry, and slightly different from the abundances used in DH06 (see our discussion in 
\S\ref{sect_mod_pres} and \S\ref{sect_discussion}).
The 4300--5000\AA\ region also contains numerous He{\,\sc i} lines.
The observed dip shortward of 2000\AA\ stems from Fe{\,\sc iii} line-blanketing, rather
than dust extinction (the continuum-only synthetic spectrum, shown in blue, does not show this dip).
The individual contributions from all important ions on that date are presented in Fig.~5, left panel,
of Brown et al. (2007). On that date, we have $T_{\rm phot}=15750$\,K, $v_{\rm phot}=6880$\,\kms, and $n=20$.
%  The model parameters for that date are: $L_{\ast}=3.0 \times 10^8 L_{\odot}$, $T_{\rm phot}=15750$\,K, 
%$R_{\rm phot}= 2.00 \times 10^{14}$\,cm, $v_{\rm phot}=6880$\,\kms, 
%$\rho_{\rm phot}=2.46 \times 10^{-13}$\,g\,cm$^{-3}$, and $n=20$. 
Such a steep density distribution is required to bring optical lines
down to their (weak) observed strength, by reducing their emission 
volumes. This is higher than the density exponent adopted in the past, but these are also the first analyses of such early-time observations 
of Type II-P SN spectra.  A steep density gradient is also used to fit the earliest observations of SN 2006bp. We discuss this choice and comment on its implications in \S\ref{sect_ion}. 

We present P06's observations of SN 2005cs on the 1st July 2005 in Fig.~\ref{05cs_early}, top left panel. 
The slope of the spectrum in the blue appears steeper than predicted by the {\sl Swift} UVOT data points, 
which are quite closely matched by our model (the observed flux in the spectrum, inaccurate below $\sim$4000\AA, is omitted;
Pastorello, priv. comm.). 
The synthetic flux is scaled by a factor 1.1. 
% The model parameters on that day are: 
%$L_{\ast}=2.3 \times 10^8 L_{\odot}$, $T_{\rm phot}=13350$\,K, $R_{\rm phot}= 2.55 \times 10^{14}$\,cm, 
%$v_{\rm phot}=6950$\,\kms, $\rho_{\rm phot}=1.08 \times 10^{-13}$\,g\,cm$^{-3}$, and $n=12$.
The SN ejecta is getting cooler ($T_{\rm phot}=13350$\,K) at the photosphere, but with a comparable velocity 
(within the uncertainties; $v_{\rm phot}=6950$\,\kms).
The density exponent $n$ has been reduced to a value of 12 (the density distribution flattens), 
motivated by the strengthening of optical line features, whose identity is the same as that on 30 June.
The UV flux is weaker, resulting primarily from the cooling of the ejecta - the line blanketing is still mostly
due to Fe{\,\sc iii}. 

Little changes as we move onto the 2nd of July 2005, as shown in Fig.~\ref{05cs_early}, top right panel,
with a model characterized by $T_{\rm phot}=13420$\,K, $v_{\rm phot}=6370$\,\kms, and $n=12$,
% The model parameters 
% are now: $L_{\ast}=2.3 \times 10^8 L_{\odot}$, $T_{\rm phot}=13420$\,K, $R_{\rm phot}= 2.54 \times 10^{14}$\,cm, 
% $v_{\rm phot}=6370$\,\kms, $\rho_{\rm phot}=1.06 \times 10^{-13}$\,g\,cm$^{-3}$, and $n=12$.  The only noticeable difference is a lower expansion velocity. At any single epoch, this velocity cannot be constrained to better than 10\%, since some lines are better fitted either with a higher 
or a lower ejecta velocity (Dessart \& Hillier 2005a). 
This 10\% range in expansion velocity ($\sim$500\kms) may 
in fact be physical, reflecting the potential presence of fluid instabilities or turbulence.
The {\sl Swift} UVOT photometry would be better fit by a faster decline in the UV than our CMFGEN model provides.
This situation improves on July 3rd, the 
following day (Fig.~\ref{05cs_early}, bottom left panel), where a single model fits both the {\sl Swift} UVOT 
photometry and the optical spectrum, with parameters $T_{\rm phot}=10850$\,K,  $v_{\rm phot}=6080$\,\kms,
and $n=10$.
% The model parameters are: $L_{\ast}=2.0 \times 10^8 L_{\odot}$, $T_{\rm phot}=10850$\,K, 
% $R_{\rm phot}= 3.26 \times 10^{14}$\,cm, $v_{\rm phot}=6080$\,\kms, 
% $\rho_{\rm phot}=7.6 \times 10^{-14}$\,g\,cm$^{-3}$, and $n=10$.
The photosphere is both cooler and slower, while the density exponent is now set 
to 10 at all remaining epochs.
The observations on July 4th resemble those of July 3rd in the optical, but the UVOT 
photometry indicates a fading of the UV flux. In the bottom right panel of Fig.~\ref{05cs_early}, 
we show the model fit to the July 4th observations is cooler, 
with $T_{\rm phot}=9300$\,K, $v_{\rm phot}=5450$\,\kms, and  $n=10$.
%$L_{\ast}=1.8 \times 10^8 L_{\odot}$, $T_{\rm phot}=9300$\,K, $R_{\rm phot}= 3.91 \times 10^{14}$\,cm, 
% $v_{\rm phot}=5450$\,\kms, $\rho_{\rm phot}=6.8 \times 10^{-14}$\,g\,cm$^{-3}$, and $n=10$.

These early observations show a strong UV flux that 
dominates over the optical flux, with low extinction. 
Optical lines are H{\,\sc i}, He{\,\sc i}, O{\,\sc ii} and 
N{\,\sc ii}, with blanketing in the UV from Fe{\,\sc iii}. The density 
distribution is steep, flattening over a few days to the canonical $n \sim 10$ exponent predicted for blast wave solutions
(Imshennik \& Nadezhin 1988; Shigeyama \& Nomoto 1990, Chevalier 1982; Ensman \& Burrows 1992, 
Blinnikov et al. 2000). 
We surmise that the steep density profile may reflect work done by the intense UV radiation at and shortly-after 
shock breakout (see \S\ref{sect_ion}).
Our models support a common origin for both the UV and the optical light
observed, with emission from the regions at and above the photosphere. 
No additional contribution to the UV and optical fluxes is present from an external source,
such as an interaction between the SN ejecta and the CSM. 

\subsection{The intermediate times: 5th of July until 14th of July 2005}
\label{int_05cs}

% date    Lstar       TPHOT         RPHOT (10^14cm)   VPHOT              RHO_PHOT      NRHO
% 07/05/05   &  1.8 &     8620   &    4.93    &  4780  &    5.6  &  10   \\ % cs0_1_l1b_B1_v1_5
% (th of July: Model cs3_3 has been replaced by r3_ba as the latter is somewhat better in the blue.
% \clearpage
\begin{figure*}
\plottwo{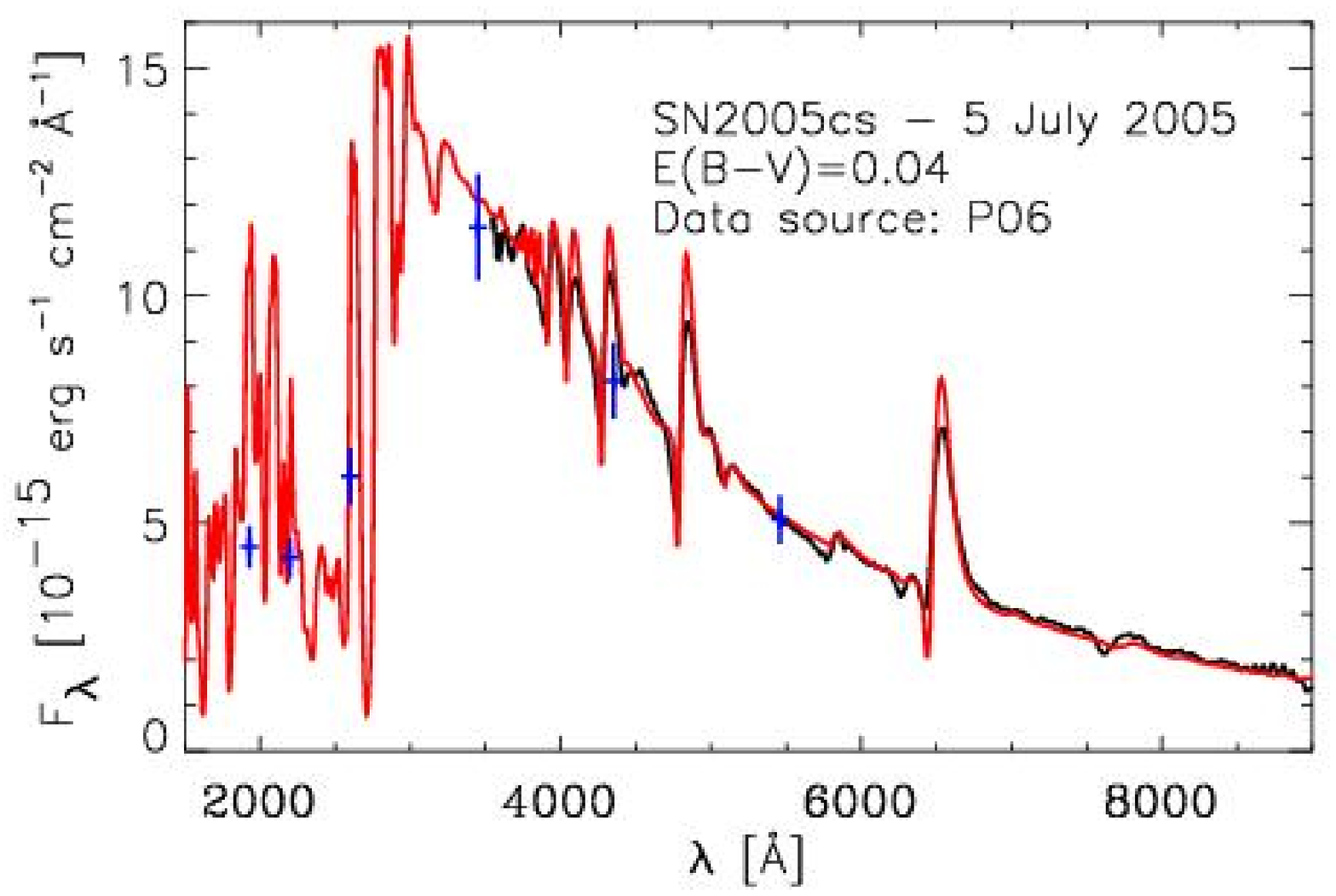}{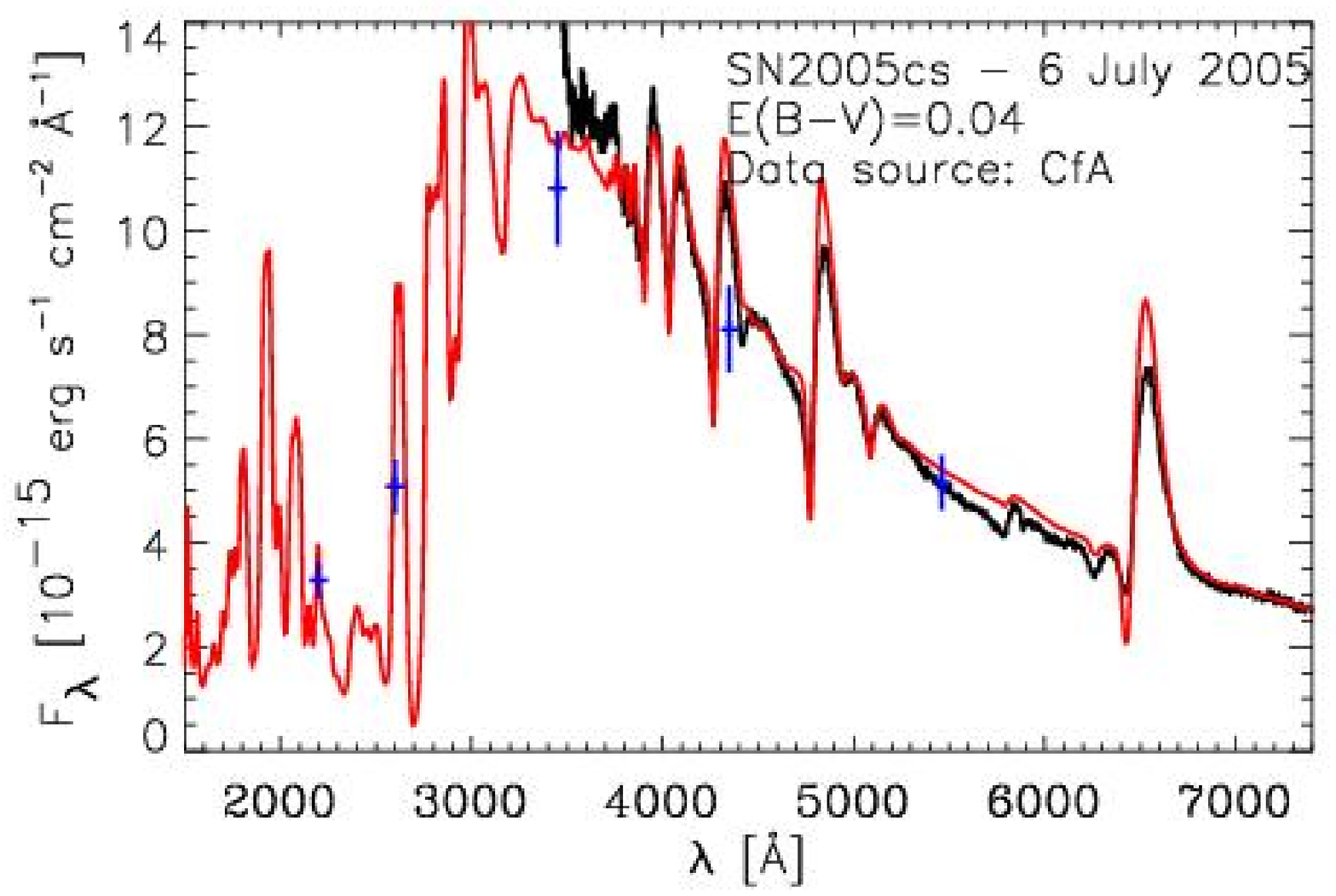}
\plottwo{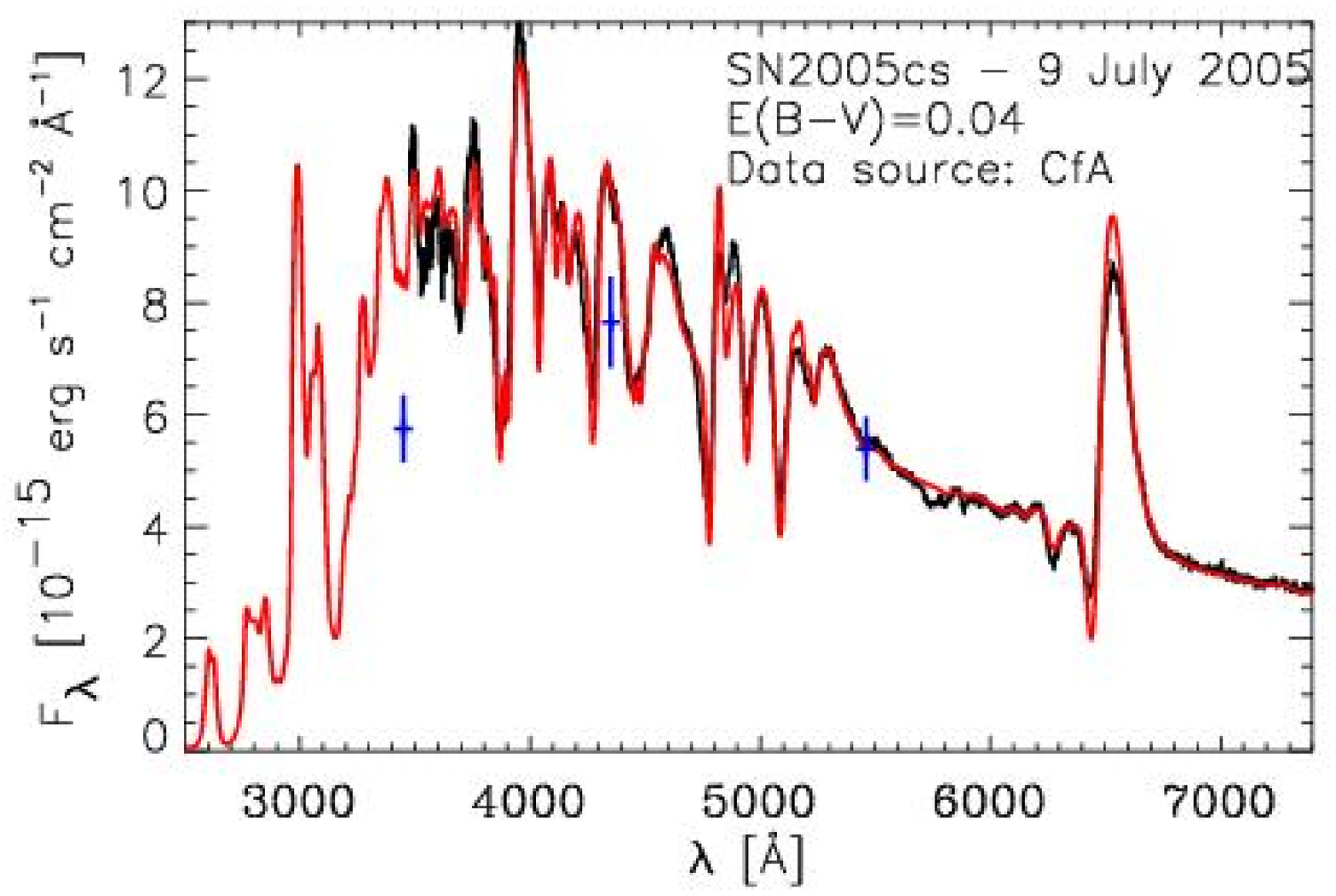}{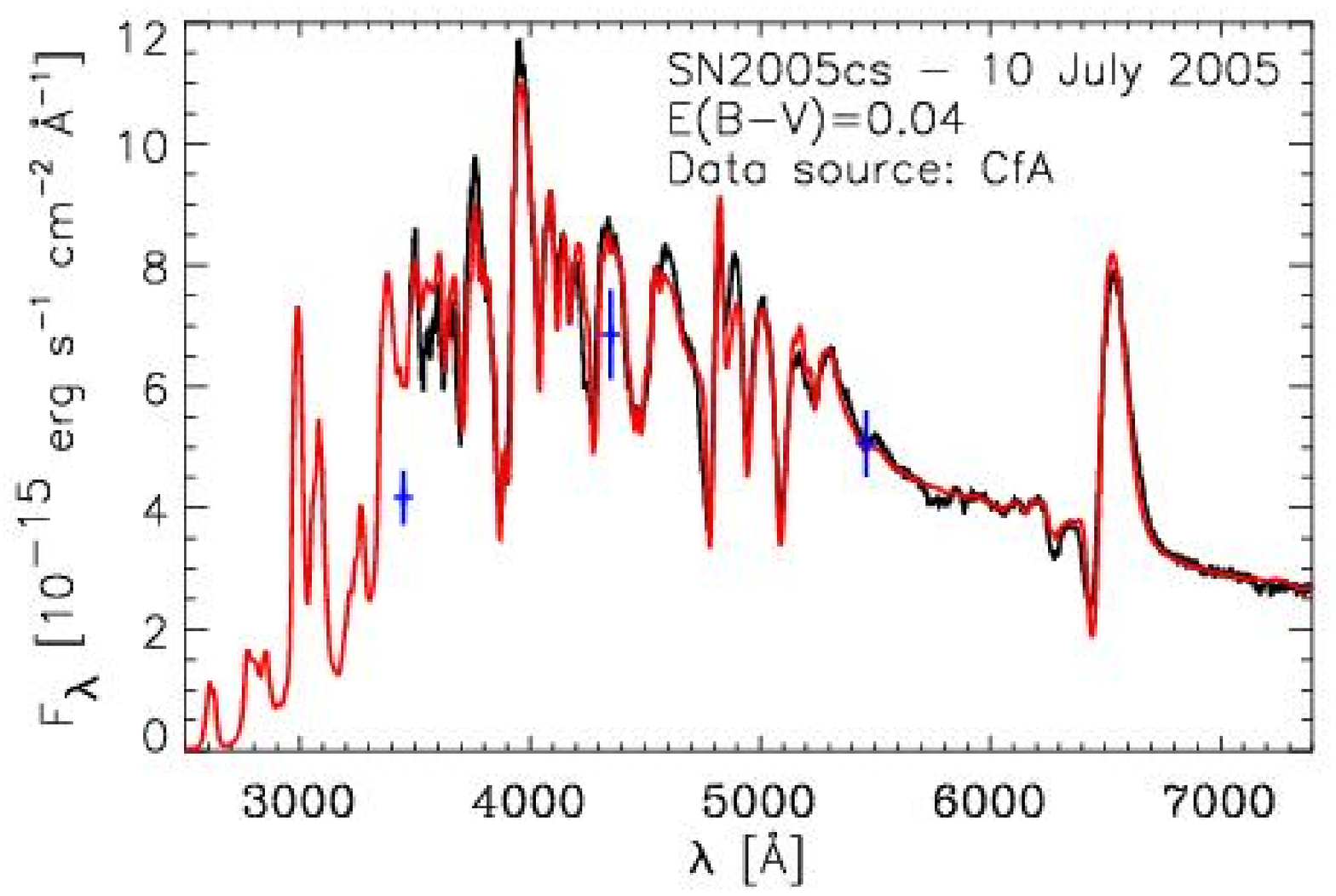}
\plottwo{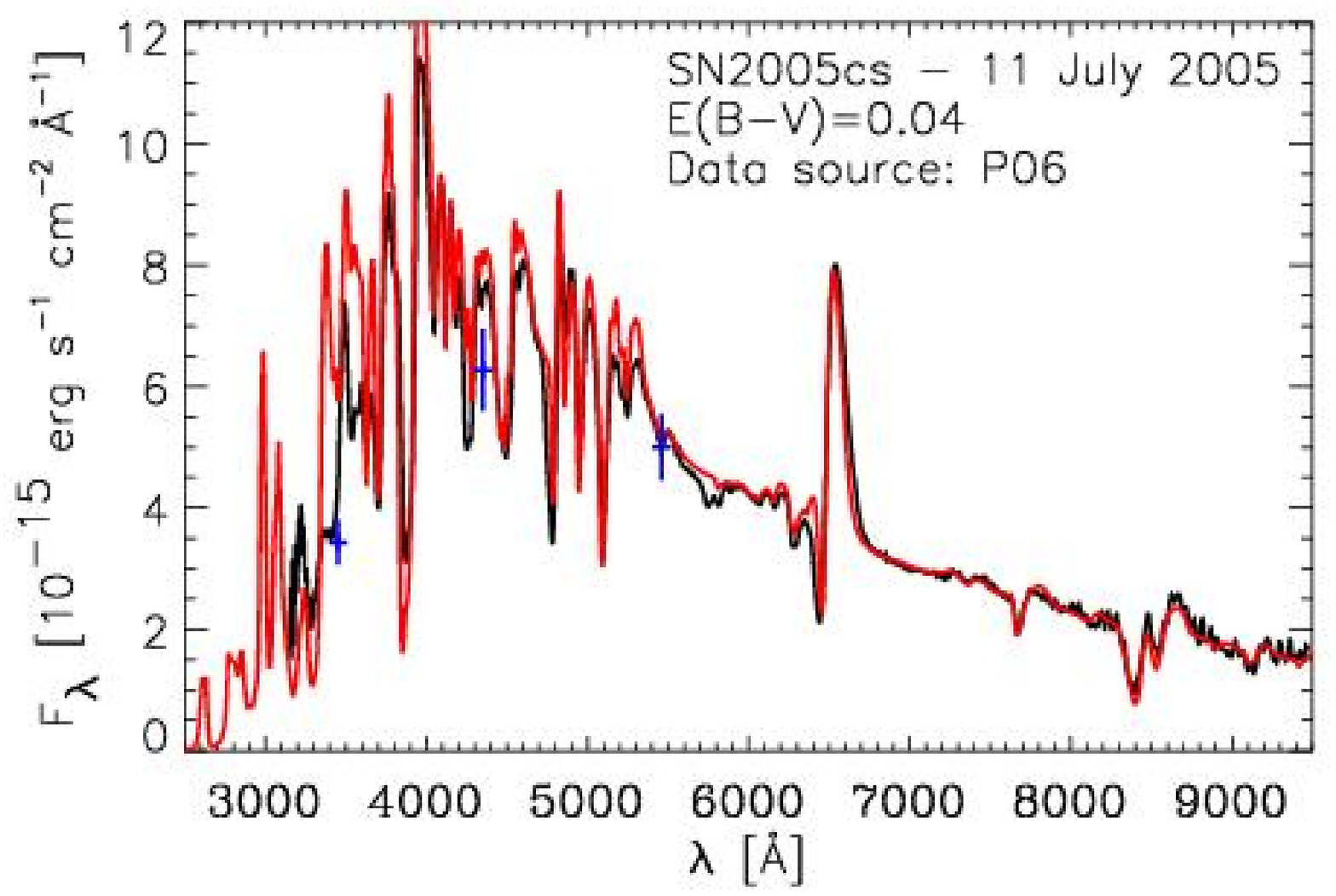}{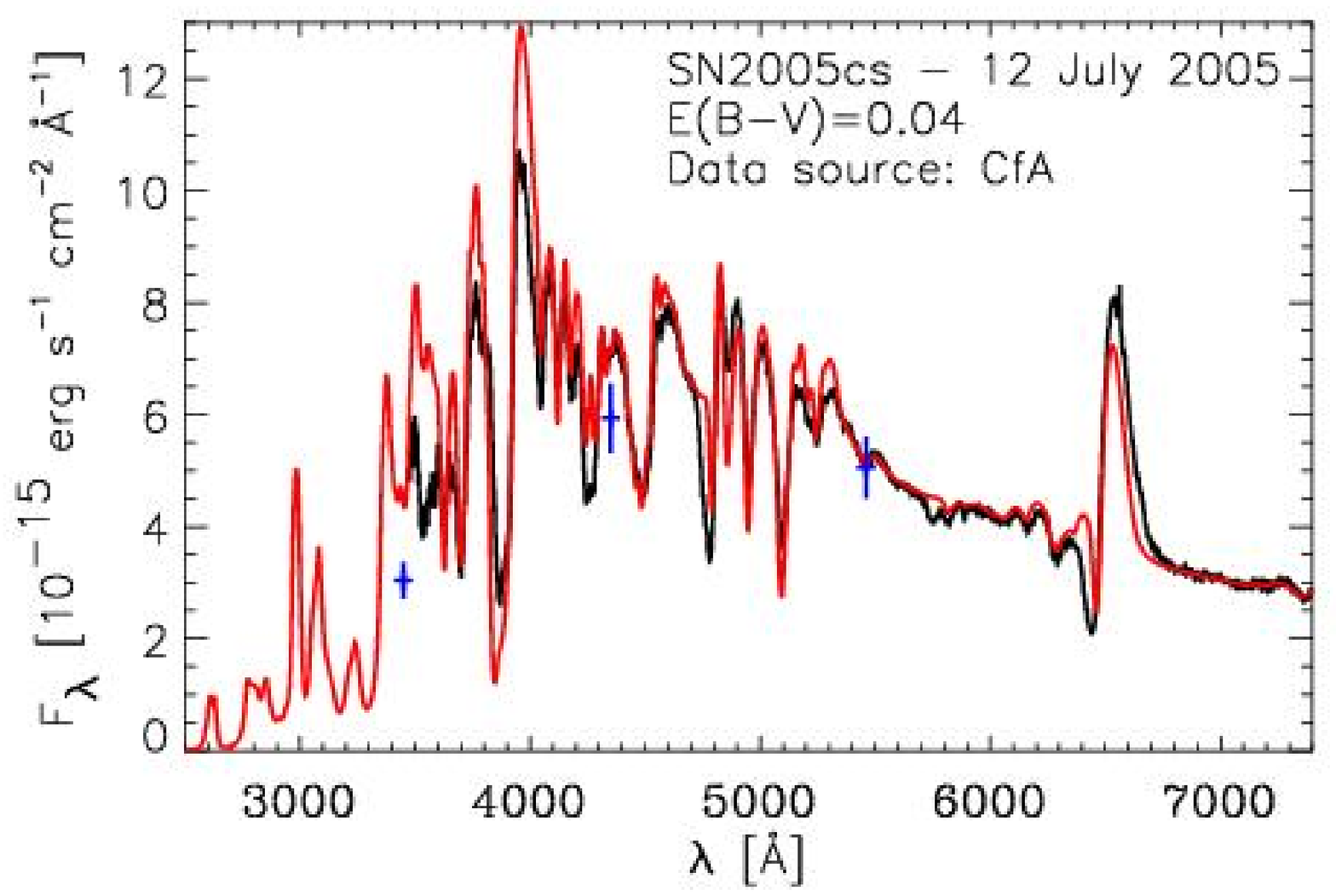}
% \plottwo{sn2005cs_spectra/sn2005cs-20050705cs0_1_l1b_B1_v1_5_uvot.eps}{sn2005cs_spectra/sn2005cs-20050706cs1_7_abund_uvot.eps}
% \plottwo{sn2005cs_spectra/sn2005cs-20050709r3_ba_uvot.eps}{sn2005cs_spectra/sn2005cs-20050710cs3_5_uvot.eps}
% \plottwo{sn2005cs_spectra/sn2005cs-20050711r3_e_abund_uvot.eps}{sn2005cs_spectra/sn2005cs-20050712r3_abund_uvot.eps}
\caption{
Same as Fig.~\ref{05cs_0630}, but for the SN 2005cs observations of July 5th (top left; $T_{\rm phot}=$8620\,K), 
6th (top right; $T_{\rm phot}=$8250\,K), 9th (middle left; $T_{\rm phot}=$7770\,K), 
10th (middle right; $T_{\rm phot}=$7120\,K), 11th (bottom left; $T_{\rm phot}=$7300\,K), 
12th (bottom right; $T_{\rm phot}=$6980\,K), 2005. A full set of model parameters is given 
in Table~\ref{tab_model_05cs}. 
[See the electronic edition of the Journal for a color version of this figure, 
and see \S\ref{int_05cs} for discussion.]
}
\label{05cs_int}
\end{figure*}
% \clearpage

  In this epoch, the flux level
decreases dramatically in the UV and metal lines, mostly of Fe{\,\sc ii}, 
appear in the optical.
This is the onset of the hydrogen-recombination epoch leading to the
``Plateau'' phase of such Type II SNe.  At the same time, He{\,\sc i}\,5875\AA\ weaken, so that the feature at $\sim$5900\AA\ is due only to Na{\,\sc i}\,D at late times.

These spectral properties are visible in Fig.~\ref{05cs_int}, top left panel, in which observations
on the 5th of July are fitted with a model characterized by $T_{\rm phot}=8620$\,K, 
$v_{\rm phot}=4780$\,\kms, and $n=10$.
% $L_{\ast}=1.8 \times 10^8 L_{\odot}$, $T_{\rm phot}=8620$\,K, $R_{\rm phot}= 4.93 \times 10^{14}$\,cm, 
% $v_{\rm phot}=4780$\,\kms, $\rho_{\rm phot}=5.6 \times 10^{-14}$\,g\,cm$^{-3}$, and $n=10$.
The O{\,\sc i}\,7770\AA\ line is now clearly seen, following the cooling and the recession of the 
photosphere. The SN 2005cs ejecta have lower velocities than SN 1999em:  this allows
the clear identification of this line profile, which was severely corrupted by the atmospheric extinction in the spectra of SN 1999em (DH06).
Similarly, while Si{\,\sc ii}\,6355\AA\ was predicted to overlap with H$\alpha$ in the spectrum of SN 1999em (Figs.~3 and 5 in Dessart \& Hillier 2005a),
we clearly observe it here as a separate line blueward of the H$\alpha$ trough (see also P06).   This line, in combination with unidentified N{\,\sc ii}/O{\sc ii} lines at similar 
Doppler velocities in the blue wing of H$\beta$ and He{\,\sc i}\,5875\AA, led to speculations on the presence of inhomogeneities above the photosphere (Leonard et al 2002ab). 
At early epochs, it seems these subtle spectral features can be explained quite naturally 
by numerous lines of O{\,\sc ii} (see above), N{\,\sc ii}, He{\,\sc i}, and Si{\,\sc ii}. 
At later times, the presence of a cool dense shell at the outer edge of Type II-P SN ejecta is expected
to produce real absorption dips in the trough of H$\alpha$ (and He{\,\sc i}\,10830\AA; see Chugai et al. 2007).
The fit to observations on July 6th is shown on the top right panel of Fig.~\ref{05cs_int}, with little
change from the previous day except for a modest decline in temperature (e.g., $T_{\rm phot}=8250$\,K).

% date    Lstar       TPHOT         RPHOT (10^14cm)   VPHOT              RHO_PHOT      NRHO
% 0714 &  1.1    &     6700   &          5.47    &     3930      &       28.2    &       10  & & & & & &  \\ % r3_g_abund
% \clearpage
\begin{figure*}
\plottwo{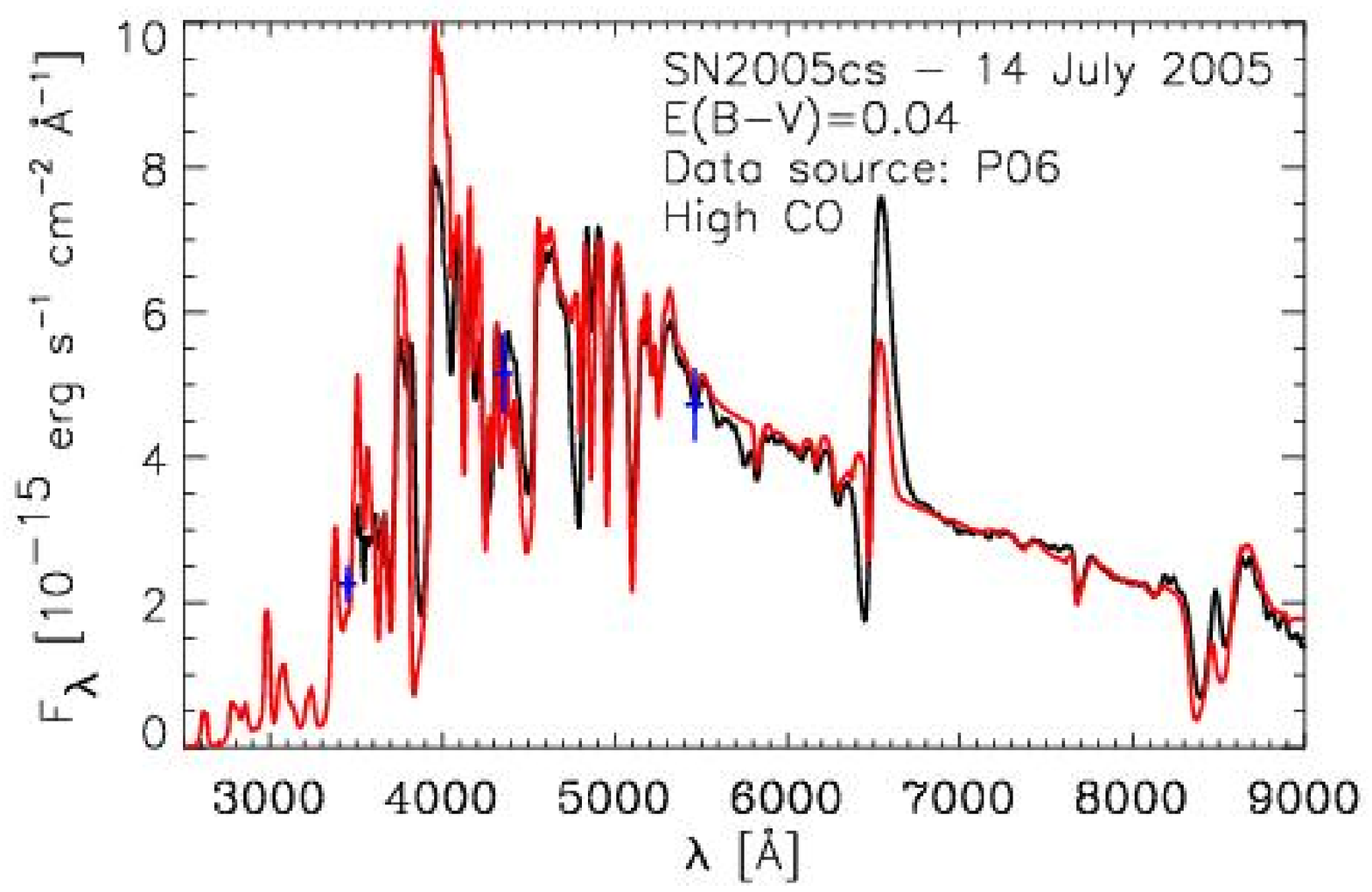}{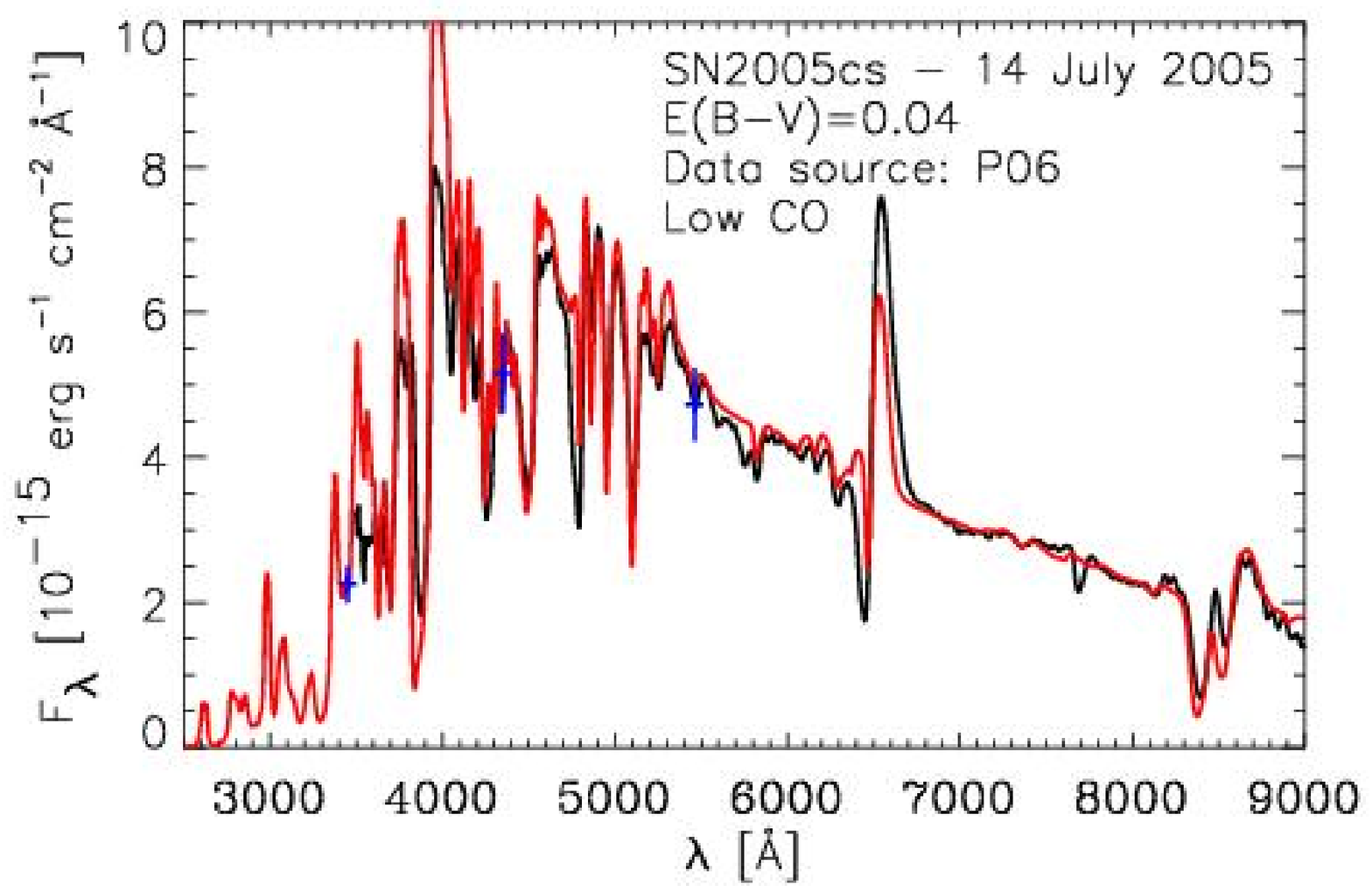}
% \plottwo{sn2005cs_spectra/sn2005cs-20050714r3_g_abund_uvot.eps}{sn2005cs_spectra/sn2005cs-20050714r3_g_uvot.eps}
\caption{
Same as Fig.~\ref{05cs_0630}, but this time for P06's observations of SN 2005cs on the 14th of July, 2005,
and using a model with enhanced (left) or day-one (right) carbon and oxygen abundances. 
[See the electronic edition of the Journal for a color version of this figure, 
and see \S\ref{int_05cs} for discussion.]
}
\label{05cs_0714}
\end{figure*}
% \clearpage

By the 9th of July (Fig.~\ref{05cs_int}, middle left panel), the UV flux is now an order of magnitude lower 
than the optical flux. The Ca{\,\sc ii}\,H\&K doublet appears as a strong P-Cygni line profile at the blue edge 
of the optical spectrum, strengthening with time, together with the Fe{\,\sc ii} lines present
throughout the UV range (as a forest of overlapping lines that give an apparently continuous
light blocking effect - see Fig.~5 in Dessart \& Hillier 2005a) and up to $\sim$5200\AA.
Ni{\,\sc ii} (shortward of 2500\AA) and Ti{\,\sc ii} (in the 3000\AA\ region) also contribute 
significant line blanketing. The model properties for the 9th of July are $T_{\rm phot}=7770$\,K,
$v_{\rm phot}=4710$\,\kms, and $n=10$.
% $L_{\ast}=1.5 \times 10^8 L_{\odot}$, $T_{\rm phot}=7770$\,K, $R_{\rm phot}= 5.97 \times 10^{14}$\,cm, 
% $v_{\rm phot}=4710$\,\kms, $\rho_{\rm phot}=4.7 \times 10^{-14}$\,g\,cm$^{-3}$, and $n=10$.
 
Subsequent evolution is quite slow 
from the 10th (Fig.~\ref{05cs_int}, middle right panel), 11th (Fig.~\ref{05cs_int}, bottom left panel), 
12th (Fig.~\ref{05cs_int}, bottom right panel), and 14th (Fig.~\ref{05cs_0714}). 
Over this 5-day period, the photospheric layers cool slowly and modestly
(7120\,K, 7300\,K, 6980\,K, 6700\,K, respectively), and slow down (4580\,\kms, 4440\,\kms, 4320\,\kms, 3930\,\kms).
On the 14th, the P06 spectrum extends beyond 9000\AA\ and reveals the presence of
the Ca{\,\sc ii}\,8500\AA\ multiplet, providing a good check that the ionization state of
the SN ejecta is well modeled. Fitting the O{\,\sc i}\,7770\AA\ line requires
a higher oxygen abundance (O/He=0.01), which we adopt for SN 2005cs starting 
on the 11th of July. We show a comparison of the enhanced and standard oxygen abundance
CMFGEN models in the left and right panels of Fig.~\ref{05cs_0714}, to show the effect on the predicted strength of the O{\,\sc i}\,7770\AA\ line.
Note finally that the H$\alpha$ line strength (as well as H$\beta$, although
less discernible) is underestimated by a factor of two in both absorption and emission, the more so
for higher carbon/oxygen abundances (corresponding to lower hydrogen mass fractions).

% date    Lstar       TPHOT         RPHOT (10^14cm)   VPHOT              RHO_PHOT      NRHO
% 0728 &  1.0    &     6140   &          7.84    &     2240      &      125.6    &       10  & & & & & &  \\ % r4_b
% \clearpage
\begin{figure}
\plotone{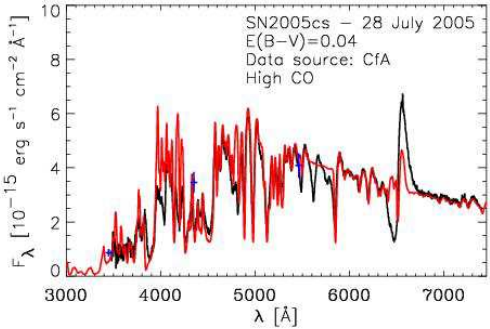}
% \plotone{sn2005cs_spectra/sn2005cs-20050728r4_b_uvot.eps}
\caption{
Same as Fig.~\ref{05cs_0630}, but this time for the observations of SN 2005cs on the 28th of July, 2005.
[See the electronic edition of the Journal for a color version of this figure, 
and see \S\ref{late_05cs} for discussion.]
}
\label{05cs_0728}
\end{figure}
% \clearpage

\subsection{28 July 2005}
\label{late_05cs}

The observation of the 28th of July 2005 is the last in our SN 2005cs sample.
This is about one month after the first observation taken and 
well into the plateau phase. We show our fits to these observations in Fig.~\ref{05cs_0728}
for which the model parameters are $T_{\rm phot}=6140$\,K, $v_{\rm phot}=2240$\,\kms, and $n=10$.
% $L_{\ast}=1.0 \times 10^8 L_{\odot}$, $T_{\rm phot}=6140$\,K, $R_{\rm phot}= 7.84 \times 10^{14}$\,cm, 
% $v_{\rm phot}=2240$\,\kms, $\rho_{\rm phot}=1.25 \times 10^{-12}$\,g\,cm$^{-3}$, and $n=10$.
The photosphere has now receded to slow-moving ejecta layers close to the bottom of the hydrogen shell.

In practice, given the low luminosity and low velocity of this SN envelope, the (outer) hydrogen shell
of the progenitor may stretch even further down to the slower velocity and more C/O enriched
layers. Unfortunately, the CfA observations do not cover the O{\,\sc i} and C{\,\sc i} lines longward of 7500\AA, 
which would provide a measure of the oxygen and carbon mass fractions. In DH06, and by comparison 
with the surface chemistry of the progenitor, we inferred significant enhancements
in carbon and oxygen abundances in the photospheric layers of SN 1999em at 45 days after explosion.
The main line diagnostics are O{\,\sc i}\,7775\AA\ and O{\,\sc i}\,7984\AA, while redder
transitions at 8446\AA\, and 9264\AA\, overlap with Ca{\,\sc ii}, Mg{\,\sc ii}, and C{\,\sc i} lines,
and are therefore less reliable.
Useful C{\,\sc i} diagnostics are the features at $\sim$9100\AA, 9400\AA, and 9600\AA,
due to transitions of the (3p-3s) series at 9061\AA, 9062\AA, 9078\AA, 9088\AA, 9112\AA, 9405\AA, 
9603\AA, 9621\AA, and 9658\AA.  In the near-IR, the C{\,\sc i}\,1.07\,$\mu$m is an additional
diagnostic, although it overlaps with P$\gamma$, and potentially with He{\,\sc i}\,1.083\,$\mu$m.  Observations in this wavelength range would be especially valuable.

Observing the C/O abundance change would be important and warrants thorough 
late time spectroscopic observations in the photospheric phase of Type II SNe, extending 
at least up to 1$\mu$m.  
A census of lines contributing in the optical is shown in Fig.~8 of DH06.
Fe{\,\sc ii} blankets completely whatever UV flux emerges from the thermalization
layers below the photosphere. Ti{\,\sc ii} and Fe{\,\sc ii} leave a complex set of features
in the optical, most noticeably resulting in a pronounced dip in the B-band region.
Na{\,\sc i}\,D is reproduced here by enhancing its abundance by a factor of four above 
its solar metallicity value (Grevesse \& Sauval 1998). 
However, the predicted line remains narrower than observed 
and time-dependent effects were found to lead to a strong and broad line profile
without any change in sodium abundance. The underestimate of the H$\alpha$
line strength suggests that this is the epoch when our steady-state approach 
departs significantly from the more correctly-computed time-dependent conditions 
in the SN ejecta (Dessart \& Hillier 2007a,b).

% \clearpage
\begin{deluxetable*}{cccccrcccc}
%\rotate
\tablewidth{16cm}
\tabletypesize{\scriptsize}
\tablecaption{Model Characteristics for SN2006\lowercase{bp}.
\label{tab_model_06bp}}
\tablehead{
\colhead{Julian Date}&
\colhead{Day}&
\colhead{$\Delta t ^a$}&
\colhead{$L_{{\rm CMF},R_0}$}& 
\colhead{$L_{{\rm OBS},R_{\rm Max}}$}& 
\colhead{$T_{\rm phot}$}& 
\colhead{$R_{\rm phot}$}&
\colhead{$v_{\rm phot}$}&
\colhead{$\rho_{\rm phot}$}&
\colhead{$n$}
\\
\colhead{}&      
\colhead{}&      
\colhead{Days}&      
\multicolumn{2}{c}{(10$^8$ $L_{\odot}$)}&
\colhead{(K)}&    
\colhead{(10$^{14}$\,cm)}&  
\colhead{(km\,s$^{-1}$)}&    
\colhead{(10$^{-14}$\,g\,cm$^{-3}$)}&  
\colhead{}    
}
\startdata
%          t - t_discovery   L_CMF_Base L_OBS_RMAX      T_phot        R_phot     V_phot     rho_phot  N_RHO  
2453836.6 &  2006-04-11  &   1.5 &    15   &  16.3 &    20600    &   2.386   &   14100  &   51.7  &  50     \\ % tst_n20_5_B            
2453837.6 &  2006-04-12  &   2.5 &    15   &  15.1 &    18120    &   3.165   &   12550  &   31.6  &  50     \\ % tst_n20_6_v1           
2453842.6 &  2006-04-17  &   7.5 &     9   &   6.9 &    11800    &   5.541   &   11300  &    8.3  &  20     \\ % n7_j20b_v4_s1_l3_v2    
2453844.6 &  2006-04-19  &   9.5 &     9   &   6.6 &    11400    &   5.868   &   11050  &    6.5  &  16     \\ % n7_j16b_v4_s1_l3_v1_B  
2453845.8 &  2006-04-20  &  10.7 &     9   &   6.2 &    10700    &   6.485   &   10300  &    4.5  &  12     \\ % n7_j12b_v4_s1_l3_v1_B  
2453846.7 &  2006-04-21  &  11.6 &     9   &   6.2 &    10700    &   6.485   &   10300  &    4.5  &  12     \\ % n7_j12b_v4_s1_l3_v1_B  
2453849.6 &  2006-04-24  &  14.5 &     5   &   3.6 &     9350    &   6.345   &    7920  &    4.2  &  10     \\ % n16_n10_s0_v1_B_new2   
2453850.6 &  2006-04-25  &  15.5 &     2.7 &   1.8 &     8150    &   6.580   &    8750  &    4.0  &  10     \\ % n16_n10_s0_v1_B_new2_3 
2453855.7 &  2006-04-30  &  20.6 &     1.5 &   1.1 &     7320    &   6.000   &    6450  &    7.3  &  10     \\ % n16_n10_s0_v1_B_new6   
2453856.8 &  2006-05-01  &  21.7 &     1.5 &   1.1 &     7100    &   6.147   &    6330  &    8.8  &  10     \\ % n16_n10_s0_v1_B_new5   
2453857.6 &  2006-05-02  &  22.5 &     1.5 &   1.1 &     7100    &   6.147   &    6330  &    8.8  &  10     \\ % n16_n10_s0_v1_B_new5 
2453858.7 &  2006-05-03  &  23.6 &     2.0 &   1.4 &     6800    &   7.26    &    5810  &   10.1  &  10     \\ % n16_n10_s0_v1_B_new4  
2453859.6 &  2006-05-04  &  24.5 &     2.0 &   1.4 &     6800    &   7.26    &    5810  &   10.1  &  10     \\ % n16_n10_s0_v1_B_new4   
2453862.7 &  2006-05-07  &  27.6 &     2.0 &   1.7 &     6420    &   8.90    &    4550  &   17.9  &  10     \\ % nj4_1v1_new3_abund     
2453867.7 &  2006-05-12  &  32.6 &     2.0 &   1.7 &     6420    &   8.90    &    4550  &   17.9  &  10     \\ % nj4_1v1_new3_abund     
2453876.7 &  2006-05-21  &  41.6 &     2.0 &   1.8 &     6300    &   9.28    &    4320  &   30.3  &  10     \\ % nj4_1v1_new2_abund     
2453891.7 &  2006-06-05  &  56.6 &     2.0 &   1.8 &     6160    &   9.73    &    3760  &   50.7  &  10     \\ % nj4_1v1_new1_abund     
2453907.6 &  2006-06-21  &  72.5 &     2.0 &   1.9 &     6050    &  10.18    &    3160  &   79.4  &  10     \\ % nj4_1v1_new_abund      
\enddata
\tablecomments{
For each date in our sample of observations, we provide the following 
CMFGEN model parameters: Base comoving-frame luminosity $L_{{\rm CMF},R_0}$ and 
emergent observer-frame luminosity $L_{{\rm OBS},R_{\rm Max}}$ (in \lsun), photospheric conditions describing
the electron temperature $T_{\rm phot}$ (in K), the radius $R_{\rm phot}$ (in 10$^{14}$\,cm), 
the velocity $v_{\rm phot}$ (in \kms), and the density (in 10$^{-14}$\,g\,cm$^{-3}$), 
together with the density exponent $n$ characterizing the density law 
$\rho(r) = \rho_{\rm phot} (R_{\rm phot}/r)^n$.   
(See \S\ref{sect_mod_pres} for a presentation of the modeling approach).
$^a$: $\Delta t$ is the time elapsed since discovery.
}
\end{deluxetable*}
% \clearpage

\section{SN 2006\lowercase{bp}}
\label{sect_06bp}

    SN 2006bp was discovered in NGC\,3953, an SB(r)bc galaxy, by Koichi Itagaki
(Teppo-cho, Yamagata, Japan) around 9.60 April 2006 (JD 2453835.1), and located at 
R.A. = 11h53mn55s.74, Decl. = +52$^{\circ}$21'09".4 (equinox 2000.0), 
which is 62" east and 93" north of the center of NGC3953. 
Quimby et al. (2006) obtained an optical spectrum on March 11.11 UT with 
the 9.2-m Hobby-Eberly Telescope (HET), reporting the observation of a blue continuum,
a narrow absorption at 592\,nm (associated with Na{\,\sc i}\,D), and line features 
at 583 nm (associated with He{\,\sc i}\,5875\AA), and 627\,nm (associated with H$\alpha$).   
SN 2006bp was not detected in unfiltered ROTSE-IIIb images taken on Apr. 9.15 
(limiting mag about 16.9), but it was found at about 15.3 mag on images taken on Apr. 10.15.
SN 2006bp was caught promptly after the explosion, as confirmed from our analysis of the spectra. 
Immler et al. (2007) report the 4.5$\sigma$ detection of X-ray 
emission at the level of 1.8 $\pm$ 0.4 $\times$ 10$^{39}$\,erg\,s$^{-1}$ up until 12 days after discovery,
indicating circumstellar interaction with a pre-SN RSG wind characterized by \mdot\,$\sim 10^{-5}$\msunyr.
This X-ray luminosity is less than 1\% of the UV and optical luminosity in the combined UV and optical ranges-- and differs from the strongly interacting Type IIn SNe
\footnote{Type IIn spectra are characterized by narrow lines for all ions that contribute
and most conspicuously H{\,\sc i} and Fe{\,\sc ii}. Examples of such Type IIn SNe are 
1994W (Chugai et al. 2004), 1995G (Pastorello et al. 2002), 1995N (Fox et al. 2000), 
1997eg (Salamanca et al. 2002), 1998S (Pooley et al. 2002). Type IIn SNe are strong radio 
and X-ray emitters, associated with the interaction between the SN ejecta and the CSM, 
and with an X-ray luminosity that can rival the optical luminosity, as in 1995N (Fox et al. 2000).}.
Model properties are given in Table~\,\ref{tab_model_06bp} and corresponding synthetic spectral
fits to observations are presented in Figs.\,\ref{06bp_0411}--\ref{06bp_late}.

%          L      T_phot        R_phot     V_phot     rho_phot  N_RHO  
% 0411 &    15   &    20600    &   2.386   &   14100  &   51.7  &  50     \\ % tst_n20_5_B            

\subsection{First week of evolution of SN 2006bp}
\label{early_06bp}
% \clearpage
\begin{figure}
\epsscale{.8}
\plotone{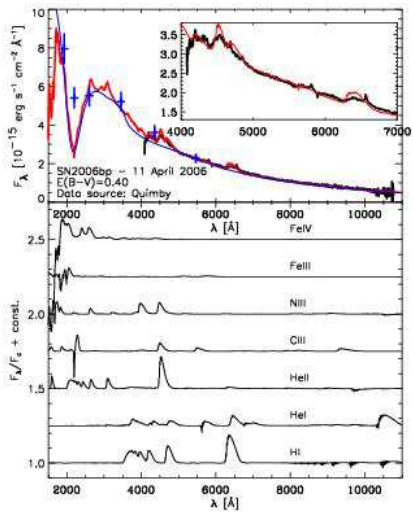}
\epsscale{1}
% \plotone{sn2006bp_spectra/sn2006bp_0411_tst_n20_5_B_line.eps}
\caption{
{\it Top:} Comparison between the reddened ($E(B-V)=0.4$) synthetic flux (red: full; 
blue: continuum only; $T_{\rm phot}= 20,600$\,K) and SN 2006bp observations on the 11th of April 2006, 
including the optical spectrum of Quimby (2007; black) and {\sl Swift} UVOT fluxes (blue crosses).
Note that, despite the larger extinction (responsible for the 2200\AA\ dip), 
the UV flux level is comparable to that observed for SN 2005cs at discovery 
(shown in Fig.~\ref{05cs_0630}), suggesting a much higher ejecta temperature and ionization. 
{\it Bottom:} Rectified synthetic spectra for the model shown at top, but including bound-bound
transitions of individual species (see label on the right).
Apart from the standard H{\,\sc i} and He{\,\sc i} lines, note the presence 
of He{\,\sc ii}\,4686\AA, N{\,\sc iii}\,4001--4099\AA\ and  N{\,\sc iii}\,4638\AA, 
C{\,\sc iii}\,2290\AA\ and C{\,\sc iii}\,4647\AA. 
[See the electronic edition of the Journal for a color version of this figure, 
and see \S\ref{early_06bp} for discussion.]
}
\label{06bp_0411}
\end{figure}
% \clearpage

  The first spectral observation of SN 2006bp, shown in Fig.~\ref{06bp_0411}, is impressive 
in several regards. It has a nearly featureless spectrum, with a very steep slope.
The dip at $\sim$2200\AA\ is not caused by line blanketing, as in the first spectrum of SN 2005cs 
(Fe{\,\sc iv} is the main blocking ion in the UV, and it does so blueward of 1800\AA; its effect on the synthetic continuum SED is shown). 
The dip stems from interstellar extinction, with an inferred value $E(B-V)$=0.4. 
We also see, at all epochs, Na{\sc i}\,D as a narrow absorption feature,  
indicating significant column density in the line of sight to SN 2006bp.
Hydrogen Balmer lines are present, but very weak
and very broad, and limited to H$\alpha$, H$\beta$, and H$\gamma$. He{\,\sc i}\,5875\AA\ is 
just barely identifiable. It is surprising that lines of N{\,\sc ii} and O{\,\sc ii} identified
in the first spectra taken of SN 2005cs are not seen here. 
Given the high reddening, fitting the steep slope of the SED requires much higher temperatures and ionization 
conditions (Fig.~\ref{fig_gammas}). Consequently, CMFGEN predicts, although somewhat too strong, 
the presence of He{\,\sc ii}\,4686\AA\ (together with a small, $\sim$20\% contribution
from C{\,\sc iii}\,4650\AA), a line that is more typical of the photosphere/wind of 
early-type O and Wolf-Rayet stars (Hillier \& Miller 1999; Dessart et al. 2000; Crowther et al. 2002;
Hillier et al. 2003), as well as features arising from N{\,\sc iii}\,4001--4099\AA,
N{\,\sc iii}\,4638\AA, and C{\,\sc iii}\,2290\AA\ (see bottom panel of Fig.~\ref{06bp_0411}). 
In practice, we used $T_{\rm phot}= 20,600$\,K, $v_{\rm phot}=14,100$\,\kms, and $n=50$.
Numerous directions in the parameter space were explored to reproduce the steepness
of the SED, the small strength of P-Cygni profiles,  and the strong blueshift of P-Cygni peak emission.  
These included varying the reddening and temperature, extent of the photosphere, ejecta chemistry,
and density exponent. The most critical was the density exponent.
% $L_{\ast}= 1.5 \times 10^9 L_{\odot}$, $T_{\rm phot}= 20,600$\,K, 
% $R_{\rm phot}= 2.386 \times 10^{14}$\,cm, $v_{\rm phot}=14,100$\,\kms, and 
% $\rho_{\rm phot}= 5.17 \times 10^{-13}$\,g\,cm$^{-3}$. 
Quimby et al. (2007) discuss the presence of weak and narrow lines in this optical spectrum,
associated with He{\sc ii}\,4200\AA, He{\,\sc ii}\,4686\AA, and C{\sc iv}\,5808\AA. They suggest 
that they emanate from material located just outside the SN ejecta that is 
flash ionized at shock breakout.
This gives additional support to our identification here of He{\sc ii}\,4686\AA\ emission 
from the SN ejecta itself.
Note the photospheric velocity of SN 2005cs inferred from the first observed 
spectrum at 5 days after explosion is only 6880\,\kms, compared to 10\,430\,\kms 
at a comparable epoch for SN 1999em; see \S\ref{sect_05cs} and DH06).
We discuss further the properties of SN 2006bp at that age
in \S\ref{sect_ion}.

% \clearpage
\begin{figure*}
\plottwo{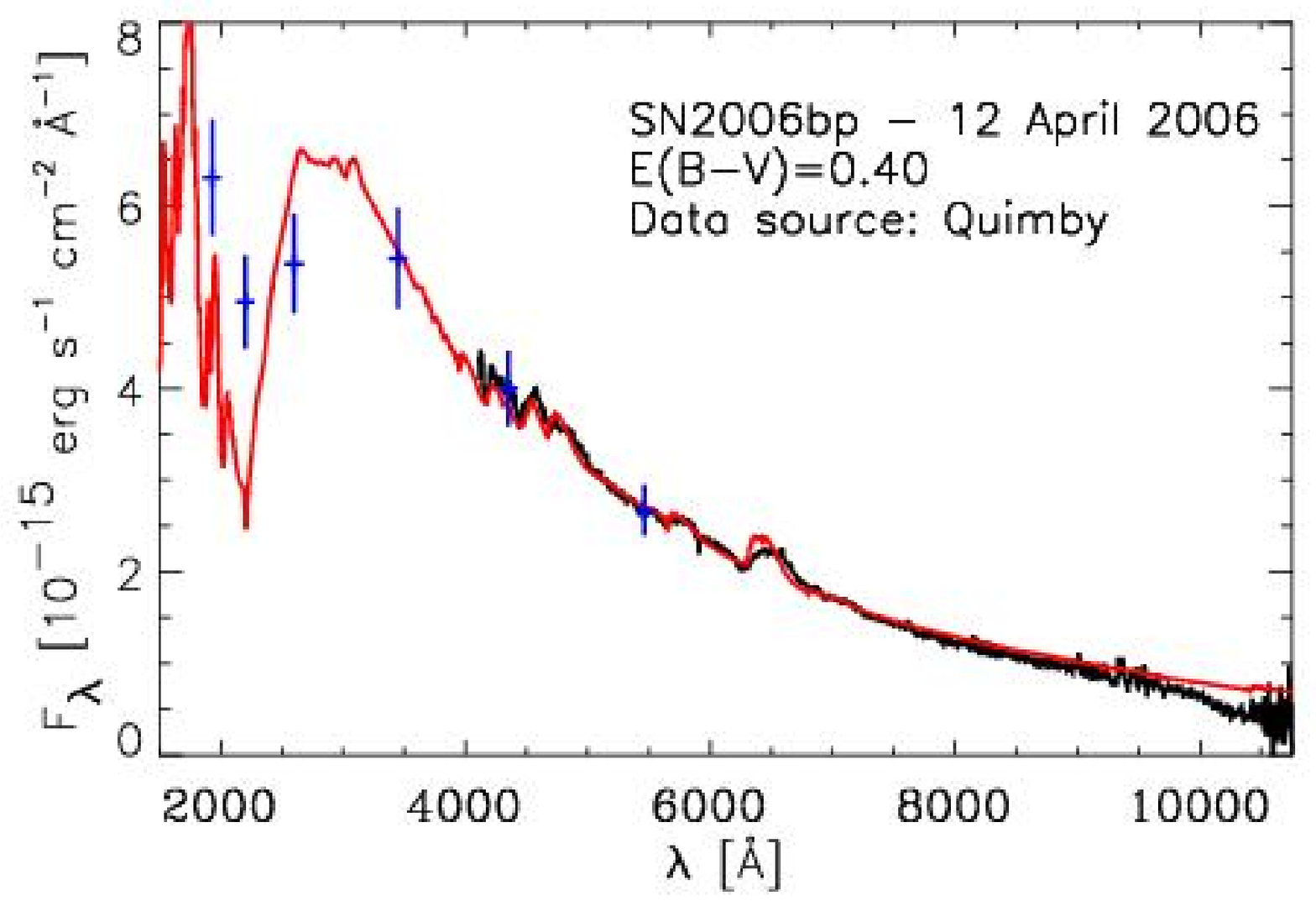}{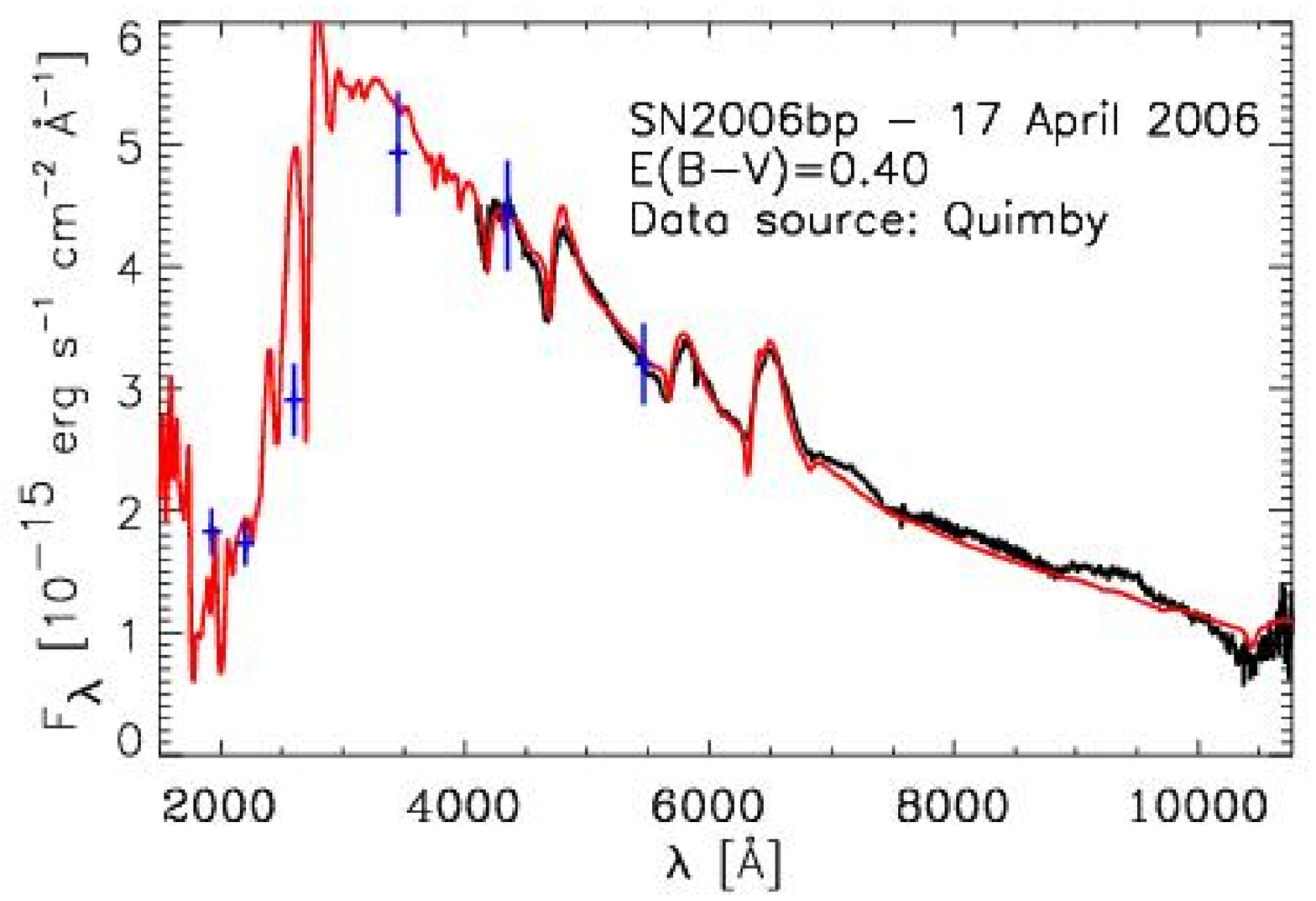}
\plottwo{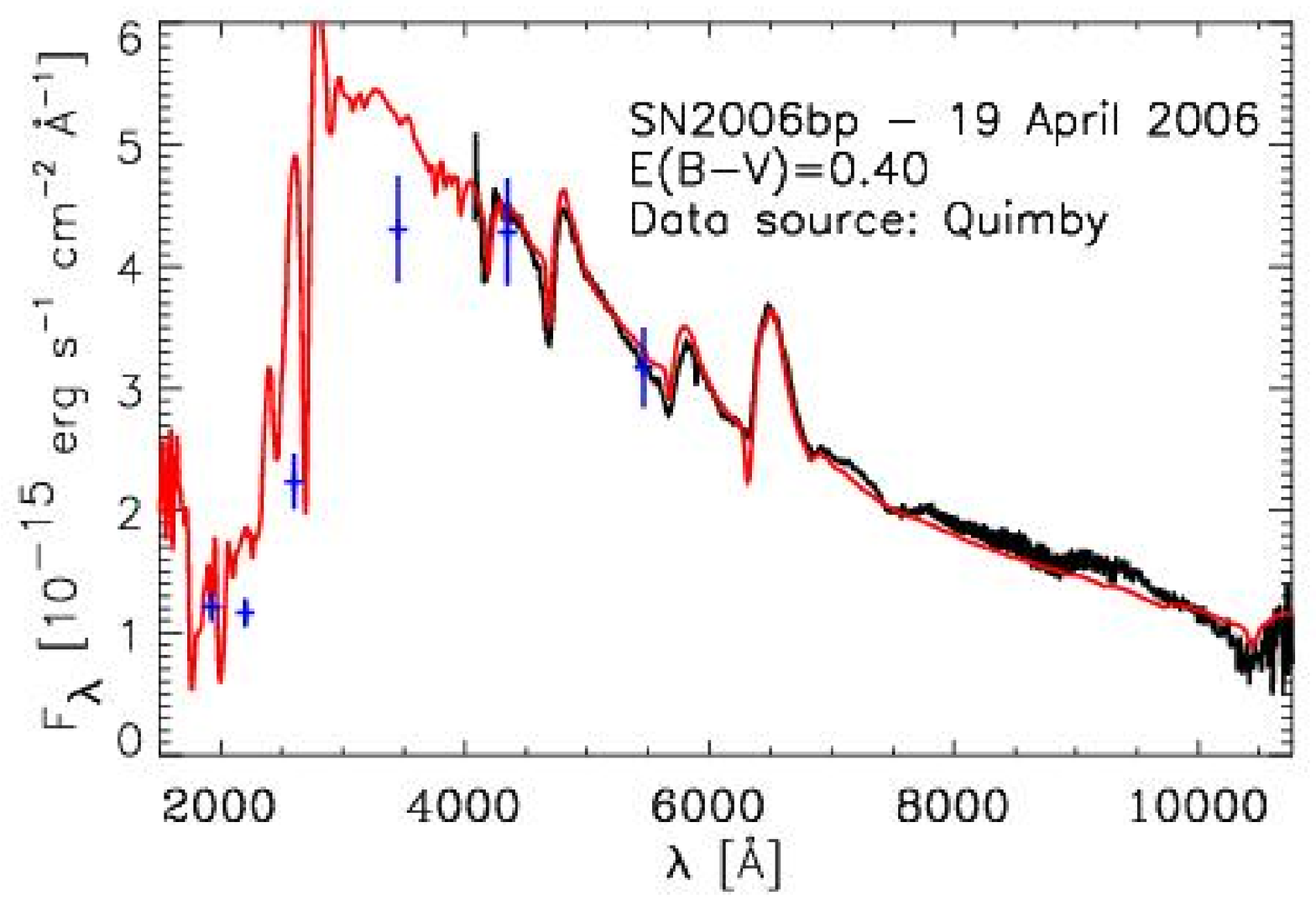}{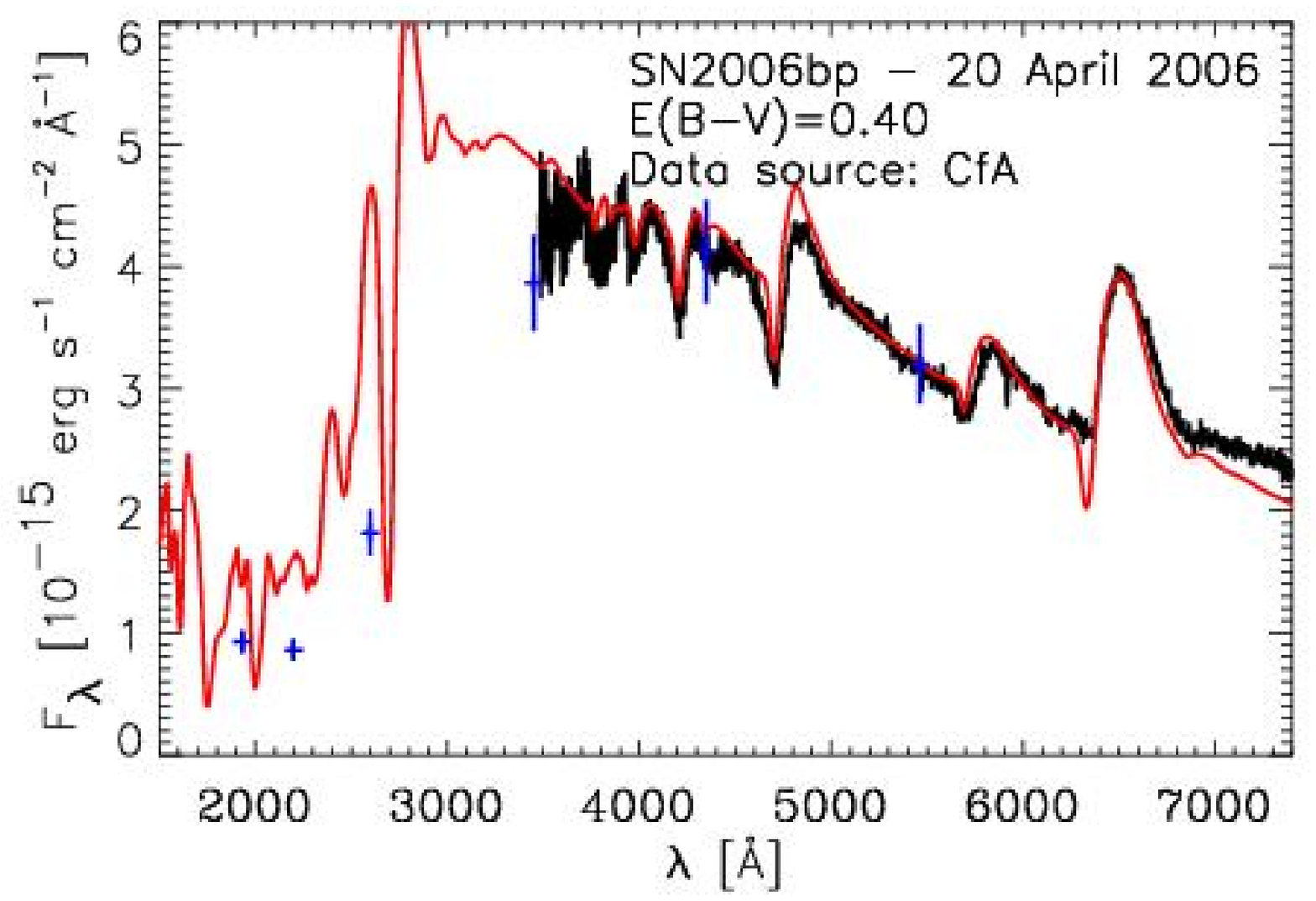}
% \plottwo{sn2006bp_spectra/sn2006bp_0412_tst_n20_6_v1.eps}{sn2006bp_spectra/sn2006bp_0417_n7_j20b_v4_s1_l3_v2.eps}
% \plottwo{sn2006bp_spectra/sn2006bp_0419_n7_j16b_v4_s1_l3_v1_B.eps}{sn2006bp_spectra/sn2006bp_0420_n7_j12b_v4_s1_l3_v1_B.eps}
\caption{
Same as Fig.~\ref{06bp_0411}, but for the SN 2006bp observations of April 12th (top left; $T_{\rm phot}=18120$\,K), 
17th (top right; $T_{\rm phot}=11800$\,K), 19th (bottom left; $T_{\rm phot}=11400$\,K), 
and 20th (bottom right; $T_{\rm phot}=10700$\,K), 2006. A full set of model parameters is given 
in Table~\ref{tab_model_06bp}. 
[See the electronic edition of the Journal for a color version of this figure, 
and see \S\ref{early_06bp} for discussion.]
}
\label{06bp_early}
\end{figure*}
% \clearpage

We could only reproduce this spectrum with CMFGEN by increasing the density exponent to 50 (see also \S\ref{sect_ion}).
Because the density distribution is so steep, line and continuum formation coincide
and occupy a very confined region of space at such early times. 
As time progresses, this region recedes, in the co-moving frame,
to deeper layers where the density profile is flatter, but the spectrum formation region remains confined, 
which prevents the observer from simultaneously probing the outer regions.
Changes in the exponent that describes the density distribution are best inferred through consecutive, rather than isolated, spectroscopic observations (Dessart \& Hillier 2005a).
The peak emission of, e.g., the H$\alpha$ line is blueward
of its rest wavelength, corresponding to a velocity shift of $\sim$8000\,\kms initially, but
decreases with time to reach $\sim$1000\,\kms on the 21st of June. This is consistent with a large
density exponent (Dessart \& Hillier 2005a), although our overestimate of the observed 
H$\alpha$-peak blueshift on the 11th of April (Fig.~\ref{06bp_0411}) suggests that $n$ may not be
as high as 50, and that the weakness of optical lines may not be caused entirely by 
the steep density fall-off. 

We show, in the top left panel of Fig.~\ref{06bp_early}, 
CMFGEN fits to observations of SN 2006bp on the 12th of April 2006.
The spectrum looks very similar to that on the 11th, the fit quality is superior, and the model
differs in having a lower photospheric temperature and velocity.
%          L      T_phot        R_phot     V_phot     rho_phot  N_RHO  
% 0412 &    15   &    18120    &   3.165   &   12550  &   31.6  &  50     \\ % tst_n20_6_v1           
The model parameters are $T_{\rm phot}= 18,120$\,K, $v_{\rm phot}=12,550$\,\kms, and $n=50$.
%: $L_{\ast}= 1.5 \times 10^9 L_{\odot}$, $T_{\rm phot}= 18,120$\,K, 
%$R_{\rm phot}=  3.165 \times 10^{14}$\,cm, $v_{\rm phot}=12,550$\,\kms, 
%$\rho_{\rm phot}= 3.16 \times 10^{-13}$\,g\,cm$^{-3}$, and $n=50$.
The synthetic UV flux also closely matches the flux-equivalent {\sl Swift} UVOT
magnitudes, with a nearly featureless spectrum strongly extinguished by the 
high reddening in the line of sight of $E(B-V)=0.4$. 
% Little line blanketing operates at such early times at and above the photosphere. 
Along a low extinction line of sight, the flux at $\sim$2000\AA\ would have been about ten times 
as large, reaching a level of $\sim$10$^{-13}$\,erg\,cm$^{-2}$\,s$^{-1}$\,\AA$^{-1}$, 
i.e., about a hundred times greater than the optical flux at that time!

The HET observations on the 15th of April 2006 were of very low signal and were 
excluded from our sample. For the 17th of April 2006, we show CMFGEN fits in the top right panel of Fig.~\ref{06bp_early}.
The optical range shows hydrogen Balmer lines and He{\,\sc i}\,5875\AA, weak N{\,\sc ii} features
in the blue edge of He{\,\sc i}\,5875\AA\ and H$\beta$, as well as a weak feature that we attribute to 
O{\,\sc ii} lines in the range 4300--4600\AA\ (see above discussion for SN 2005cs; Baron et al. 2007).
The UV flux no longer dominates the optical and decreases rapidly from the combined effects of cooling 
(due to radiative losses and expansion) and increased line-blanketing. 
The optical lines now appear stronger and are well fitted with a density exponent of 20.
%          L      T_phot        R_phot     V_phot     rho_phot  N_RHO  
% 0417 & 	   9   &    11800    &   5.541   &   11300  &    8.3  &  20     \\ % n7_j20b_v4_s1_l3_v2    
The model parameters for the 17th of April 2006 are $T_{\rm phot}= 11,800$\,K, 
$v_{\rm phot}=11,300$\,\kms, and $n=20$.
%: $L_{\ast}= 9 \times 10^8 L_{\odot}$, 
% $T_{\rm phot}= 11,800$\,K, $R_{\rm phot}=  5.541 \times 10^{14}$\,cm, 
% $v_{\rm phot}=11,300$\,\kms, $\rho_{\rm phot}= 8.3 \times 10^{-14}$\,g\,cm$^{-3}$, and $n=20$.

Two days later, on the 19th of April 2006, the spectrum looks very similar (bottom left panel of 
Fig.~\ref{06bp_early}), with optical lines somewhat stronger and a diminishing UV flux.
%          L      T_phot        R_phot     V_phot     rho_phot  N_RHO  
% 0419 & 	   9   &    11400    &   5.868   &   11050  &    6.5  &  16     \\ % n7_j16b_v4_s1_l3_v1_B  
The model parameters are $T_{\rm phot}= 11,400$\,K, $v_{\rm phot}=11,050$\,\kms, and $n=16$.
% : $L_{\ast}= 9 \times 10^8 L_{\odot}$, $T_{\rm phot}= 11,400$\,K, 
% $R_{\rm phot}=  5.868 \times 10^{14}$\,cm, $v_{\rm phot}=11,050$\,\kms, 
% $\rho_{\rm phot}= 6.5 \times 10^{-14}$\,g\,cm$^{-3}$, and $n=16$.
Note the decreasing density exponent from 50 to 16 in just one week, and the 
decreasing photospheric velocity and temperature. Because the density distribution flattens, the
outer ionized regions contribute more to the photospheric optical depth, shifting it outward
to a lower density (Table~\ref{tab_model_06bp}).
The same holds on the following day, the 20th of April 2006 (Fig.~\ref{06bp_early}, bottom right panel), for which the
%          L      T_phot        R_phot     V_phot     rho_phot  N_RHO  
% 0420 & 	   9   &    10700    &   6.485   &   10300  &    4.5  &  12     \\ % n7_j12b_v4_s1_l3_v1_B  
model parameters are $T_{\rm phot}= 10,700$\,K, $v_{\rm phot}=10,300$\,\kms, and $n=12$.
% : $L_{\ast}= 9 \times 10^8 L_{\odot}$, $T_{\rm phot}= 10,700$\,K, $R_{\rm phot}=  6.485 \times 10^{14}$\,cm, 
% $v_{\rm phot}=10,300$\,\kms, $\rho_{\rm phot}= 4.5 \times 10^{-14}$\,g\,cm$^{-3}$, and $n=12$.
The best-fit density exponent has a value of 12, and changes little after this time to reach the canonical value of 10.
The spectral evolution is very slow and no new lines are identified in the spectrum. N{\,\sc ii} and O{\,\sc ii} lines have, however, disappeared,
following the recombination of the ejecta.   

\begin{figure*}
\plottwo{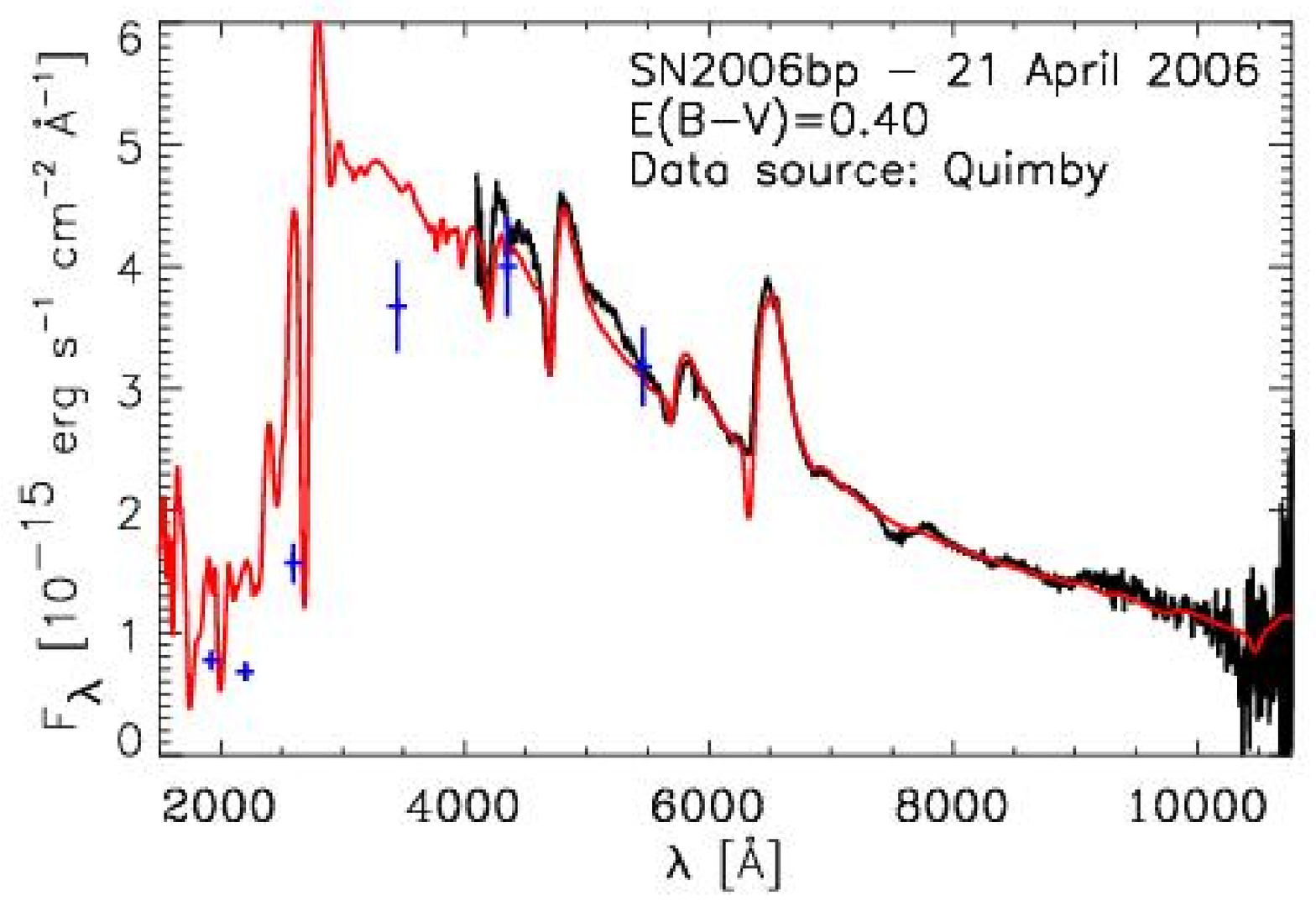}{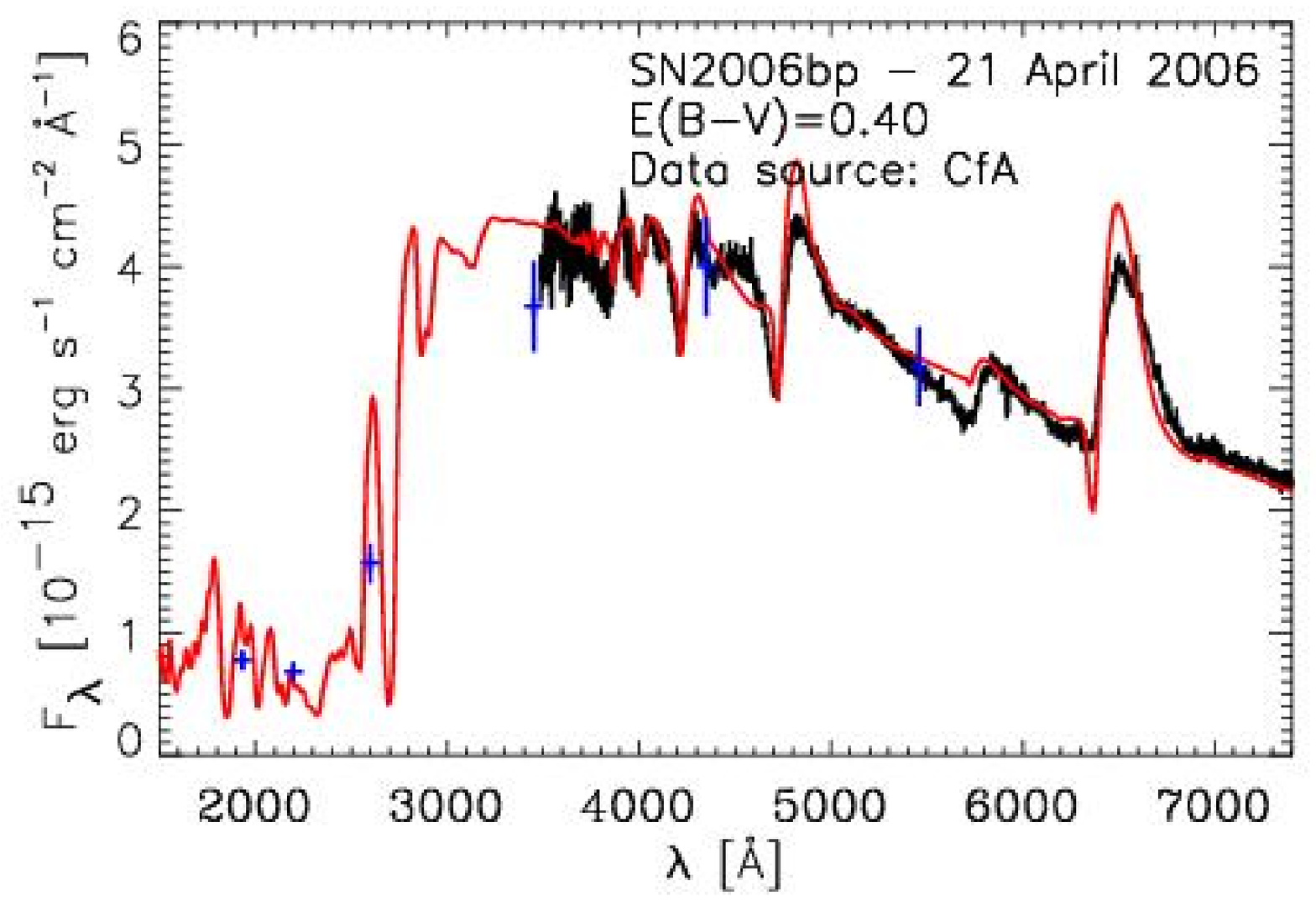}
% \plottwo{sn2006bp_spectra/sn2006bp_0421_n7_j12b_v4_s1_l3_v1_B_QUIMBY.eps}{sn2006bp_spectra/sn2006bp_0421_n16_n10_s0_v1_B_new2_1_n15_v1_CFA.eps}
\caption{
Same as Fig.~\ref{06bp_0411}, but for the SN 2006bp observations on the 21st of April 2006, 
using the data of Quimby et al. (2007; left) or the CfA data (right).
The models used in each panel differ mostly in their temperature/ionization, with $T_{\rm phot}= 10700$\,K 
(left; Table~\ref{tab_model_06bp}) and $T_{\rm phot}= 9700$\,K (right). 
[See the electronic edition of the Journal for a color version of this figure, 
and see \S\ref{early_06bp} for discussion.]
}
\label{06bp_0421}
\end{figure*}

On the 21st of April 2006, we have two sources of observations. The data of Quimby et al. (2007) 
starts only at $\sim$4200\AA\
but extends out to 10700\AA, revealing a weak O{\,\sc i}\,7770\AA\ and the absence of the Ca{\,\sc ii} multiplet
at 8500\AA\ (left panel of Fig.~\ref{06bp_0421}). The {\sl Swift} UVOT data points are well matched 
by the synthetic UV flux distribution, although we overestimate the U-band flux somewhat.
%          L      T_phot        R_phot     V_phot     rho_phot  N_RHO  
% 0421 & 	   9   &    10700    &   6.485   &   10300  &    4.5  &  12     \\ % n7_j12b_v4_s1_l3_v1_B  
The model parameters are $T_{\rm phot}= 10,700$\,K, $v_{\rm phot}=10,300$\,\kms, and $n=12$.
%: $L_{\ast}= 9 \times 10^8 L_{\odot}$, $T_{\rm phot}= 10,700$\,K, $R_{\rm phot}=  6.485 \times 10^{14}$\,cm, 
%$v_{\rm phot}=10,300$\,\kms, $\rho_{\rm phot}= 4.5 \times 10^{-14}$\,g\,cm$^{-3}$, and $n=12$.
In the right panel of Fig.~\ref{06bp_0421}, we present a similar model with a photospheric temperature
$\sim$1000\,K lower, allowing a better match to the UV range, but a poorer match to the optical lines.
This cooler model seems to have the right ionization for the UV flux, but has too low an ionization to 
explain the optical. We surmise that this stems from time-dependent effects, present at all times, but more
visible in spectral lines of species (here, helium) that are about to change their ionization state in the ejecta
(Dessart \& Hillier 2007a,b). 
This cooler model fits much better the weak Fe{\,\sc ii}\,5169\AA\ that indicates
the start of the recombination of Fe{\sc iii} to Fe{\sc ii}, together with the recombination of hydrogen. So this date closes the early epoch of SN 2006bp corresponding to fully ionized conditions
in the photosphere region. 

% \clearpage
% \clearpage

\subsection{The recombination epoch of SN 2006bp}
\label{int_06bp}

   From this epoch onwards, the ejecta in the photospheric layers is too cool to radiate 
a substantial UV flux, and line blanketing, due in particular to Fe{\,\sc ii} lines, 
blocks whatever UV light is produced, re-processing it to 
longer wavelengths. Extinction plays only a secondary role.

\begin{figure*}
\plottwo{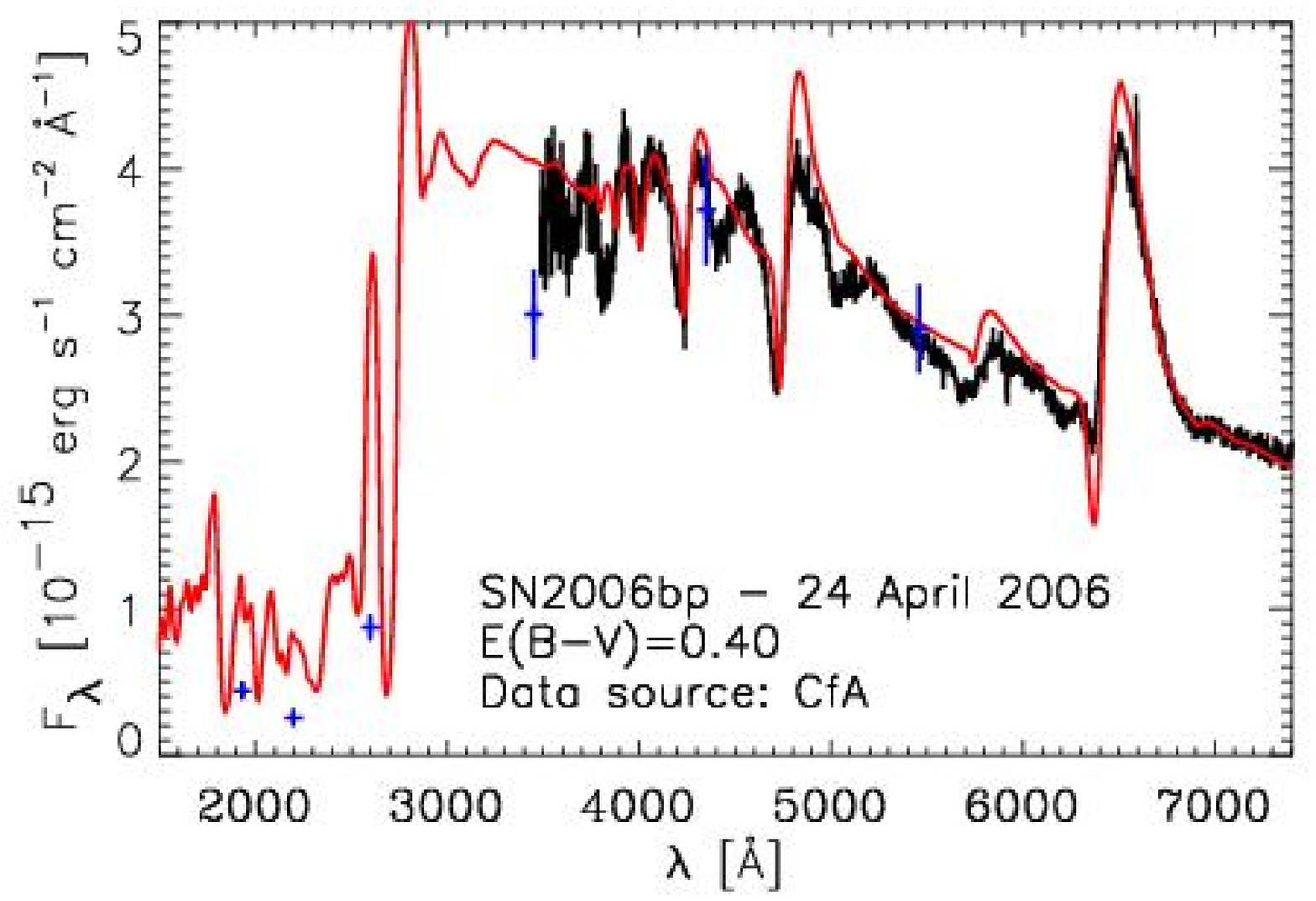}{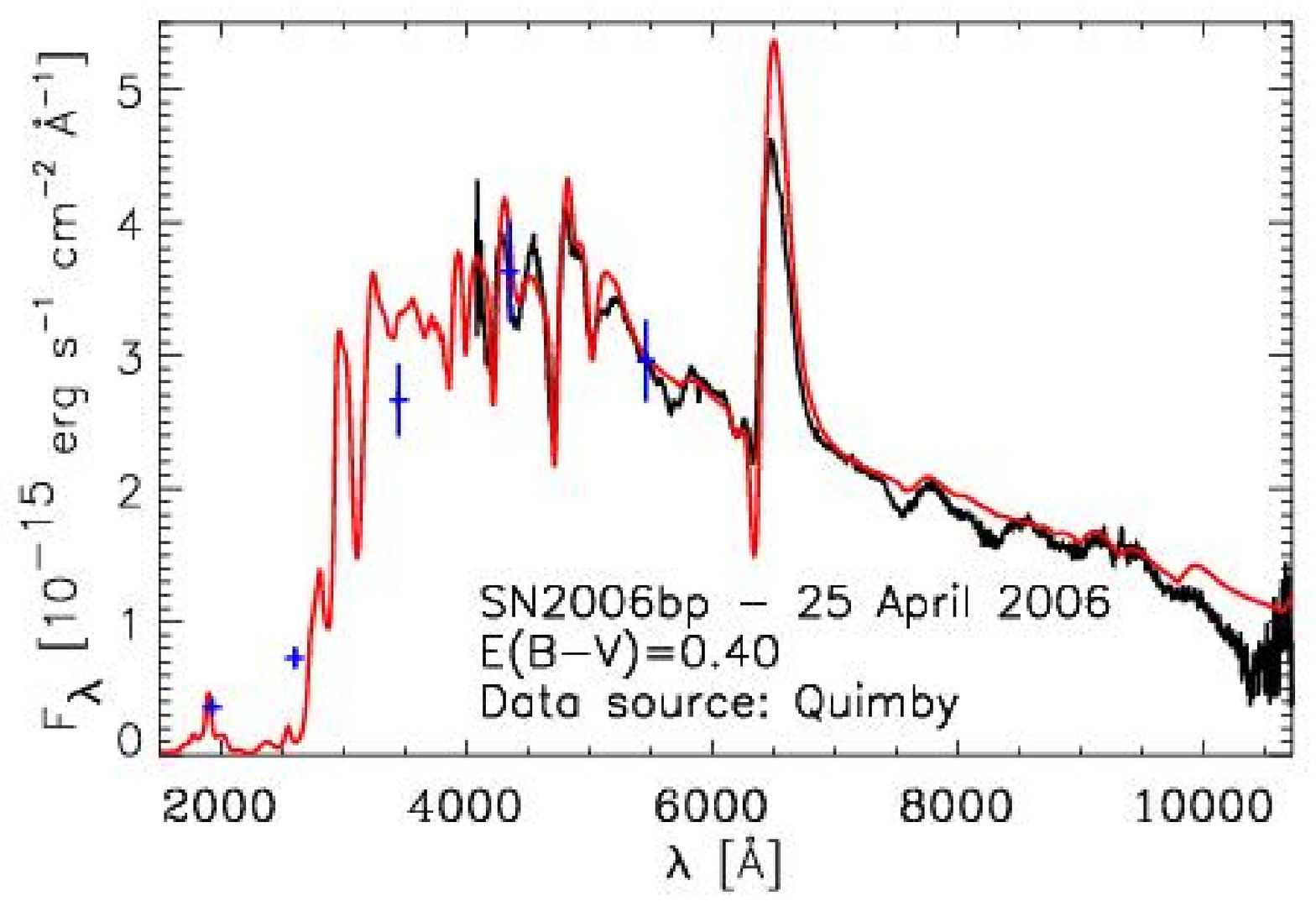}
\plottwo{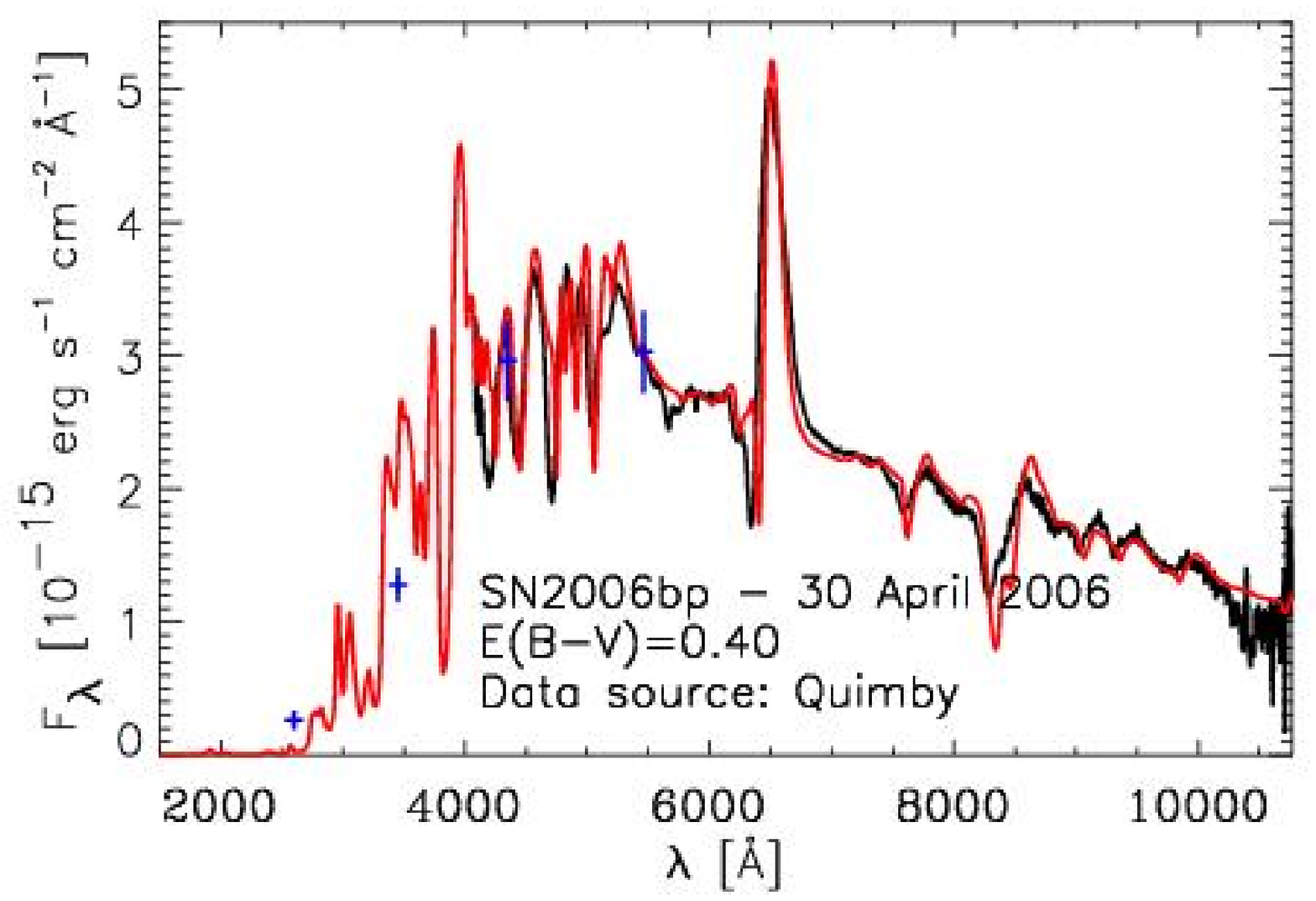}{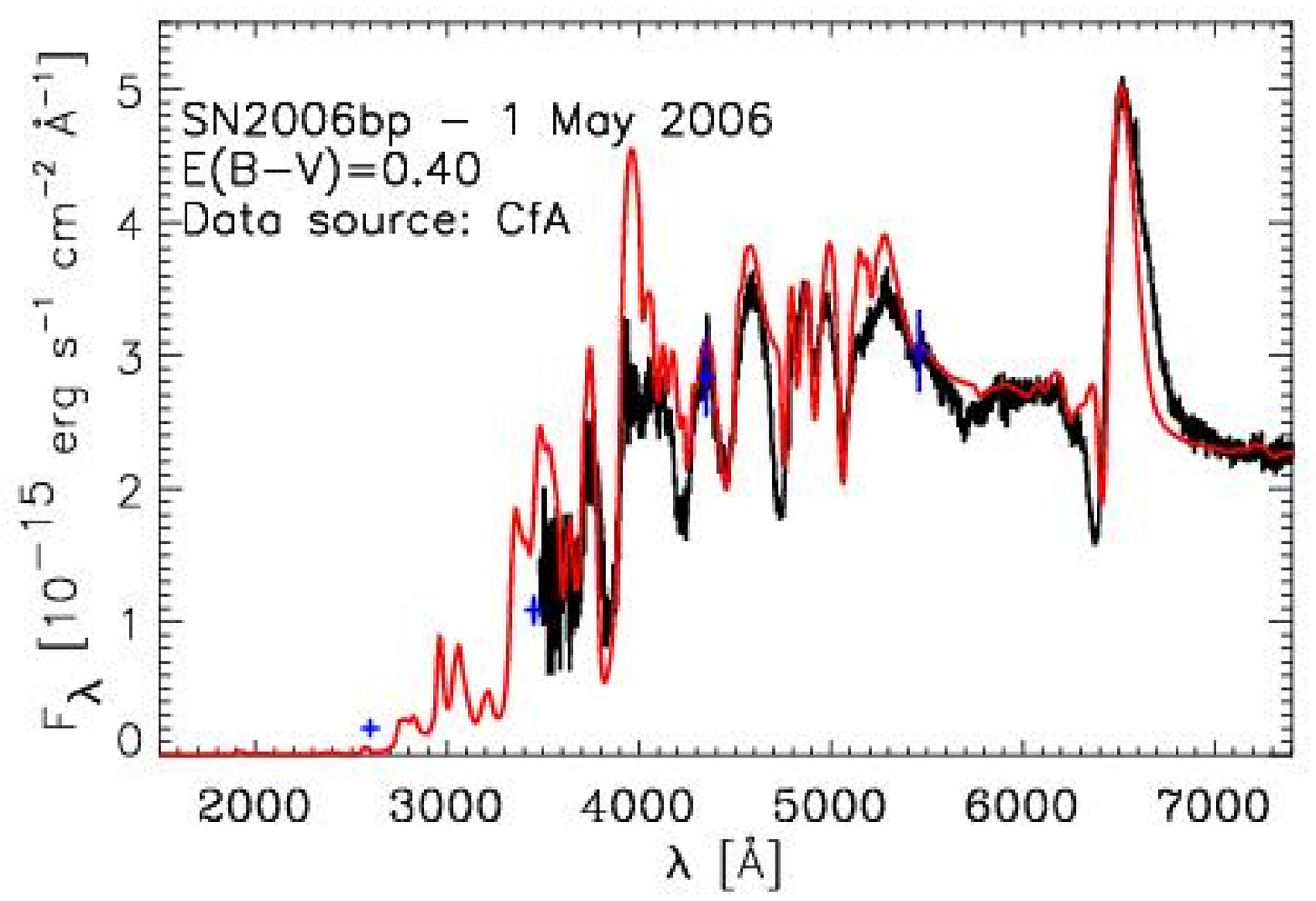}
\plottwo{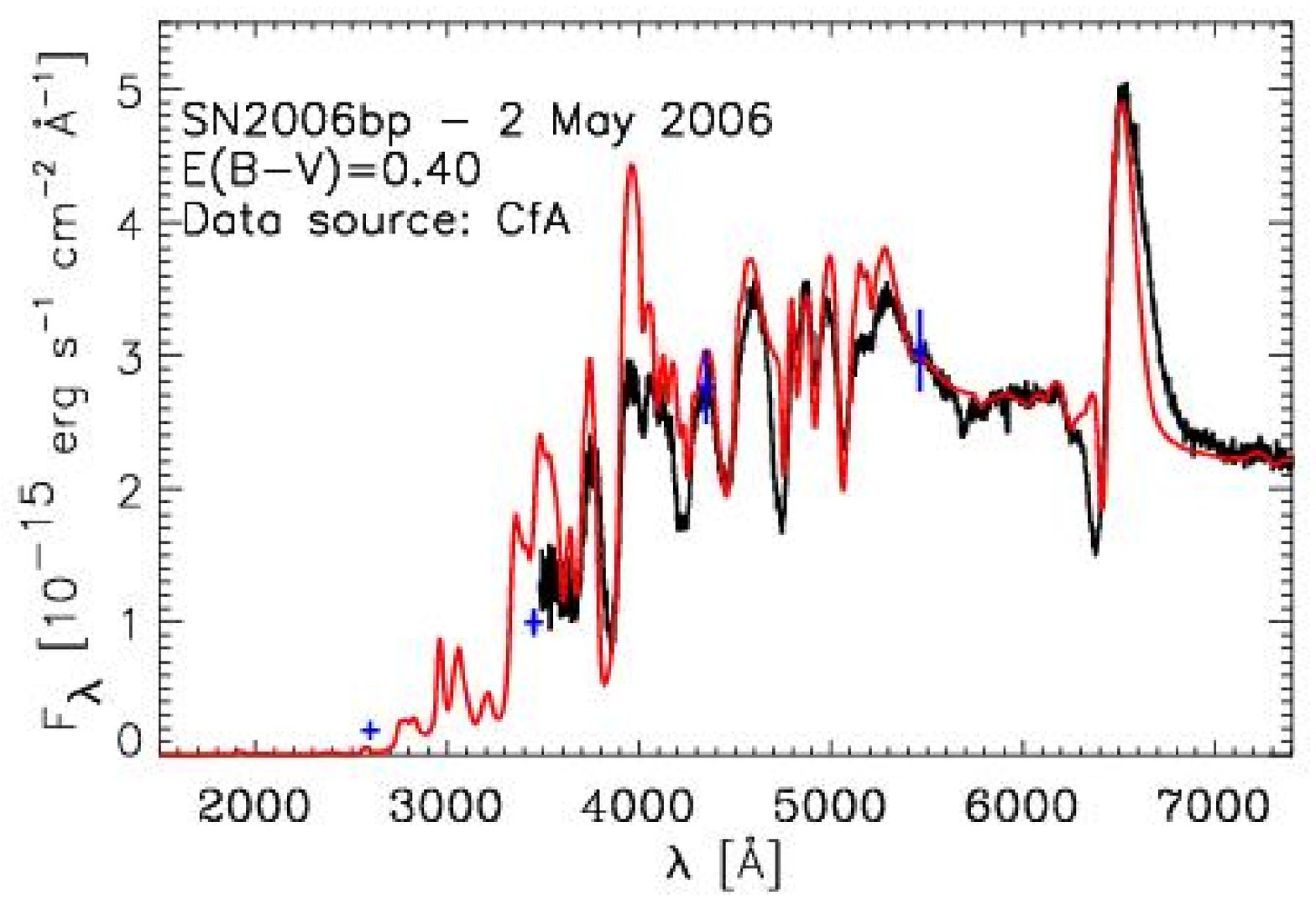}{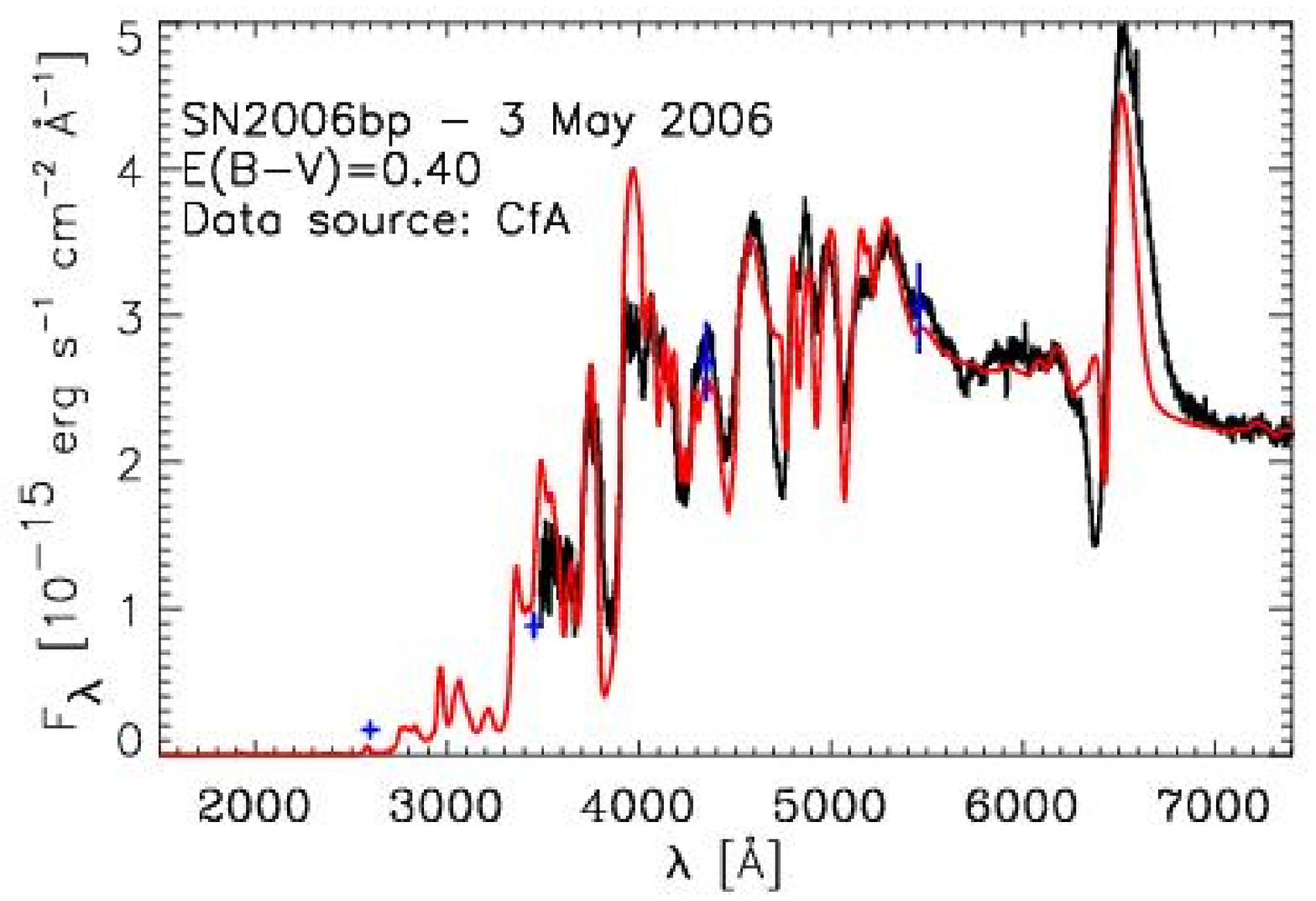}
% \plottwo{sn2006bp_spectra/sn2006bp_0424_n16_n10_s0_v1_B_new2.eps}{sn2006bp_spectra/sn2006bp_0425_n16_n10_s0_v1_B_new2_3.eps}
% \plottwo{sn2006bp_spectra/sn2006bp_0430_n16_n10_s0_v1_B_new6.eps}{sn2006bp_spectra/sn2006bp_0501_n16_n10_s0_v1_B_new5.eps}
% \plottwo{sn2006bp_spectra/sn2006bp_0502_n16_n10_s0_v1_B_new5.eps}{sn2006bp_spectra/sn2006bp_0503_n16_n10_s0_v1_B_new4.eps}
\caption{
Same as Fig.~\ref{06bp_0411}, but for the SN 2006bp observations of April 24th (top left; $T_{\rm phot}=$9350\,K), 
25th (top right; $T_{\rm phot}=$8150\,K), 30th (middle left; $T_{\rm phot}=$7320\,K), and May 1st 
(middle right; $T_{\rm phot}=$7100\,K), 2nd (bottom left; $T_{\rm phot}=$7100\,K), 
and 3rd (bottom right; $T_{\rm phot}=$6800\,K), 2006. 
A full set of model parameters is given in Table~\ref{tab_model_06bp}. 
[See the electronic edition of the Journal for a color version of this figure, 
and see \S\ref{int_06bp} for discussion.]
}
\label{06bp_int}
\end{figure*}

   On the 24th of April 2006 (Fig.~\ref{06bp_int}, top left panel), in addition to the strengthening hydrogen Balmer 
lines and the weakening He{\,\sc i}\,5875\AA, we see the appearance of spectral features due to 
Fe{\,\sc ii} lines, around 4400\AA\ and 5200\AA\ (the main contributor there is Fe{\,\sc ii}\,5169\AA)
and Mg{\,\sc ii}\,4480\AA.
%          L      T_phot        R_phot     V_phot     rho_phot  N_RHO  
% 0424 & 	   5   &     9350    &   6.345   &    7920  &    4.2  &  10     \\ % n16_n10_s0_v1_B_new2   
The model parameters are $T_{\rm phot}= 9350$\,K, $v_{\rm phot}=7920$\,\kms, and $n=10$.
%: $L_{\ast}= 5 \times 10^8 L_{\odot}$, $T_{\rm phot}= 9350$\,K, 
%$R_{\rm phot}=  6.345 \times 10^{14}$\,cm, 
%$v_{\rm phot}=7920$\,\kms, $\rho_{\rm phot}= 4.2 \times 10^{-14}$\,g\,cm$^{-3}$, and $n=10$.
Note that we set the density exponent to 10 for this and all following dates.

% \clearpage
% \clearpage

On the 25th of April 2006 (Fig.~\ref{06bp_int}, top right panel), the lines of Fe{\,\sc ii} strengthen and we 
see the appearance of Si{\,\sc ii}\,6355\AA, although not as clearly as for SN 2005cs due to the higher 
photospheric velocities.
It is not clear if the observed dip at around 5800\AA\ is due to weakening He{\,\sc i}\,5875\AA\ (as supported 
by time-dependent calculations which predict a higher ejecta ionization sustained over a longer time) 
or due to the strengthening Na{\,\sc i}\,D. In the red part of the spectrum, we also see the unambiguous 
presence of O{\,\sc i}\,7770\AA\ and the weak appearance of the Ca{\,\sc ii}\,8500\AA\ multiplets, which are 
not predicted by CMFGEN at this epoch. 

H$\alpha$ is somewhat overestimated both in emission and absorption, which may be
related to the assumption of steady state or to the
constant density exponent adopted at a given epoch.
%          L      T_phot        R_phot     V_phot     rho_phot  N_RHO  
% 0425 & 	   2.7 &     8150    &   6.580   &    8750  &    4.0  &  10     \\ % n16_n10_s0_v1_B_new2_3 
The model parameters are $T_{\rm phot}= 8150$\,K, $v_{\rm phot}=8750$\,\kms, and $n=10$.
%: $L_{\ast}= 2.7 \times 10^8 L_{\odot}$, $T_{\rm phot}= 8150$\,K, $R_{\rm phot}=  6.58 \times 10^{14}$\,cm, 
%$v_{\rm phot}=8750$\,\kms, $\rho_{\rm phot}= 4.0 \times 10^{-14}$\,g\,cm$^{-3}$, and $n=10$.

The spectrum of SN 2006bp on the 30th of 
April 2006 (middle left panel of Fig.~\ref{06bp_int}) shows evidence of recombination in the ejecta.
In particular, Fe{\,\sc ii} lines are now clearly visible throughout the $V$-band, affecting the
H$\alpha$ trough (with a small contribution from Si{\,\sc ii}\,6355\AA). 
Ca{\,\sc ii} lines at 3800\AA\ and 8500\AA\ are also strong and broad. Na{\,\sc i}\,D is 
the likely contributor to the dip at 5800\AA. At this epoch, we employ enhanced oxygen and carbon abundance compared
to the starting conditions (C/He=0.0017 and O/He=0.01 by number), while nitrogen is depleted (N/He=0.0068) and nitrogen lines are not seen.
In the red part of the spectrum, we observe numerous C{\,\sc i} lines due to multiplets 
around 9100\AA, 9400\AA, and 9630\AA. We also predict additional contributions from O{\,\sc i} 
at 7990\AA, 8450\AA, and 9260\AA, but all much weaker than the strong O{\,\sc i}\,7770\AA.
%          L      T_phot        R_phot     V_phot     rho_phot  N_RHO  
% 0430 & 	   1.5 &     7320    &   6.000   &    6450  &    7.3  &  10     \\ % n16_n10_s0_v1_B_new6   
The model parameters are $T_{\rm phot}= 7320$\,K, $v_{\rm phot}=6450$\,\kms, and $n=10$.
%: $L_{\ast}= 1.5 \times 10^8 L_{\odot}$, $T_{\rm phot}= 7320$\,K, $R_{\rm phot}=  6.0 \times 10^{14}$\,cm, 
%$v_{\rm phot}=6450$\,\kms, $\rho_{\rm phot}= 7.3 \times 10^{-14}$\,g\,cm$^{-3}$, and $n=10$.

For the CfA observations of the 1st and 2nd of May 2006bp (middle right and bottom left panels in 
Fig.~\ref{06bp_int}), we employ the same model with $T_{\rm phot}= 7100$\,K, $v_{\rm phot}=6330$\,\kms,
and $n=10$.
%          L      T_phot        R_phot     V_phot     rho_phot  N_RHO  
% 0501 & 	   1.5 &     7100    &   6.147   &    6330  &    8.8  &  10     \\ % n16_n10_s0_v1_B_new5   
% $L_{\ast}= 1.5 \times 10^8 L_{\odot}$, $T_{\rm phot}= 7100$\,K, $R_{\rm phot}=  6.147 \times 10^{14}$\,cm, 
% $v_{\rm phot}=6330$\,\kms, $\rho_{\rm phot}= 8.8 \times 10^{-14}$\,g\,cm$^{-3}$, and $n=10$.
The spectral range misses the red part of the spectrum where
the oxygen, carbon, and calcium lines of interest would be found.
The spectral evolution is slow: we see a slight fading of the flux in the blue, 
with a modest recession of the photosphere, as shown by the narrower line profiles.

We now discuss the late-time observations of SN 2006bp, an epoch that corresponds, in our steady-state modeling
with CMFGEN, to conditions where hydrogen has recombined at and above the photosphere, 
and where the density of free electrons is low.
% \clearpage
\begin{figure*}
\plottwo{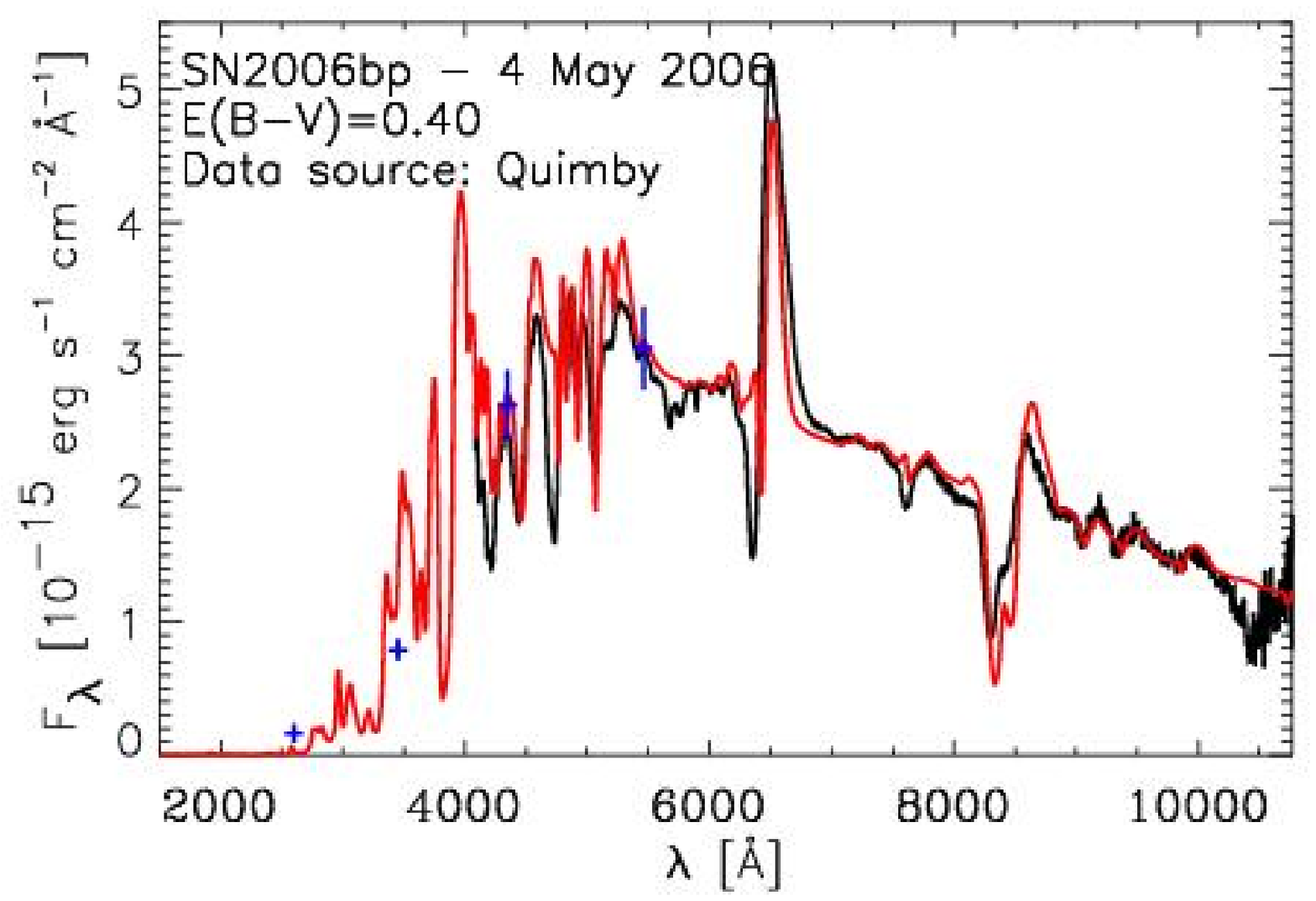}{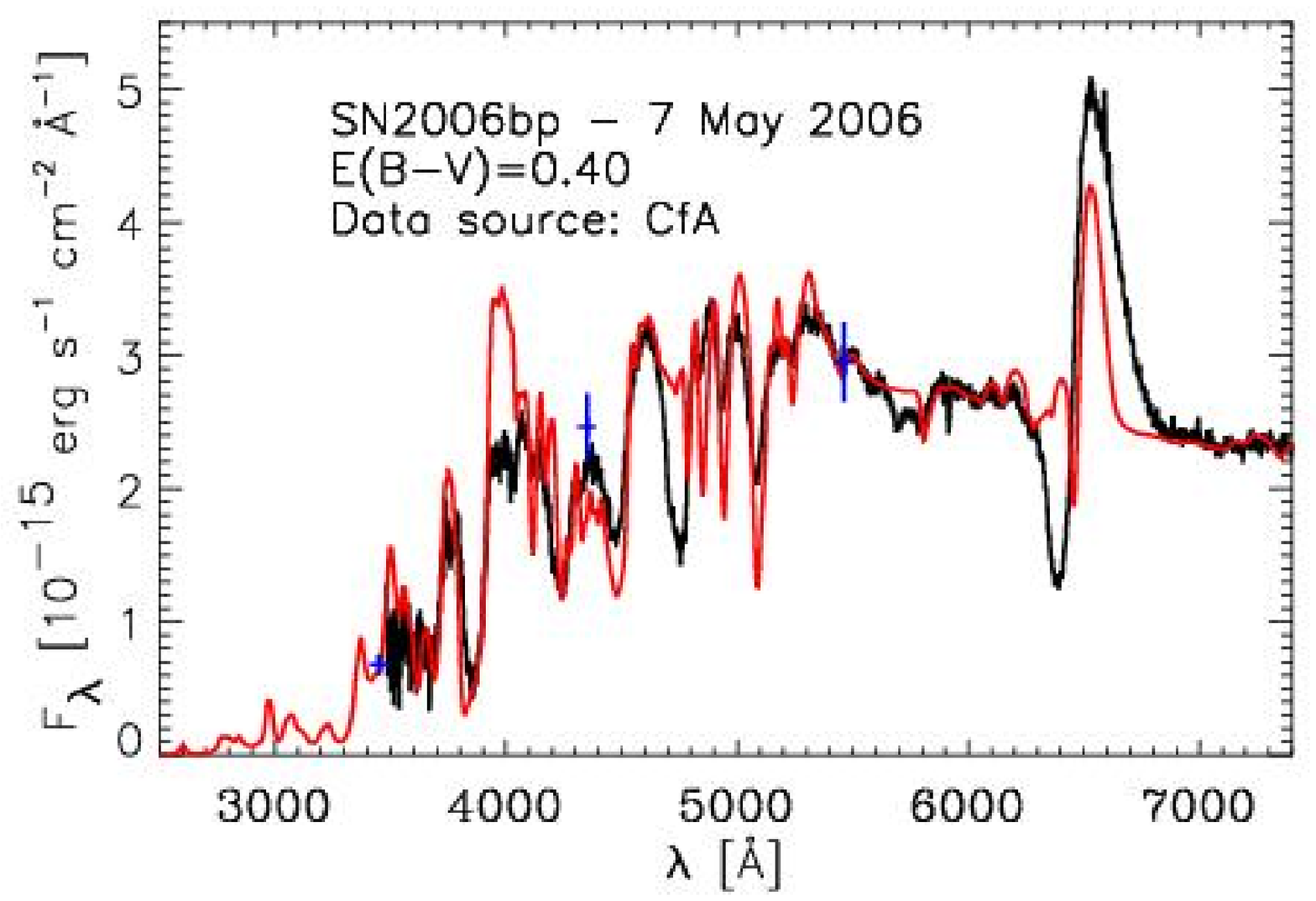}
\plottwo{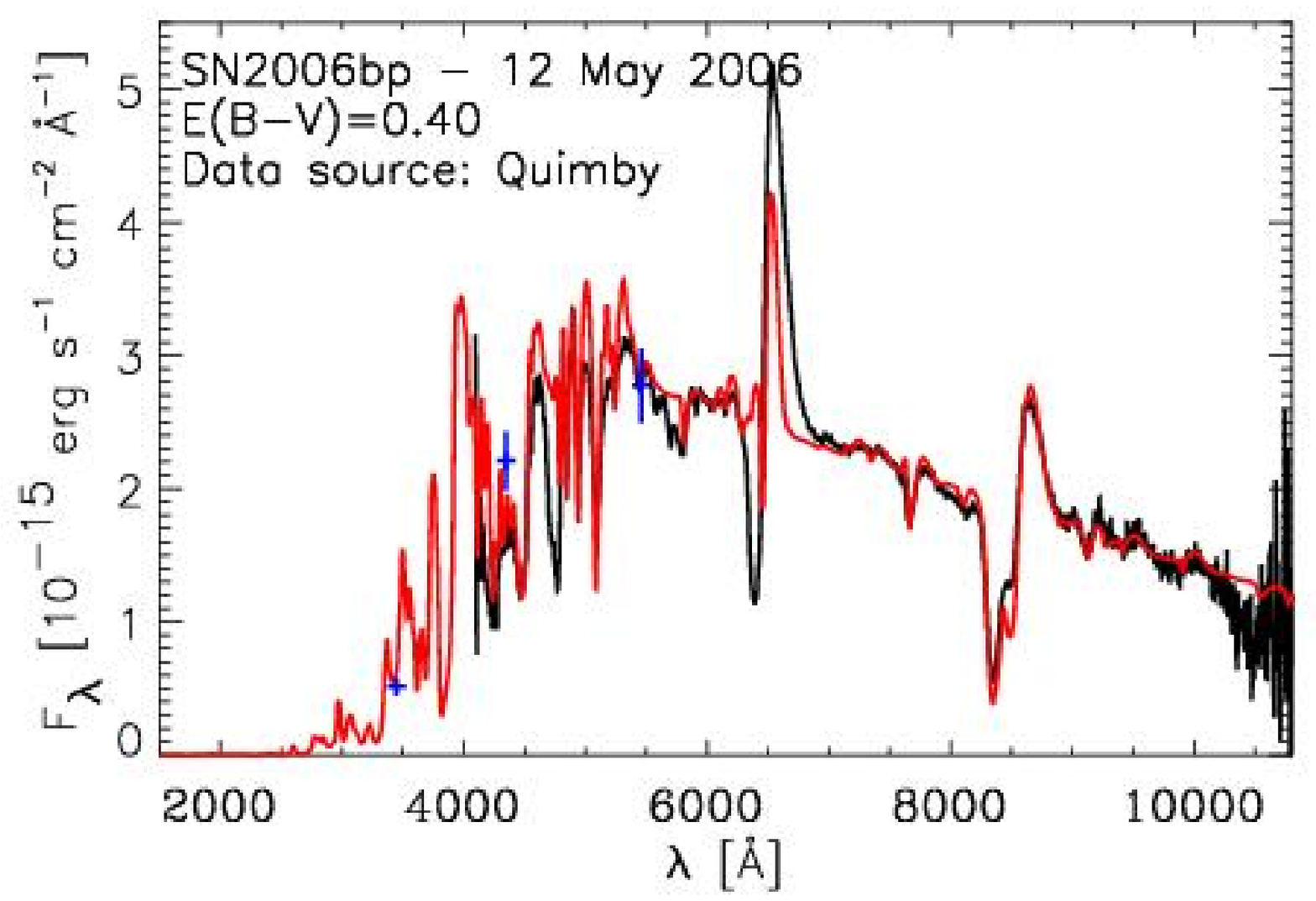}{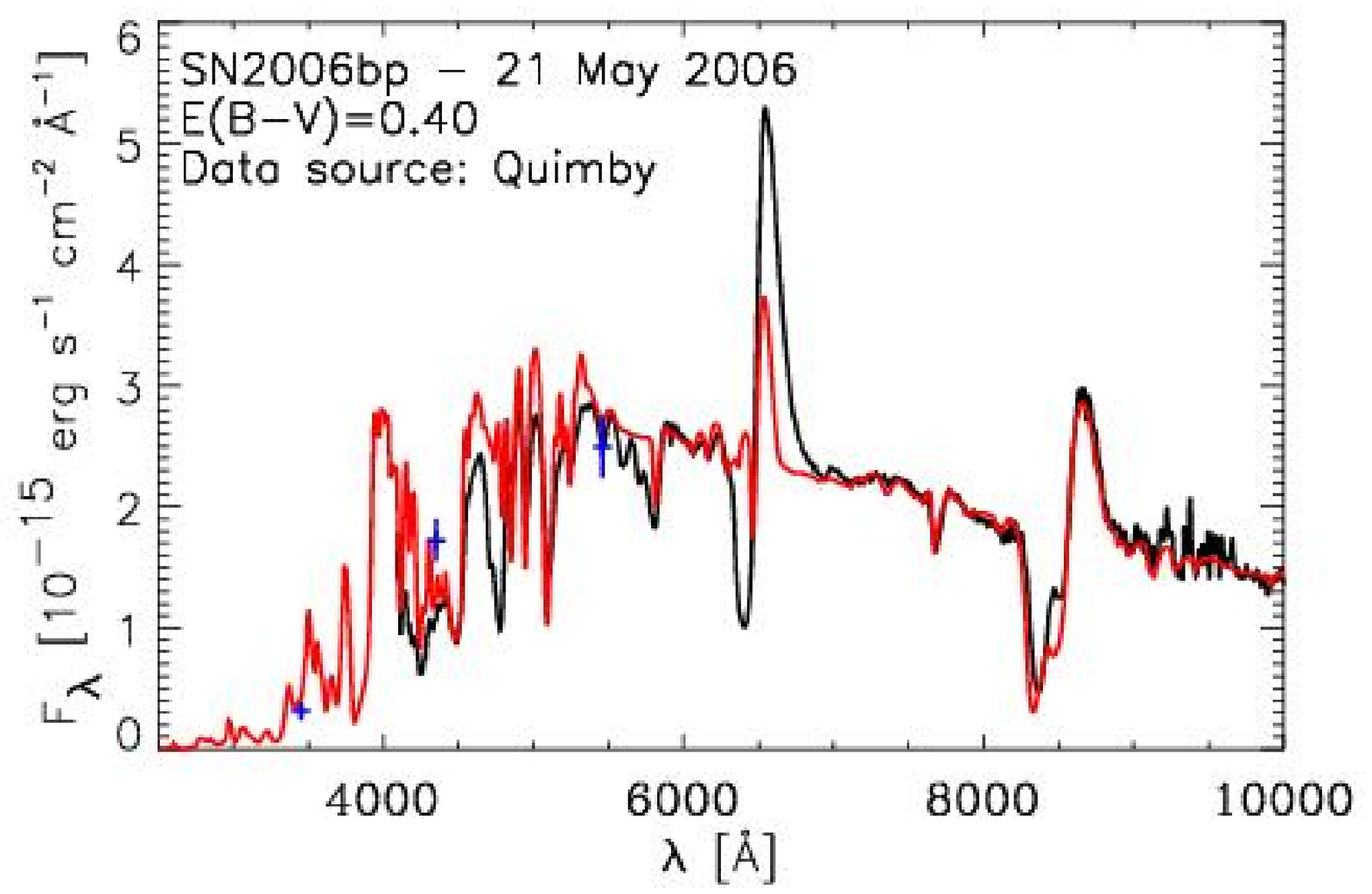}
\plottwo{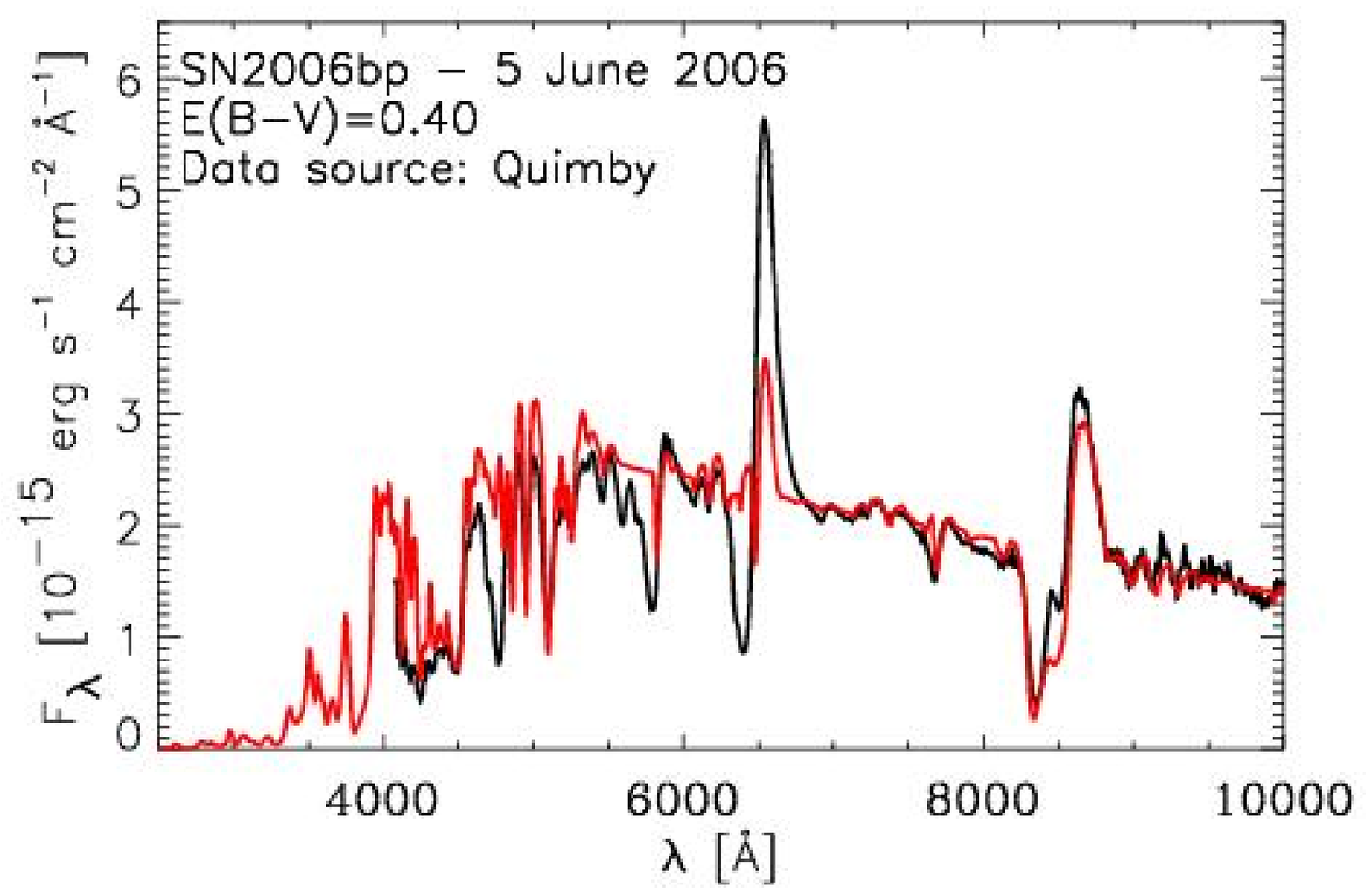}{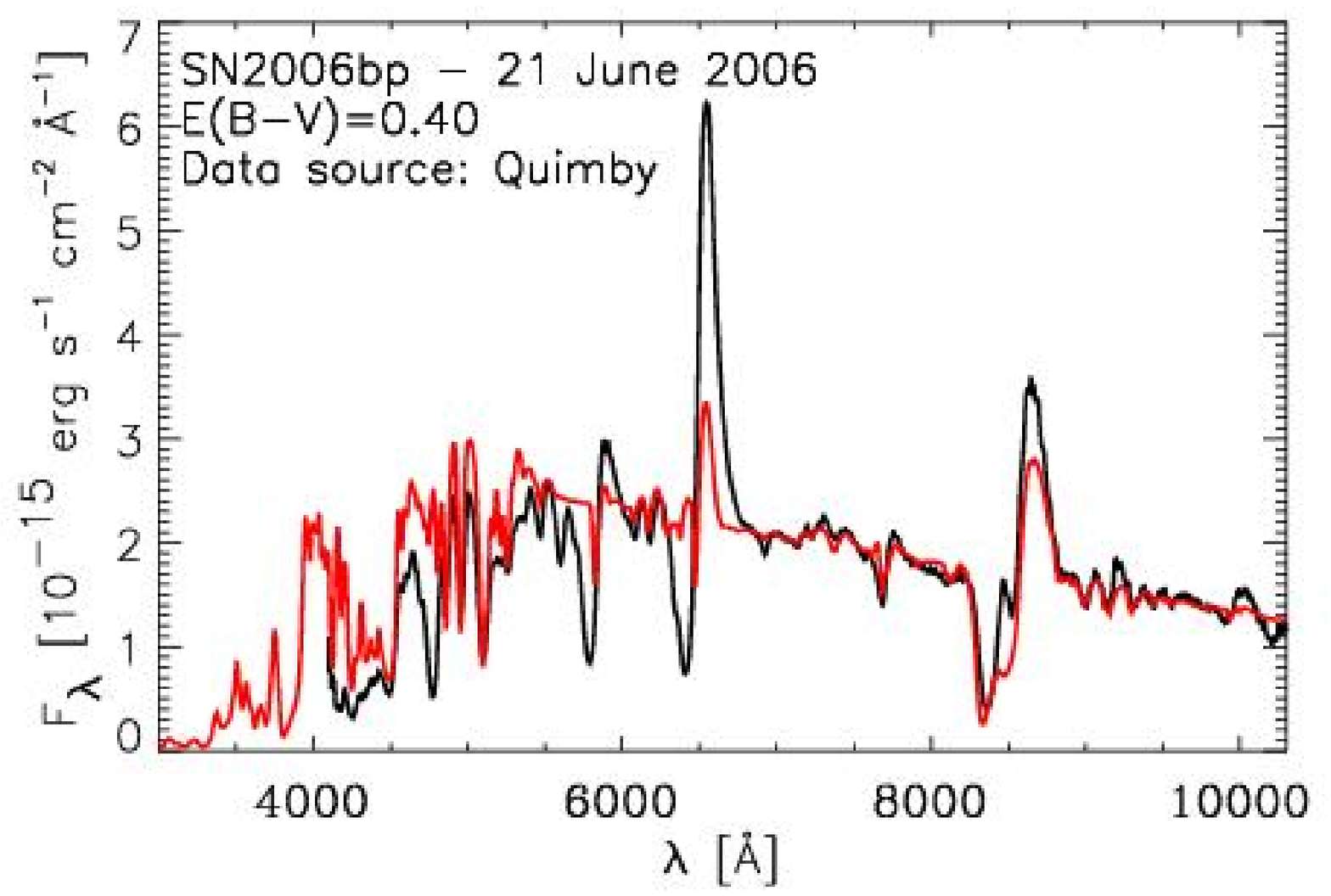}
% \plottwo{sn2006bp_spectra/sn2006bp_0504_n16_n10_s0_v1_B_new4_QUIMBY.eps}{sn2006bp_spectra/sn2006bp_0507_nj4_1v1_new3_abund.eps}
% \plottwo{sn2006bp_spectra/sn2006bp_0512_nj4_1v1_new3_abund.eps}{sn2006bp_spectra/sn2006bp_0521_nj4_1v1_new2_abund.eps}
% \plottwo{sn2006bp_spectra/sn2006bp_0605_nj4_1v1_new1_abund.eps}{sn2006bp_spectra/sn2006bp_0621_nj4_1v1_new_abund.eps}
\caption{
Same as Fig.~\ref{06bp_0411}, but for the SN 2006bp observations of May 4th (top left; $T_{\rm phot}=$6800\,K), 
7th (top right; $T_{\rm phot}=$6420\,K), 12th (middle left; $T_{\rm phot}=$6420\,K), 21st 
(middle right; $T_{\rm phot}=$6300\,K), and June 5th (bottom left; $T_{\rm phot}=$6160\,K), 
and 21st (bottom right; $T_{\rm phot}=$6050\,K), 2006. 
A full set of model parameters is given in Table~\ref{tab_model_06bp}. 
[See the electronic edition of the Journal for a color version of this figure, 
and see \S\ref{int_06bp} for discussion.]
}
\label{06bp_late}
\end{figure*}
% \clearpage

\subsection{Late evolution in the photospheric phase of SN 2006bp}
\label{late_06bp}

 We present in Fig.~\ref{06bp_late} a montage of synthetic spectra fitted to observations
of SN 2006bp on May 4th (top left), 7th (top right), 12th (middle left), 21st (middle right),
and June 5th (bottom left) and 21st (bottom right), 2006. In this elapsed time, the model 
photospheric temperature (velocity) decreases steadily from 6800\,K (5810\,\kms) 
to 6050\,K (3160\,\kms), while the luminosity and the density exponent remain constant 
(a full set of model parameters is given in Table~\ref{tab_model_06bp}).

 Throughout this period, the UV flux fades and the SED peaks in the optical. The continuum energy distribution is 
very well fitted at all times, becoming more and more affected by metal line blanketing, whose
effect appears first in the optical with Fe{\,\sc ii} lines. Contributions 
from Ti{\,\sc ii} lines in the 4200\AA\ region strengthen with time and cause the $B$-band magnitude to plummet.
By the 21st of June, Fe{\,\sc ii} lines even affect the flux in the red part of the spectrum, around 7300\AA,
while O{\,\sc i} and C{\,\sc i} lines continue to contribute numerous narrow features in the red.
Ca{\,\sc ii}\,H\&K is overestimated early on, while the Ca{\,\sc ii} multiplet at 8500\AA\ 
is generally well fitted.
We note the striking evolution of the H$\alpha$ strength, observed with a fixed strength
over this 6-week period, while steady-state CMFGEN models predict an ever decreasing strength
and width (always underestimated) for that line (Fig.~\ref{06bp_late}). 
This is caused by time-dependent effects,
as proposed by Utrobin \& Chugai (2005) and Dessart \& Hillier (2007a,b), and can be remedied
by accounting for the time-dependent terms in the statistical and radiative equilibrium 
equations. Other lines tend to be too narrow, in particular Na{\,\sc i}\,D.
The good fit to the central part of the line profile indicates that the photospheric velocity
is well matched, but the extent of the corresponding emission line regions is systematically
underestimated.
So, while we provide fits to observations at those late epochs, 
time dependent models would yield different parameters and synthetic spectra 
more compatible with the observations. Since time dependence does not profoundly alter the continuum
energy distribution, our synthetic fits may still be useful for distance determinations. It remains to be checked
to what extent the effects introduced by time dependence affect the correction factors used in the EPM,
the true magnitude of the flux dilution, etc (Dessart \& Hillier 2005b). In practice, we exclude the
observations of SN 2006bp on the 5th and on the 21st of June for the determination of its distance.

\section{Distances to Type II-P SN\lowercase{e}}
\label{sect_dist}

  In this section, we compute distances and explosion times of SNe 2005cs and 2006bp 
using our photometric observations and synthetic spectra\footnote{Here, we associate 
the time of explosion with that of shock breakout, which
may be about a day after core bounce, but corresponds well to the time when the entire
envelope is turned into SN ejecta.}.
We use two methods, which emerged as the most suitable
from the detailed discussion in Dessart \& Hillier (2005b) and DH06. The first approach is to apply the
Expanding Photosphere Method (EPM; Kirshner \& Kwan 1974; Eastman \& Kirshner 1989; Schmidt et al. 1994;
Eastman et al. 1996; Hamuy et al. 2001; Leonard et al. 2002a; Elmhamdi et al. 2003), but using correction factors
and color temperatures calculated directly from our tailored models at each epoch (as in DH06). This approach 
is superior to using analytical formulae describing the dependence of correction factors
on color temperature, since those are computed
for models that may not match the observed spectrum. This approach avoids the physical scatter that stems from such a diversity among SN II.
The second approach is to use our synthetic spectra to directly fit observations,
as described in DH06, in a similar fashion to the SEAM (Baron et al. 2004).
For both, the observed magnitudes shown in Table~\ref{tab_phot_05cs} and \ref{tab_phot_06bp} 
are used, either directly or interpolated linearly between adjacent dates to correspond to the date of the
spectroscopic observation. This does not introduce any sizable error given the very slow 
change (nearly constant from day to day within the errors) of $B$, $V$, and $I$ 
magnitudes in the early photospheric phase of Type II-P SNe.

We have described at length the EPM in Dessart \& Hillier (2005b), complementing the previous work of 
Eastman et al. (1996), providing some insights into the
method and its limitations, as well as a new and independently determined set of correction factors 
based on a large database of CMFGEN models.
Correction factors determined with CMFGEN
are comparable to, albeit systematically 10-20\% larger than, those of Eastman et al., although
the origin of this small, but important, difference remains unclear. In DH06, we applied the EPM
to SN 1999em and found that, unless the correction factors and associated color temperatures
are determined with tailored models, the EPM distance and the explosion time were inconsistent when different bandpass-sets were used (amongst $BV$, $BVI$, and $VI$). Since the photospheric radius is nearly constant across the optical range (Dessart \& Hillier 2005b) the inferred distance ought to be the same whatever optical set is used. 

Another source of uncertainty, although smaller, is the inference of the photospheric velocity, $v_{\rm phot}$,
usually approximated by adopting the Doppler velocity at maximum absorption in 
weak and seemingly isolated P-Cygni profiles (note that sphericity and projection effects, subtleties in the 
line transfer problem, and diversity in atomic properties lead to a non-trivial relation
between the photospheric velocity and the velocity at maximum absorption in a given line; 
see Dessart \& Hillier 2005b for a discussion on this issue). 
In our approach, we adopt the value from our CMFGEN model.
Although our model fits are satisfactory, they sometimes overestimate or underestimate
profile widths, matching some profiles very well while doing a poorer job on others at the same time.
There is likely a genuine range of velocities at the photosphere due to fluid instabilities and 
density variations, so that a 10\% inaccuracy in the photospheric velocity at a given time may in fact exist. 
However, we expect the sense and the magnitude of the uncertainty in our inferences to behave randomly.
We do not consistently overestimate or underestimate v$_{\rm phot}$, 
or the flux in a given spectral region, so that 
observations at many times should reduce the uncertainty.

Reddening is another ingredient requiring special care in distance estimates. In our approach,
we use synthetic SEDs to constrain the reddening $E(B-V)$ within $\pm0.05$. The dust properties in other galaxies 
may not be well reproduced by the Cardelli law, which we employ, and this introduces an additional uncertainty, which is significant for the  strongly extinguished SN 2006bp (for such a discussion, see, e.g., Elias-Rosa et al. 2006).

For both objects, we have a good constraint on the explosion time from prior optical
detection/non-detection.
However, compact progenitor stars like blue supergiants brighten in the optical much more quickly than RSG stars
with their extended envelopes, whose SED may peak in the UV a few days after explosion (the
difference comes from the energy penalty associated with the ejecta expansion). Hence,
optical non-detection/detection may not constrain the explosion date to better than one day.
Searching for the UV and X-ray signatures at shock breakout would be more valuable in this respect.

\subsection{SN 2005\lowercase{cs}}
\label{sect_dist_05cs}

%   date     halpha  6562.79 hbeta 4861.3 hgamma 4340.46 hei 5875    FeII 5169     Model
%   0630      7500                6500        6000.        6800        -1.         6880    
%   0701      7000                6000        6000         6000       -1.	   6950
%   0702      6800                6000        5500         5600       -1.	   6370
%   0703      6200                5800        5500         5500       -1.	   6080
%   0704      6500                6000        5300         5000       -1.	   6080
%   0705      6400                6000        5300         5000       -1.	   5570
%   0706      6200                5900        5200         4500       -1.	   5230
%   0709      5900                5300        5000         -1.        4800.	   4710
%   0710      5800                5200        5000         -1.        4800.	   4580
%   0711      5500                5000        -1.          -1.        4100	   4440
%   0712      5300                4800        -1.          -1.        4000	   4320
%   0714      5200                4300        -1.          -1.        3800	   3930
%   0728      4000                -1.         -1.          -1.        2700	   2240

   The high quality spectral fits to the 13 photospheric-phase observations of SN 2005cs 
presented in the preceding sections are well suited to determine the distance
to the SN and its host galaxy M\,51a (NGC\,5194). 

% \clearpage
\begin{deluxetable*}{ccrrrrrrrrr}
%\rotate
\tablewidth{16cm}
\tabletypesize{\scriptsize}
\tablecaption{EPM quantities for, and EPM-based distance to, SN2005\lowercase{cs}
\label{tab_epm_05cs}}
\tablehead{
\colhead{Julian Date}&
\colhead{Day}&
\multicolumn{3}{c}{Angular Size}& 
\multicolumn{3}{c}{Correction Factor}& 
\multicolumn{3}{c}{Color Temperature} 
\\
\colhead{}&
\colhead{}&
\multicolumn{3}{c}{(10$^8$\,km\,Mpc$^{-1}$)}& 
\multicolumn{3}{c}{}& 
\multicolumn{3}{c}{(K)}
\\ 
\colhead{}&
\colhead{}&
\colhead{$\theta_{BV}$}&
\colhead{$\theta_{BVI}$}&
\colhead{$\theta_{VI}$}&
\colhead{$\xi_{BV}$}&
\colhead{$\xi_{BVI}$}&
\colhead{$\xi_{VI}$}&
\colhead{$T_{BV}$}&
\colhead{$T_{BVI}$}&
\colhead{$T_{VI}$}
}
\startdata
  2453552.25   &  2005-06-30  &  3.006 &  3.028 & 3.028 &  0.562 & 0.597 &  0.617 & 17835 &  16813 & 16122   \\ % cs0_1_l1b1_rho_v1_B	     
  2453553.25   &  2005-07-01  &  3.778 &  3.822 & 3.822 &  0.496 & 0.532 &  0.552 & 16272 &  15280 & 14649   \\ % cs0_1_l1b1_rho3_B_v1     
  2453554.50   &  2005-07-02  &  3.645 &  3.778 & 3.822 &  0.496 & 0.527 &  0.552 & 16242 &  15400 & 14619   \\ % cs0_1_l1b1_rho3_B	     
  2453555.25   &  2005-07-03  &  4.505 &  4.527 & 4.527 &  0.466 & 0.501 &  0.527 & 14438 &  13597 & 12935   \\ % cs0_1_l1b_B1_v1	     
  2453556.00   &  2005-07-04  &  5.139 &  5.139 & 5.104 &  0.431 & 0.486 & 0.526  & 13498 &  12327 & 11456   \\ % cs0_1_l1b_B1_v1_2
  2453557.75   &  2005-07-05  &  5.783 &  5.748 & 5.692 &  0.403 & 0.471 & 0.518  & 12792 &  11441 & 10525   \\ % cs0_1_l1b_B1_v1_5
  2453558.25   &  2005-07-06  &  5.982 &  5.960 & 5.872 &  0.416 & 0.486 &  0.542 & 12184 &  10951 & 10020   \\ % cs1_7		     
  2453561.25   &  2005-07-09  &  6.599 &  6.643 & 6.709 &  0.471 & 0.486 &  0.491 & 10200 &   9989 &  9899   \\ % cs3_3		     
  2453562.50   &  2005-07-10  &  6.776 &  6.820 & 6.864 &  0.552 & 0.506 &  0.491 &  8997 &   9448 &  9689   \\ % cs3_5		     
  2453563.50   &  2005-07-11  &  6.291 &  6.423 & 6.555 &  0.567 & 0.516 &  0.486 &  9118 &   9599 & 10080   \\ % r3_e_abund		     
  2453564.25   &  2005-07-12  &  6.445 &  6.577 & 6.754 &  0.622 & 0.537 &  0.491 &  8547 &   9268 &  9929   \\ % r3_abund		     
  2453566.00   &  2005-07-14  &  6.291 &  6.379 & 6.423 &  0.888 & 0.657 &  0.542 &  6923 &   7975 &  9178   \\ % r3_g_abund		     
  2453580.25   &  2005-07-28  &  7.084 &  7.084 & 7.216 &  1.359 & 0.858 &  0.642 &  5240 &   6262 &  7525   \\ % r4_b                     
\hline
\multicolumn{2}{c}{}& 
\multicolumn{3}{c}{$BV$ set}& 
\multicolumn{3}{c}{$BVI$ set}& 
\multicolumn{3}{c}{$VI$ set}\\
\hline 
    \multicolumn{11}{c}{Using First 12 Dates}     \\
\hline 
  \multicolumn{2}{c}{D (Mpc)}             &  
  \multicolumn{3}{c}{9.0 $\pm$ 0.5}       &  
  \multicolumn{3}{c}{8.9 $\pm$ 0.5}       &  
  \multicolumn{3}{c}{8.7 $\pm$ 0.6}       \\
  \multicolumn{2}{c}{t$_{\rm explosion}$ (JD)} &  
  \multicolumn{3}{c}{2453547.6 $\pm$ 0.5}      &  
  \multicolumn{3}{c}{2453547.6 $\pm$ 0.5}      &  
  \multicolumn{3}{c}{2453547.7 $\pm$ 0.6} \\
\hline
    \multicolumn{11}{c}{Using All 13 Dates}     \\
\hline 
  \multicolumn{2}{c}{D (Mpc)}             &  
  \multicolumn{3}{c}{9.0 $\pm$ 0.5}       &  
  \multicolumn{3}{c}{8.9 $\pm$ 0.5}       &
  \multicolumn{3}{c}{8.7 $\pm$ 0.5}      \\
  \multicolumn{2}{c}{t$_{\rm explosion}$ (JD)} &
  \multicolumn{3}{c}{2453547.6 $\pm$ 0.5}      &
  \multicolumn{3}{c}{2453547.6 $\pm$ 0.5}      &
  \multicolumn{3}{c}{2453547.7 $\pm$ 0.5}      \\
\enddata
\tablecomments{Summary of EPM quantities for SN 2005cs. The quantity $\theta/v$  is computed assuming a reddening $E(B-V)=0.04$.
(See text for discussion.)}
\end{deluxetable*} 

\begin{deluxetable}{lcccccc}
%\rotate
\tablewidth{9cm}
\tabletypesize{\scriptsize}
\tablecaption{Distance to SN2005\lowercase{cs} using a direct spectral-fitting approach.
\label{tab_seam_05cs}}
\tablehead{
\colhead{t$_{\rm explosion}$}&
\colhead{d}&
\colhead{$E(B-V)$}&
\colhead{$v_{\rm scale}$}&
\colhead{Set}&
\colhead{$n_{\rm obs}$}
\\
\colhead{(JD)}&
\colhead{(Mpc)}&
\colhead{}&
\colhead{($v_{\rm phot}$)}&
\colhead{}&
\colhead{}
}
\startdata
       2453547.8 &   8.9 &    0.04 &      1.0 &  $BV$   &   13 \\
       2453547.8 &   8.8 &    0.04 &      1.0 &  $BVI$  &   13 \\
       2453547.9 &   8.7 &    0.04 &      1.0 &  $VI$   &   13 \\
\hline		        	      	          
       2453547.8 &   8.8 &    0.04 &      1.0 &  $BVI$  &   12 \\
       2453548.0 &   8.5 &    0.04 &      1.0 &  $BVI$  &   11 \\
       2453548.3 &   8.2 &    0.04 &      1.0 &  $BVI$  &   10 \\
       2453548.6 &   7.7 &    0.04 &      1.0 &  $BVI$  &    9 \\
       2453548.7 &   7.5 &    0.04 &      1.0 &  $BVI$  &    8 \\
\hline		        	      	          
       2453547.8 &   8.0 &    0.04 &      0.9 &  $BVI$  &   13 \\
       2453547.8 &   8.8 &    0.04 &      1.0 &  $BVI$  &   13 \\
       2453547.8 &   9.7 &    0.04 &      1.1 &  $BVI$  &   13 \\
\hline		        	      	          
       2453547.8 &   9.3 &    0.00 &      1.0 &  $BVI$  &   13 \\
       2453547.8 &   8.8 &    0.04 &      1.0 &  $BVI$  &   13 \\
       2453547.7 &   8.2 &    0.10 &      1.0 &  $BVI$  &   13 \\
\hline
    2453547.2$^a$&   9.4 &    0.04 &      1.0 &  $BV$   &   13 \\
    2453547.8$^a$&   8.9 &    0.04 &      1.0 &  $BV$   &   13 \\
    2453548.4$^a$&   8.3 &    0.04 &      1.0 &  $BV$   &   13 \\
\enddata 
\tablecomments{Summary of SEAM results for SN 2005cs, as well as tests on dependencies with reddening, photospheric velocity,
and number of observations used in the sample. In this approach, the error on the distance is on the order of 5--10\%
and corresponds to the minimum distance-modulus scatter between dates.
$^a$: In this case, the distance is obtained by assuming the time of explosion, as shown.
JD 2453548.4 corresponds to the latest non-detection prior to discovery (P06).
(See text for discussion.)}
\end{deluxetable} 
% \clearpage

Focusing first on the EPM, we calculated the correction factors and color temperatures
for all dates in our SN 2005cs sample. As in DH06, we employ three bandpass
sets, using \{B,V\}, \{B,V,I\}, and \{V,I\}. We log in  
Table~\ref{tab_epm_05cs} the color temperature and associated correction factors for each model.
Both quantities follow the same trend discussed in Dessart \& Hillier (2005b) or Eastman et al. (1996), 
with comparable values to those found at comparable epochs for SN 1999em (DH06).
At late times corresponding to the recombination epoch, the correction factors rise 
for \{B,V\} and \{B,V,I\}, driven by the
strong Fe{\,\sc ii} and Ti{\,\sc ii} line blanketing in the $B$-band. The resulting
fading of the $B$-band magnitude is, thus, not related to a variation in the 
flux-dilution magnitude.
Also given in Table~\ref{tab_epm_05cs} are the angular sizes resulting from the EPM, at each date.
Note that these are determined with an estimated accuracy of $\sles$10\%, which is a minimum error
that accounts for the mismatch between synthetic and observed spectra.
The SN 2005cs photosphere approximately doubles in radius within one month of explosion,
a comparable increase to that inferred for SN 1999em.
% Note that Tak{\'a}ts \& Vink{\'o} (2006) find a challenging enhancement in this angular size 
% of about 10--20 over the same period, which, in light of our CMFGEN model results, 
% would suggest an outward migration, in the Lagrangean (or co-moving) frame, of the 
% photosphere with time. 
Using the model photospheric velocity (Table~\ref{tab_model_05cs}) 
together with these angular sizes, we use the method of least squares to fit the equation 
$\theta/v_{\rm phot} = (t-t_{\rm explosion})/D$, taking time as the independent variable, 
and list the resulting distance and explosion dates in Table~\ref{tab_epm_05cs}.

Using either the first 12 dates (to test the effect of a shorter time baseline of two weeks,  
we exclude the last date, corresponding to the recombination epoch) or all 13 dates (encompassing four weeks), 
we find a distance of 8.9$\pm$0.5\,Mpc and an explosion that 
occurred on JD 2453547.6$\pm$0.5, i.e., 2.8 days prior to discovery ($\sim$5 days prior 
to the first observation). The agreement between the three bandpasses is good.
Lengthening the time baseline by two weeks by adding the thirteenth date yields the same results, 
which provides further credence to the reliability of this result.
A plot of these EPM data points for the $BVI$ passband, together
with a linear fit to the data, are shown in Fig.~\ref{fig_dist} (top curve).

The second approach is to fit directly the observations with our models. Assuming homologous expansion 
and adopting the model $v_{\rm phot}$ (as in the EPM), we select an explosion time to
obtain a photospheric radius at each epoch. Each model photospheric radius 
is then scaled to match that corresponding value, together with the synthetic magnitude.
We then calculate the distance modulus and the associated standard deviation. We repeat this calculation
for a range of explosion times, using a fine time increment, and select the explosion date that 
corresponds to the smallest scatter in distance modulus among all dates.
The results are given in Table~\ref{tab_seam_05cs}.
Using all 13 dates, a reddening of 0.04, and our model photospheric velocity, we find a distance
of 8.9$\pm$0.7\,Mpc and an explosion date of JD 2453547.8, in agreement with the results
obtained with the EPM (here, the quoted error on the distance corresponds to the minimum distance-modulus scatter
between all observations for the chosen time of explosion, and is typically 5--10\%). 
We investigate, as in DH06 the effect of changing some of the parameters, limiting ourselves to
the $BVI$ set. Varying the number of observations from 13 to 8 moves forward the explosion date by about one day 
while the distance decreases by $\sim$1.4\,Mpc. Reducing the time baseline to less than two weeks does not give a satisfactory solution: observations every three days over 
the first month is optimal for distance determinations based on the SEAM or the EPM.
Increasing the photospheric velocity at all times by 10\% (to mimic a systematic 
overestimate by 10\%) places the SN further away by 0.9\,Mpc, the opposite being true 
for a 10\% systematic underestimate of photospheric velocity. 
These uncertainties are gross overestimates since our models would not give good fits to 
the observations at all times with such systematic over- or under-estimates.
Our multi-wavelength multi-epoch modeling to spectroscopic observations of SN 2005cs 
yields a reddening of 0.04. A value greater than 0.1 would be impossible to accommodate
with our models, but values of 0 or 0.1 would yield acceptable fits, albeit poorer. 
Keeping the same models at all dates, changing the reddening from 0 to 0.1 makes the distance
vary from 9.4\,Mpc down to 8.2\,Mpc. In DH06, we discussed how allowing instead for changes in 
model parameters simultaneously, to adjust to the change in reddening, actually reduces the effect 
of the reddening uncertainty.

Overall, the agreement between the EPM and the SEAM-like distances to SN 2005cs give us confidence that
our inference is accurate (further confidence on such an accuracy could come from a Cepheid distance  
to M 51, in analogy to the case of SN 1999em and NGC 1637; see Leonard et al. 2003, Baron et al. 2004, 
DH06). 
We thus adopt a distance of 8.9$\pm$0.5\,Mpc to SN 2005cs and M\,51, and an explosion
date of JD 2453547.6$\pm$0.5. This inferred explosion date is compatible within errors with the
constraint of 2453549$\pm$1 imposed by prior non-detection (P06), since even after the explosion has occurred, optical brightening is not instant.
Adopting a 10\% uncertainty in the inferred angular size (which corresponds to a minimum error associated
with fitting mismatches) leads to uncertainties
of 0.5\,Mpc and 0.5\,day for the distance and explosion date determined through the EPM. 
The additional uncertainties in photospheric velocity, reddening, and photometry, are small, so that, through
our use of many observations, our inferred distance should be accurate to within $\sles$10\%.

Our inferred distance is slightly larger than two recent EPM-like distance measurements
to SN 2005cs. Using PHOENIX version 14 (Hauschildt \& 
Baron 1999), Baron et al. (2007) obtained a SEAM distance to SN 2005cs of 7.9$^{+0.7}_{-0.6}$\,Mpc 
using all $UBVRI$ magnitudes, but at only two epochs. 
In some sense the level of agreement is satisfying: PHOENIX and CMFGEN are two very distinct codes, 
and use different assumptions in the modeling. Tak{\'a}ts \& Vink{\'o} (2006) 
used the EPM (with analytical correction factors and an estimate of the reddening $E(B-V)=0.1$) and  
the standard candle method of Hamuy \& Pinto (2002) to determine a distance of 7.59$\pm$1.02\,Mpc and 6.36$\pm$1.3\,Mpc,  
respectively.  Their EPM distance was actually an average of two determinations --- one in which the explosion time 
was fixed (d=8.34$\pm0.31$) and one in which the explosion time was a free parameter (d=$6.84 \pm 0.18$). 
The later gave a better fit to the data, but yielded an explosion time 4 days later than that inferred directly 
from the observations. Tak{\'a}ts \& Vink{\'o} (2006) find a challenging enhancement in the photospheric 
angular size of about 10--20 over the same period, which, in light of our CMFGEN model results, 
would suggest an outward migration, in the Lagrangean (or co-moving) frame, of the 
photosphere with time. At later times our results, and those of Tak{\'a}ts \& Vink{\'o},
appear to be consistent; at early times Tak{\'a}ts \& Vink{\'o} estimate much hotter
photospheric temperatures, and smaller photospheric radii.
%Our inferred distance is larger than most previous estimates. Recently, 
%Baron et al. (2007) obtained a SEAM distance to SN 2005cs of 7.9$^{+0.7}_{-0.6}$\,Mpc
%using all $UBVRI$ magnitudes, but at only two epochs.Tak{\'a}ts \& Vink{\'o} (2006) 
%used the EPM (with analytical correction factors and an estimate of the reddening $E(B-V)=0.1$) and 
%the standard candle method of Hamuy \& Pinto (2002) to determine a distance of 7.59$\pm$1.02\,Mpc and 6.36$\pm$1.3\,Mpc, 
%respectively. 

In general, our inferred distance is larger than most other estimates in the literature. 
Previously, using light-curve or spectroscopic modeling of the Type Ic SN 1994I, 
Iwamoto et al. (1994) and Baron et al. (1996) obtained a distance of 
6.9$\pm$1\,Mpc and 6$\pm$1.9\,Mpc, respectively, although the reddening may be uncertain
in the line of sight to this object. 
Georgiev et al. (1990) obtained a distance of 6.9$\pm$0.7\,Mpc using surface photometry of stellar 
associations.
Feldmeier et al. (1997) found a distance of 8.4$\pm$0.6\,Mpc
using planetary nebulae, revised downward by Ciardullo et al. (2002) to 7.6$\pm$0.6\,Mpc
with an improved reddening of 0.04. Note that this is the reddening we infer in the line of sight
to SN 2005cs.
Tonry et al (2001) infer a distance to NGC\,5195 (M 51b, a companion galaxy to M 51) 
of 7.66$\pm$1\,Mpc from surface brightness fluctuations.
Our estimate is compatible, within the errors, with that of Tonry et al. (2001), but mildly in disagreement with the other studies. 
% Finally, the distance estimate of Sandage \& Tammann (1974),
% based on giant H{\,\sc ii} region sizes, predicts a distance of 9.6\,Mpc, marginally compatible within errors
% with our results, although this time overestimating our prediction.
Obtaining a Cepheid distance to M 51 would help in pinning down the distance to that Galaxy and in providing an independent test for the accuracy of our distance estimate.

\subsection{SN 2006\lowercase{bp}}
\label{sect_dist_06bp}

% \clearpage
\begin{deluxetable*}{ccrrrrrrrrr}
%\rotate
\tablewidth{16cm}
\tabletypesize{\scriptsize}
\tablecaption{EPM quantities for, and EPM-based distance to, SN2006\lowercase{bp}
\label{tab_epm_06bp}}
\tablehead{
\colhead{Julian Date}&
\colhead{Day}&
\multicolumn{3}{c}{Angular Size}& 
\multicolumn{3}{c}{Correction Factor}& 
\multicolumn{3}{c}{Color Temperature} 
\\
\colhead{}&
\colhead{}&
\multicolumn{3}{c}{(10$^8$\,km\,Mpc$^{-1}$)}& 
\multicolumn{3}{c}{}& 
\multicolumn{3}{c}{(K)}
\\ 
\colhead{}&
\colhead{}&
\colhead{$\theta_{BV}$}&
\colhead{$\theta_{BVi'}$}&
\colhead{$\theta_{Vi'}$}&
\colhead{$\xi_{BV}$}&
\colhead{$\xi_{BVi'}$}&
\colhead{$\xi_{Vi'}$}&
\colhead{$T_{BV}$}&
\colhead{$T_{BVi'}$}&
\colhead{$T_{Vi'}$}
}
\startdata
% First line is an example from the corresponding table for 2005cs
%  2453552.25   &  06/30/05  &  3.00601 &  3.02806 & 3.02806 &  0.562 & 0.597 &  0.617 & 17835 &  16813 & 16122   \\ % cs0_1_l1b1_rho_v1_B
%
2453836.6  &  2006-04-11 &    2.345  &   2.373  &   2.352  &    0.586  &  0.663  &  0.788  &   24899 &  21736 &  17632   \\	  
2453837.6  &  2006-04-12 &    2.856  &   2.884  &   2.842  &    0.563  &  0.657  &  0.734  &   21916 &  18753 &  16511   \\   
2453842.6  &  2006-04-17 &    4.838  &   4.929  &   4.873  &    0.499  &  0.557  &  0.608  &   15330 &  14009 &  12847   \\   
2453844.6  &  2006-04-19 &    5.111  &   5.216  &   5.160  &    0.480  &  0.538  &  0.589  &   15030 &  13708 &  12547   \\   
2453845.8  &  2006-04-20 &    5.580  &   5.685  &   5.615  &    0.454  &  0.512  &  0.560  &   14489 &  13188 &  12107   \\   
2453846.7  &  2006-04-21 &    5.524  &   5.643  &   5.580  &    0.454  &  0.512  &  0.560  &   14489 &  13188 &  12107   \\   
2453849.6  &  2006-04-24 &    6.239  &   6.351  &   6.365  &    0.422  &  0.490  &  0.544  &   13548 &  12127 &  11026   \\   
2453850.6  &  2006-04-25 &    7.527  &   7.569  &   7.597  &    0.451  &  0.496  &  0.531  &   11126 &  10445 &   9864   \\   
2453855.7  &  2006-04-30 &    7.772  &   7.884  &   7.940  &    0.583  &  0.525  &  0.486  &    8963 &   9524 &  10125   \\   
2453856.8  &  2006-05-01 &    7.891  &   8.003  &   8.039  &    0.634  &  0.544  &  0.486  &    8423 &   9164 &  10025   \\   
2453857.6  &  2006-05-02 &    7.828  &   7.961  &   8.032  &    0.634  &  0.544  &  0.486  &    8423 &   9164 &  10025   \\   
2453858.7  &  2006-05-03 &    8.165  &   8.249  &   8.333  &    0.692  &  0.576  &  0.502  &    7802 &   8583 &   9524   \\   
2453859.6  &  2006-05-04 &    8.039  &   8.165  &   8.263  &    0.692  &  0.576  &  0.502  &    7802 &   8583 &   9524   \\   
2453862.7  &  2006-05-07 &    8.627  &   8.732  &   8.823  &    0.862  &  0.653  &  0.531  &    6681 &   7562 &   8743   \\   
2453867.7  &  2006-05-12 &    8.172  &   8.396  &   8.571  &    0.862  &  0.653  &  0.531  &    6681 &   7562 &   8743   \\   
2453876.7  &  2006-05-21 &    7.969  &   8.200  &   8.375  &    1.058  &  0.724  &  0.560  &    6001 &   7022 &   8343   \\   
\hline
\multicolumn{2}{c}{}& 
\multicolumn{3}{c}{$BV$ set}& 
\multicolumn{3}{c}{$BVi'$ set}& 
\multicolumn{3}{c}{$Vi'$ set}\\
\hline 
    \multicolumn{11}{c}{Using First 13 Dates}     \\
\hline 
  \multicolumn{2}{c}{D (Mpc)}                  & 
  \multicolumn{3}{c}{18.1$\pm$0.8\,Mpc}        &
  \multicolumn{3}{c}{17.8$\pm$0.8\,Mpc}        &
 \multicolumn{3}{c}{17.7$\pm$0.8\,Mpc}         \\
  \multicolumn{2}{c}{t$_{\rm explosion}$ (JD)} &
  \multicolumn{3}{c}{2453833.2$\pm$0.4}        &
  \multicolumn{3}{c}{2453833.3$\pm$0.4}        &
  \multicolumn{3}{c}{2453833.3$\pm$0.4}        \\
\hline
    \multicolumn{11}{c}{Using All 16 Dates}     \\
\hline 
  \multicolumn{2}{c}{D (Mpc)}                  &
  \multicolumn{3}{c}{17.9$\pm$0.7\,Mpc}        &
  \multicolumn{3}{c}{17.5$\pm$0.6\,Mpc}        &
  \multicolumn{3}{c}{17.4$\pm$0.6\,Mpc}       \\
  \multicolumn{2}{c}{t$_{\rm explosion}$ (JD)} &
  \multicolumn{3}{c}{2453833.3$\pm$0.3}        &
  \multicolumn{3}{c}{2453833.4$\pm$0.3}        &
  \multicolumn{3}{c}{2453833.4$\pm$0.3}        \\
\enddata
\tablecomments{Summary of EPM quantities for SN 2006bp. 
The quantity $\theta/v$ is computed assuming a reddening $E(B-V)=0.4$.
(See text for discussion.)}
\end{deluxetable*} 

\begin{deluxetable}{lcccccc}
%\rotate
\tablewidth{9cm}
\tabletypesize{\scriptsize}
\tablecaption{Distance to SN2006\lowercase{bp} using a direct spectral-fitting approach.
\label{tab_seam_06bp}}
\tablehead{
\colhead{t$_{\rm explosion}$}&
\colhead{d}&
\colhead{$E(B-V)$}&
\colhead{$v_{\rm scale}$}&
\colhead{Set}&
\colhead{$n_{\rm obs}$}
\\
\colhead{(JD)}&
\colhead{(Mpc)}&
\colhead{}&
\colhead{($v_{\rm phot}$)}&
\colhead{}&
\colhead{}
}
\startdata
2453833.4    &  17.4  &   0.4  &     1.0 & $BV$    &    16  \\  %% updated
2453833.4    &  17.1  &   0.4  &     1.0 & $BVi'$  &    16  \\  %% updated
2453833.5    &  16.7  &   0.4  &     1.0 & $Vi'$   &    16  \\  %% updated
\hline	          
2453833.5    &  16.8  &   0.4  &     1.0 & $BVi'$  &    15  \\  %% updated
2453833.5    &  16.8  &   0.4  &     1.0 & $BVi'$  &    14  \\  %% updated
2453833.3    &  17.4  &   0.4  &     1.0 & $BVi'$  &    13  \\  %% updated
2453833.3    &  17.6  &   0.4  &     1.0 & $BVi'$  &    12  \\  %% updated
2453833.1    &  18.0  &   0.4  &     1.0 & $BVi'$  &    11  \\  %% updated
2453833.1    &  18.3  &   0.4  &     1.0 & $BVi'$  &    10  \\  %% updated
2453832.9    &  18.9  &   0.4  &     1.0 & $BVi'$  &     9  \\  %% updated
\hline	          
2453833.4    &  17.1  &   0.4  &     1.0 & $BVi'$  &    16  \\  %% updated
2453833.4    &  18.8  &   0.4  &     1.1 & $BVi'$  &    16  \\  %% updated
2453833.4    &  15.4  &   0.4  &     0.9 & $BVi'$  &    16  \\  %% updated
\hline
2453833.4    &  15.9  &   0.45 &     1.0 & $BVi'$  &    16  \\  %% updated
2453833.4    &  17.1  &   0.40 &     1.0 & $BVi'$  &    16  \\  %% updated
2453833.4    &  18.3  &   0.35 &     1.0 & $BVi'$  &    16  \\  %% updated
\hline
2453833.0$^a$&  17.6  &   0.4  &     1.0 & $BVi'$  &    16  \\  %% updated
2453833.5$^a$&  16.9  &   0.4  &     1.0 & $BVi'$  &    16  \\  %% updated
2453834.0$^a$&  16.2  &   0.4  &     1.0 & $BVi'$  &    16  \\  %% updated
\enddata 
\tablecomments{Summary of SEAM results for SN 2006bp, as well as tests on dependencies with reddening, photospheric velocity,
and number of observations used in the sample. In this approach, the error on the distance is of the order of 5-10\%,
and corresponds to the minimum distance-modulus scatter between dates.
$^a$: In this case, the distance is obtained by assuming the time of explosion, as shown.
(See text for discussion.)}
\end{deluxetable} 
% \clearpage

  We now apply the methods of the previous section to SN 2006bp.
To avoid the hydrogen-recombination phase, when steady-state CMFGEN models struggle to reproduce the
hydrogen Balmer and Na{\,\sc i}\,D lines, we exclude the two extreme observations 
in our sample, those of the 5th and of the 21st of June 2006, leaving us with 16 observations
and a time coverage of 4 weeks.

We present the EPM-inferred quantities in Table~\ref{tab_epm_06bp}, giving the 
angular size, correction factors and color temperatures for the three sets of bandpasses
$BV$, $BVi'$, and $Vi'$. Because of the faster expansion of SN 2006bp compared to SN 2005cs, the photospheric angular size more than triples 
over the first month of observation. At the 
recombination epoch, the associated front, which coincides closely with the photosphere,
recedes in the co-moving frame at a rate that is comparable to the expansion rate of the ejecta.
At late times, the angular size of the photosphere is, thus, roughly constant.

\begin{figure}
\epsscale{1}
\plotone{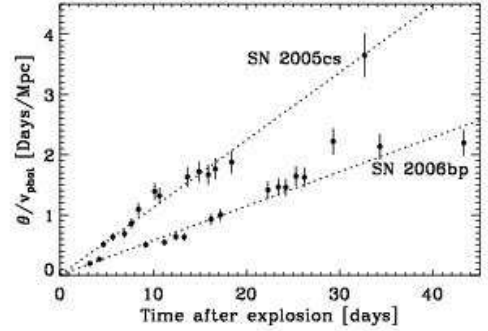}
% \plotone{dist.ps}
\caption{
Linear fit (dotted line) to the $BVI$ data shown in Table~\ref{tab_epm_05cs} for SN 2005cs (using 13 dates) 
and to the $BVi'$ data shown in Table~\ref{tab_epm_06bp} for SN2006bp (using 16 dates). 
The reciprocal of the slope is the distance $D$,
i.e. $\theta/v_{\rm phot} = (t-t_{\rm explosion}) / D$, where $\theta = R_{\rm phot}/D$.
The origin for the time is the inferred time of explosion (more specifically the shock breakout time).
The vertical bars indicate a $\pm$10\% error.
}
\label{fig_dist}
\end{figure}

Using all 16 observations in our sample and our EPM-based technique leads to a distance
of 17.5$\pm$0.6\,Mpc for the $BVi'$ set, only slightly varying with bandpass set 
(17.9$\pm$0.7\,Mpc for the $BV$ set and 17.4$\pm$0.6\,Mpc for the $Vi'$ set; 
Table~\ref{tab_epm_06bp}).
Including only the first 13 dates (to assess the effect of a shorter time baseline of one month) 
leads to a higher distance by 0.2--0.3\,Mpc 
in each band and a larger error of $\pm$0.8\,Mpc. The explosion date is also
in good agreement for either 13 or 16 dates in our sample, with a Julian date
of 2453833.4$\pm$0.4\,day. A plot of  these EPM data points for the $BVi'$ passband, together
with a linear fit to the data, are shown in Fig.~\ref{fig_dist} (bottom curve).

We also use our synthetic spectra and magnitudes to constrain directly the distance, 
in a fashion similar to the SEAM. We find a distance in agreement with 
the EPM inference, with 17.1\,Mpc for the $BVi'$ set (Table~\ref{tab_seam_06bp}). 
Uncertainties are larger
than the scatter of values across bandpasses, but should remain at the 5--10\% level,
in particular due to the large number of observations.
Errors are unlikely to be purely statistical, 
but they have some level of randomness as the deviations from a perfect match of synthetic to observed
spectra are very different for individual epochs (blue featureless 
spectrum at early time to
strongly blanketed spectrum at the recombination epoch), different spectral regions (the blue and red
wavelength regions reflect different contributions from different elements that we model with variable
levels of success).
Reducing the number of observations to 9 dates increases the distance estimate by 1.7\,Mpc.
Uncertainties in reddening and photospheric velocity have a similar effect 
to that described for SN 2005cs (\S\ref{sect_05cs}) and SN 1999em (DH06).

To conclude, we estimate a distance to SN 2006bp and its galaxy host NGC\,3953 of
17.5$\pm$0.8\,Mpc together with an explosion date of 2453833.4$\pm$0.4\,day only
1.7 days after discovery ($\sim$3 days prior to the first observation). 
This makes it the earliest-discovered Type II-P 
SN and gives further support to the novel properties of the SN ejecta inferred from
the first observed spectra (steep density distribution, highly ionized ejecta with
He{\,\sc ii}\,4686\AA\ line emission). This distance estimate is in close match to
the value of 17\,Mpc obtained by Tully (1988). This explosion date is also in agreement
with the date of April 9th, inferred by Quimby et al. (2007) from fits to optical ROTSE-III
light curve (note however that the SN may remain dim in the optical for many hours
after breakout).

% \clearpage
% \clearpage

\section{Discussion}
\label{sect_discussion}

\subsection{Ejecta properties a few days after shock breakout}
\label{sect_ion}

\begin{figure*}
\includegraphics[width=6cm]{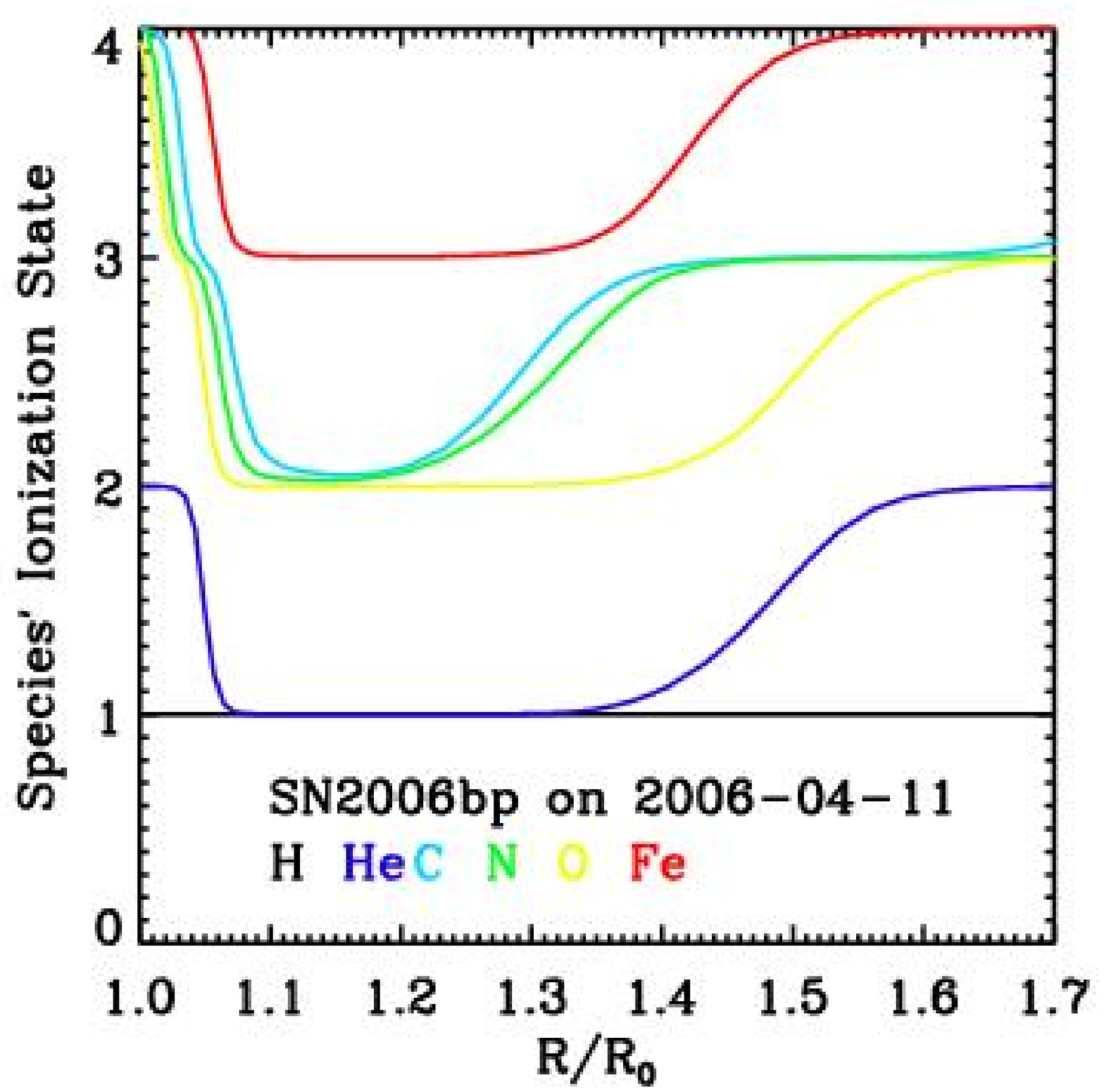}
\includegraphics[width=6cm]{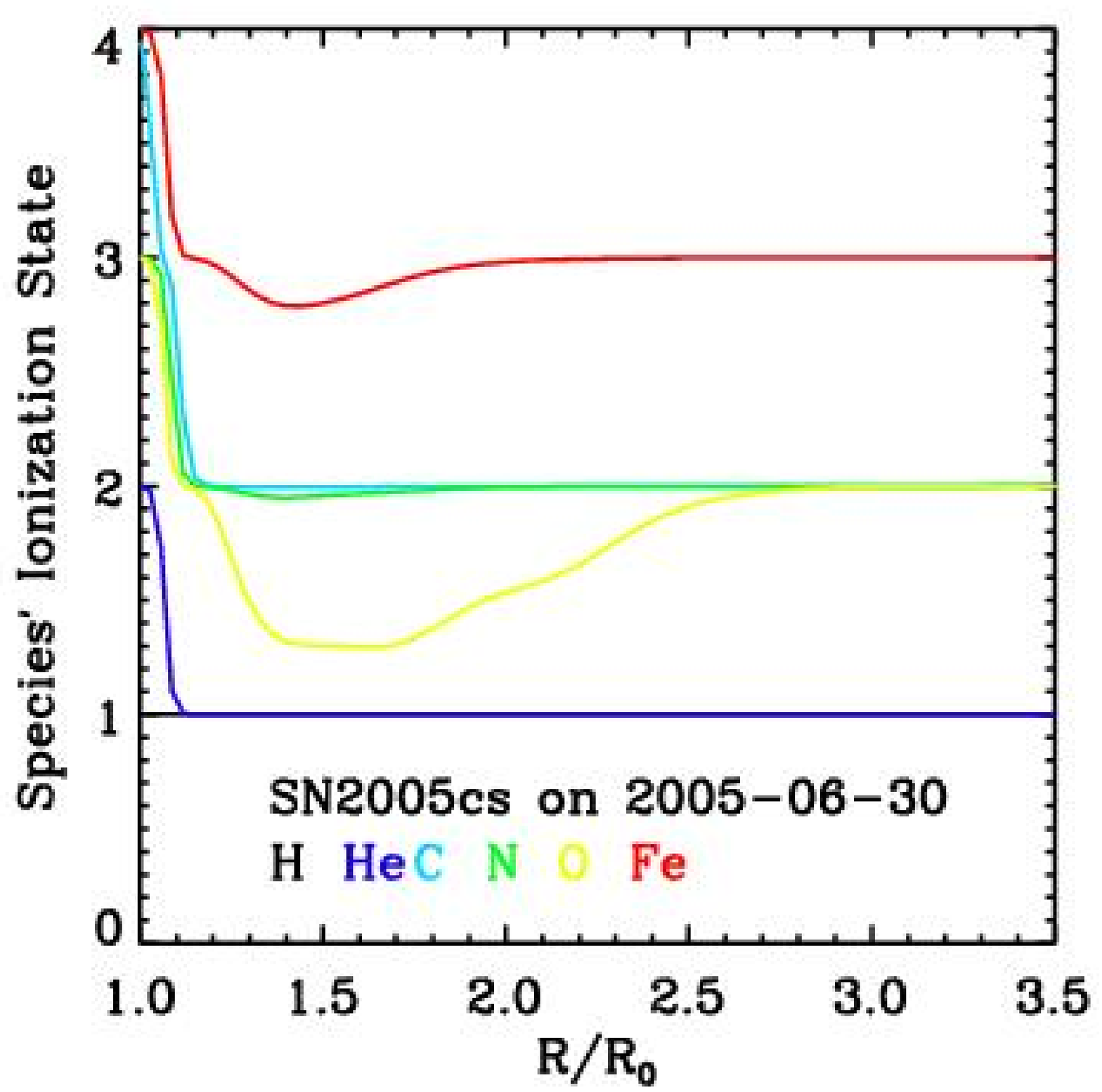}
\includegraphics[width=6cm]{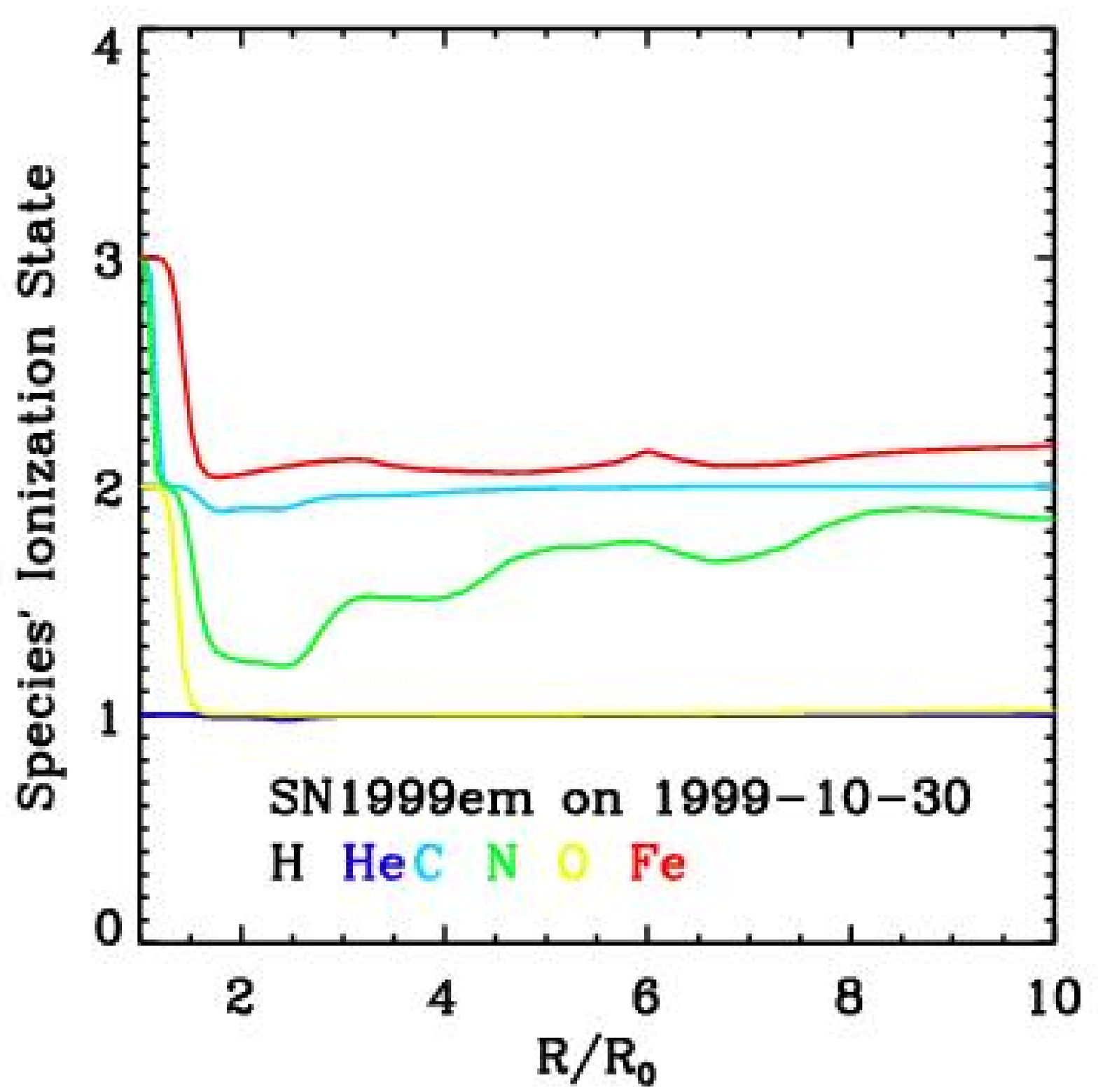}
\caption{Radial variation of the ionization state of H (black), He (blue), C (turquoise), 
N (green), O (yellow), and Fe (red)
in the ejecta of SN 2006bp (left; $R_0$=2.2$\times$10$^{14}$\,cm), SN 2005cs 
(middle; $R_0$=1.6$\times$10$^{14}$\,cm), and SN 1999em (right; $R_0$=4.176$\times$10$^{14}$\,cm)
based on the modeling of the first spectroscopic observation (which corresponds to 
3 (\S\ref{sect_06bp} and Fig.~\ref{06bp_0411}), 5 (\S\ref{sect_05cs} and Fig.~\ref{05cs_0630}), 
and 6 (DH06 and their Fig.~1) days
after explosion) and limited to the regions exterior to the layer where the 
inward integrated Rosseland optical depth is $\sim$50 on that day.
Note the systematic decrease of the ejecta ionization state as we go from SN 2006bp,
to SN 2005cs, and to SN 1999em, following the trend of elapsed time since explosion.
This clarifies the identified emission in  He{\,\sc ii}\,4686\AA\ line in the first spectrum 
of SN 2006bp.
Note also the rise in ionization at large distances (together with a rise of the temperature, 
not shown here), a non-LTE effect in these scattering-dominated environments.
[See the electronic edition of the Journal for a color version of this figure, 
and see \S\ref{sect_ion} for discussion.]
}
\label{fig_gammas}
\end{figure*}

We have presented quantitative spectroscopic analyses of two Type II-P SNe, using 
observations triggered at an exceptionally early time after explosion (3--5 days), providing 
a glimpse into the ejecta
structure relatively soon after shock breakout. We infer a very 
high-temperature/very high ionization in the corresponding photospheric layers and in particular
identify the presence of broad, ejecta-related, He{\,\sc ii}\,4686\AA\ in the first two observations of 
SN 2006bp (narrow emission in He{\,\sc ii}\,4686\AA\ is also identified by Quimby et al. (2007), related to
material outside of the SN ejecta, which also testifies for the high temperature associated with shock breakout).
In Fig.~\ref{fig_gammas}, we show the ionization level for H, He, C, N, O, and Fe
in the ejecta computed with CMFGEN and for models that fit the first spectroscopic and
photometric observations of 
SN 2006bp (left panel), SN 2005cs (middle panel), and SN 1999em (right panel).
These correspond, in the same sequence, to 3, 5, and 6 days after
explosion, as determined in this work and in DH06\footnote{The maximum radius 
differs between simulations, resulting from the different
density exponents, but in all cases is such that the radiation is freely-streaming at the 
outer grid radius (optically-thin conditions)}. 
One clearly sees in this figure the evolution to lower ionization through this sequence,
following directly the elapsed time since explosion. 
Oxygen changes from being twice- to singly-ionized at mid-heights, from SN 2006bp
to SN 1999em. Helium is singly ionized throughout the SN 1999em ejecta,
but twice ionized in the inner and outer regions of the SN 2006bp ejecta.
This trend towards much higher ionization at earlier times supports  the identification of 
He{\,\sc ii}\,4686\AA\ emission in the SN 2006bp ejecta, as well as O{\,\sc ii} lines for both 
SN 2006bp (at about 5 days after explosion) and SN 2005cs (in the discovery spectrum).
O{\,\sc ii} lines were not detected unambiguously in the first spectrum taken of SN 1999em
because it was taken too late, about 1--2 days later than SN 2005cs relative to the explosion time.

Moreover, at such early times, adopting the nitrogen abundance of DH06 now overestimates the 
strength of N{\,\sc ii} lines and underestimates the strength of O{\,\sc ii} lines,
justifying the adoption in this work of CNO abundances closer to solar, i.e. 
C/He=0.0004, N/He=0.0013, O/He = 0.0016, in agreement with the surface chemistry of blue
supergiants analyzed by Crowther et al. (2006).
By comparison, DH06 used C/He=0.00017, N/He=0.0068, O/He = 0.0001 
(all given by number; for both works we use H/He=5).
We thus concur with the findings of Baron et al. (2007) on the necessary CNO chemistry
to reproduce the observations of O{\,\sc ii} and N{\,\sc ii} lines in the early spectra of 
SN 2005cs (as well as in SN 2006bp), with a depletion of carbon and oxygen compared to
solar that is modest rather than severe.
In general, an accurate determination of CNO abundances in the fast-expanding photospheres
of Type II-P SNe is difficult since the corresponding lines are weak and 
overlap with lines of other species such as He{\,\sc i}. Having recourse to quantitative spectroscopy
of the emergent light from their supergiant progenitors, which can be found in number and 
nearby, seems a more viable route to constraining the surface chemistry of such pre-SN massive stars.

In both objects, the near featureless spectral appearance over the first few days 
requires the choice of a very high power-law density exponent of 20--50 
to dwarf the emission volume of all optical lines, as well as very high ejecta velocities 
to reproduce the large widths of emission line features. 
This is admittedly well above the value $n=10$ for the propagation of a shock wave 
through the envelope of a massive star progenitor, both obtained theoretically 
(Imshennik \& Nadezhin 1988; Shigeyama \& Nomoto 1990, Chevalier 1982; Ensman \& Burrows 1992, 
Blinnikov et al. 2000) and inferred observationally (see, e.g., Eastman \& Kirshner 1989; 
Schmutz et al. 1990; Baron et al. 2000; Dessart \& Hillier 2005a,2006a).
Nonetheless, for the outermost layers of the progenitor envelope, a much steeper density 
profile is expected, with values in the vicinity of a few tens (Blinnikov et al. 2000, 
and in particular their Fig.~1), and, under certain conditions, a small amount of mass at the 
progenitor surface may even reach relativistic speeds as the shock breaks out (Matzner \& McKee 1999).

% \clearpage
% \clearpage

Unfortunately, the radiation-hydrodynamics codes used to compute the ejecta structures 
of core-collapse SN are Lagrangean, and are thus not well designed to describe the 
dynamics and the structure of the low density surface layers. 
In those fully-ionized regions, the luminosity is super-Eddington by up to a few orders 
of magnitude, even when accounting only for the electron-scattering opacity. 
When accounting for the additional contribution of optically thick and thin lines,
we find that the radiation pressure can be very substantial, following the combination of a huge 
luminosity, a low density, and a large velocity gradient (Castor, Abbott, \& Klein 1975).
In the CMFGEN model of SN 2006bp on 30 June 2006, the radiative acceleration in the 
layers above the photosphere reaches a few thousand cm\,s$^{-2}$, which combined with a
velocity of a few 10000\,\kms and a length scale of 10$^{14}$\,cm lead to a velocity
gain of few tens of percent (assuming the acceleration remains constant). 
Sooner after shock breakout, the radiative acceleration could be even larger.
Overall, we expect that the outer optically-thin 
layers would adopt a steeper density distribution and a faster-than-linear velocity distribution with radius.
We plan to investigate these issues more quantitatively in the future, but at present, it seems that 
the earliest observations of SNe 2005cs and 2006bp support the very steep density distribution obtained
in radiation-hydrodynamics simulations of core-collapse SN ejecta soon after shock breakout.

\begin{figure}
\plotone{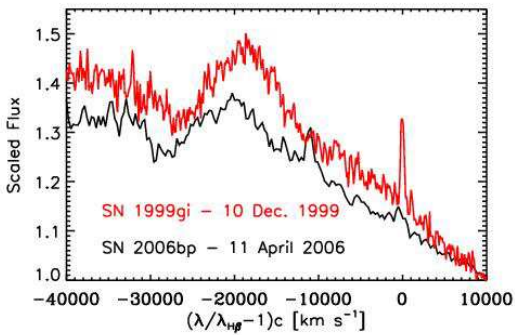}
% \plotone{he2_comp.ps}
\caption{
Comparison between the first spectroscopic observation of SN 2006bp (11th of April 2006) 
and of SN 1999gi (10 December 1999; Leonard et al. 2002b) showing the observed flux versus 
the Doppler velocity, with respect to the rest wavelength of H$\beta$ 
(Note that the inferred reddening of SN 1999gi is $E(B-V)=0.21$, so about half what we infer
for SN 2006bp; Leonard et al. 2002b). The P-Cygni profile seen in the SN 2006bp spectrum is identified
as He{\sc ii}\,4686\AA\ (Fig.~\ref{06bp_0411}), and, given the similarity in profile shape and position,
suggests a similar identification for SN 1999gi.
[See the electronic edition of the Journal for a color version of this figure, 
and see \S\ref{sect_ion} for discussion.]
}
\label{fig_he2}
\end{figure}

Before closing this section, we note that SN 1999gi is another Type II-P SN discovered very early,
perhaps only $\sim$4$^{+3.0}_{-3.1}$ days after explosion (Leonard et al. 2002b). Interestingly, 
Leonard et al. report the presence 
of a full P-Cygni profile shifted from the rest wavelength of H$\beta$ by $-30000$\,\kms in the 
first spectrum of SN 1999gi, and speculate on the potential association with ejecta inhomogeneities.
In the context of a density inhomogeneity far above the photosphere, a blueshifted absorption 
is a possibility, but a P-Cygni profile entirely blueshifted,
and by that amount, is not (see, however, Dessart \& Hillier 2005a).
We reproduce their observation in Fig.~\ref{fig_he2}, and overplot the first spectrum of SN 2006bp (red line). 
Given the close correspondence between the two spectra and the likely early detection of SN 1999gi,
we propose that, in fact, the broad line feature just blueward of H$\beta$ in the spectrum of SN 1999gi
is due to He{\sc ii}\,4686\AA, as for SN 2006bp.

% \clearpage
% \clearpage

\subsection{Energetics}
\label{sect_ener}

\begin{figure*}
\plottwo{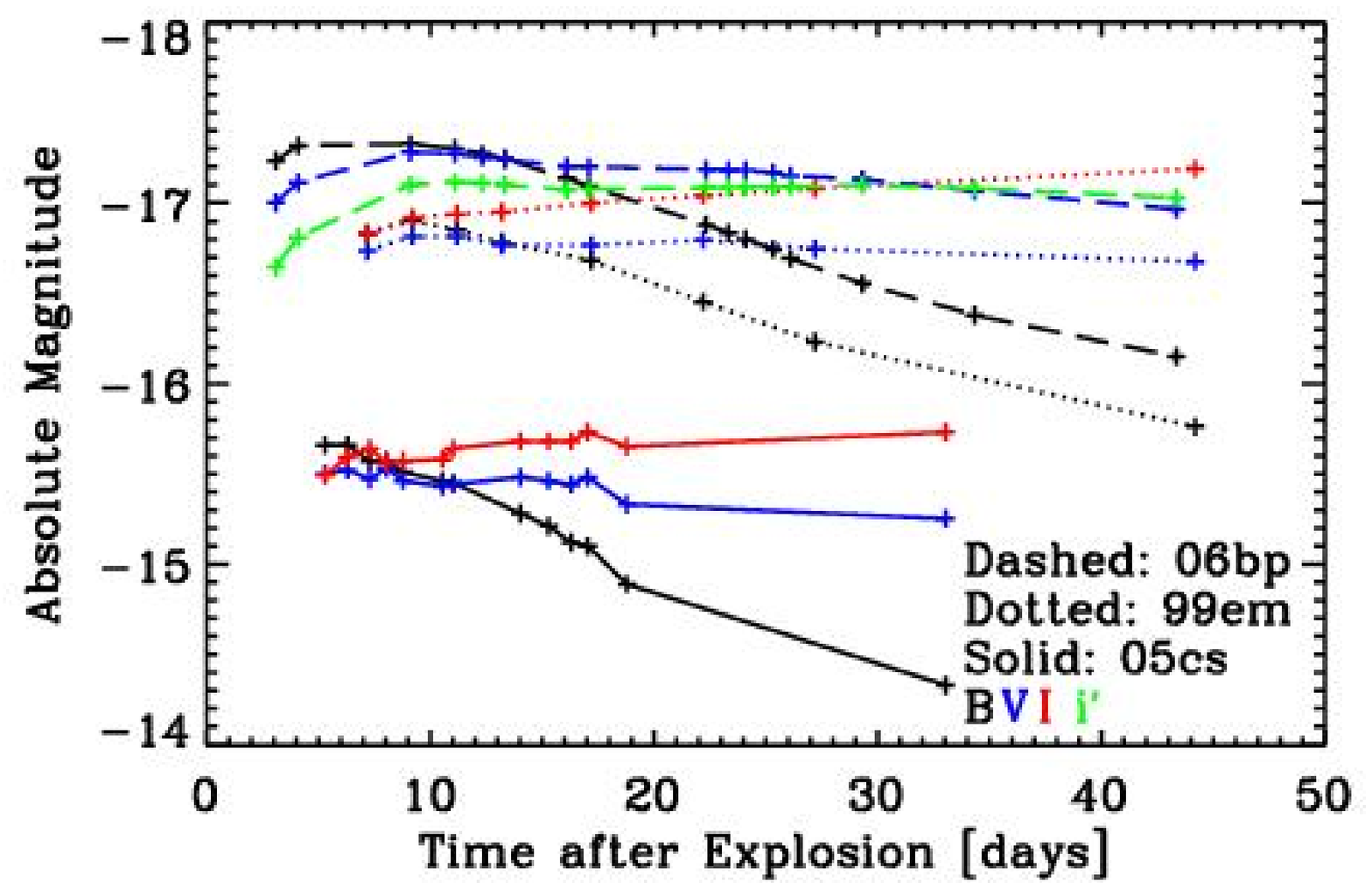}{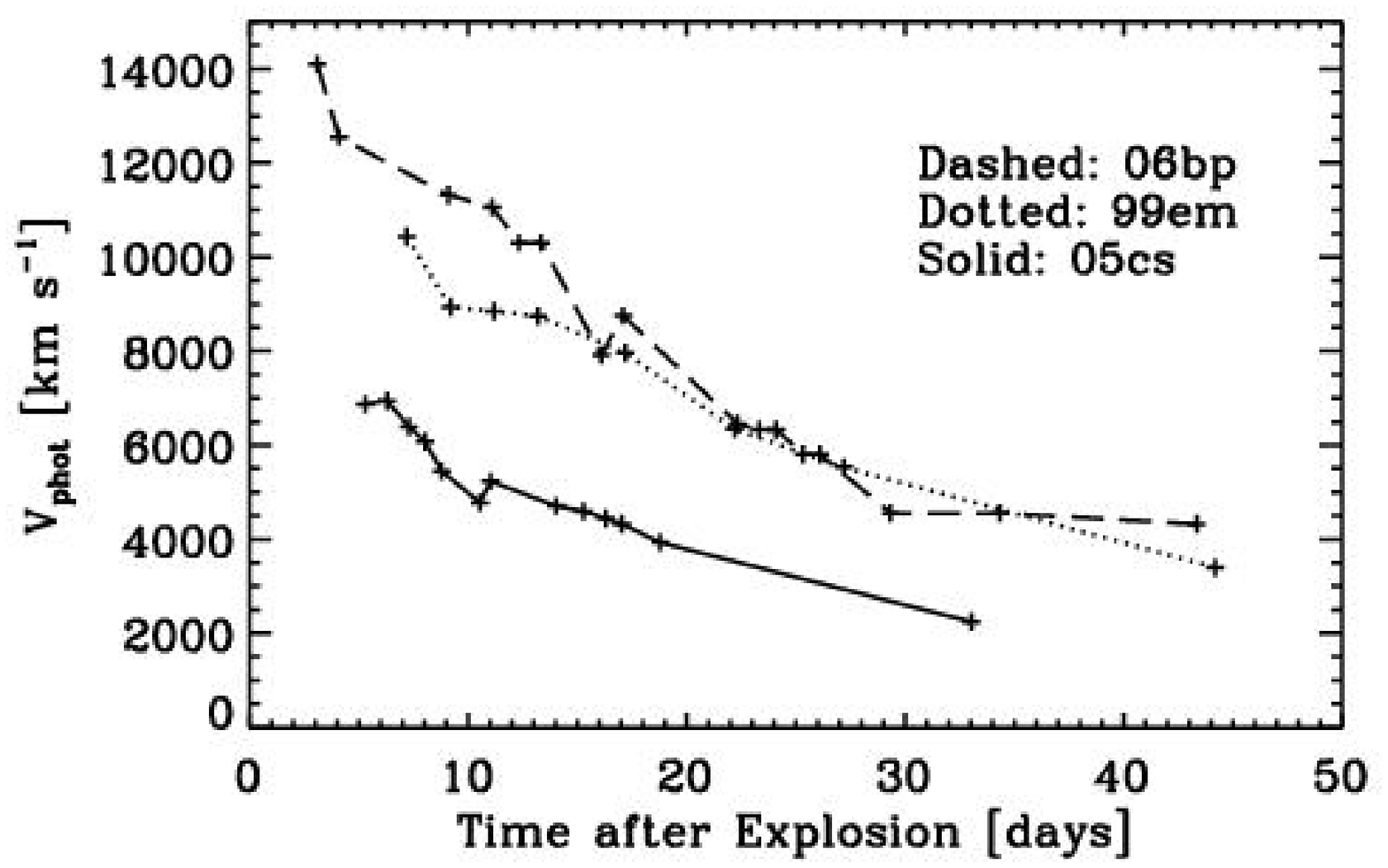}
\plottwo{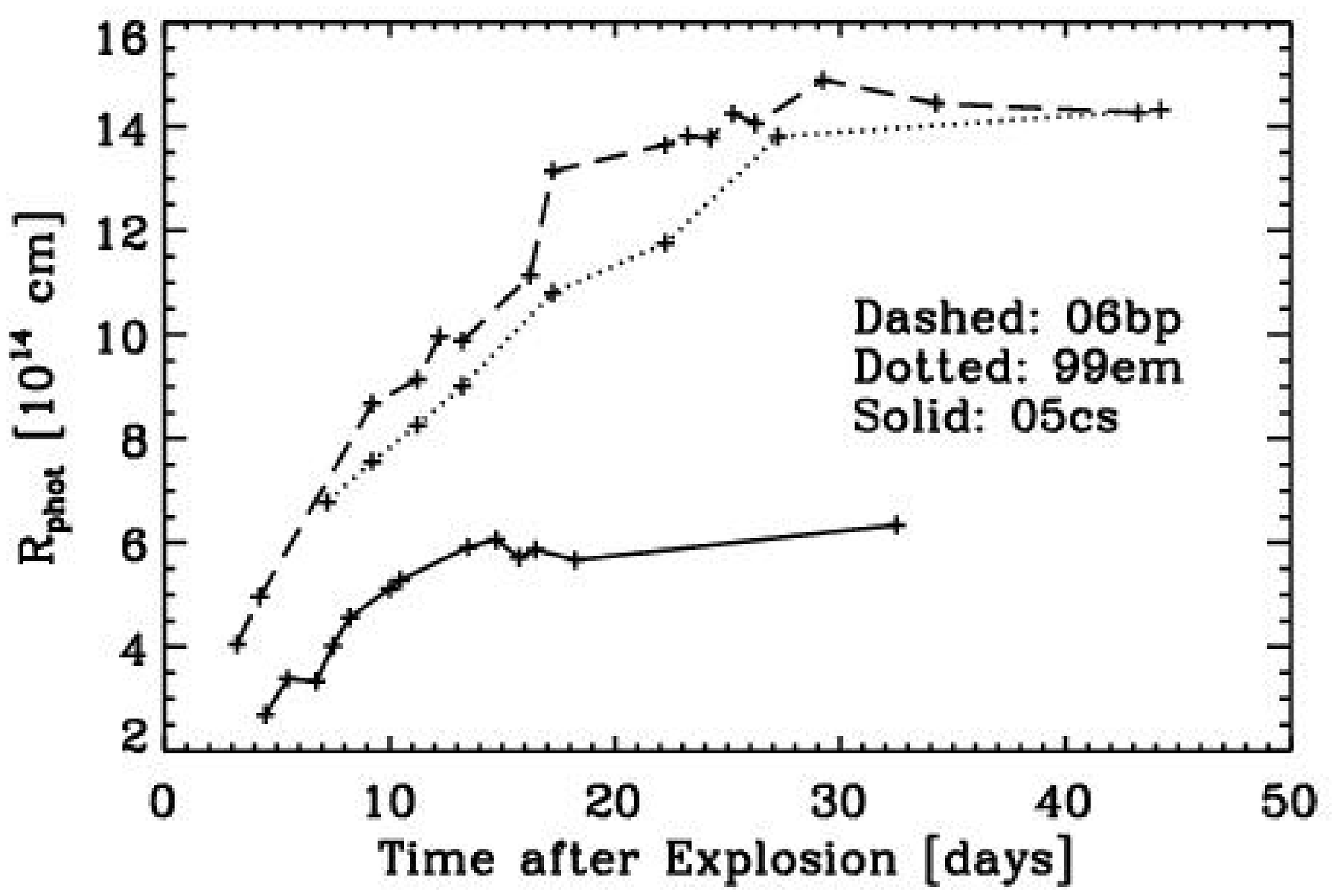}{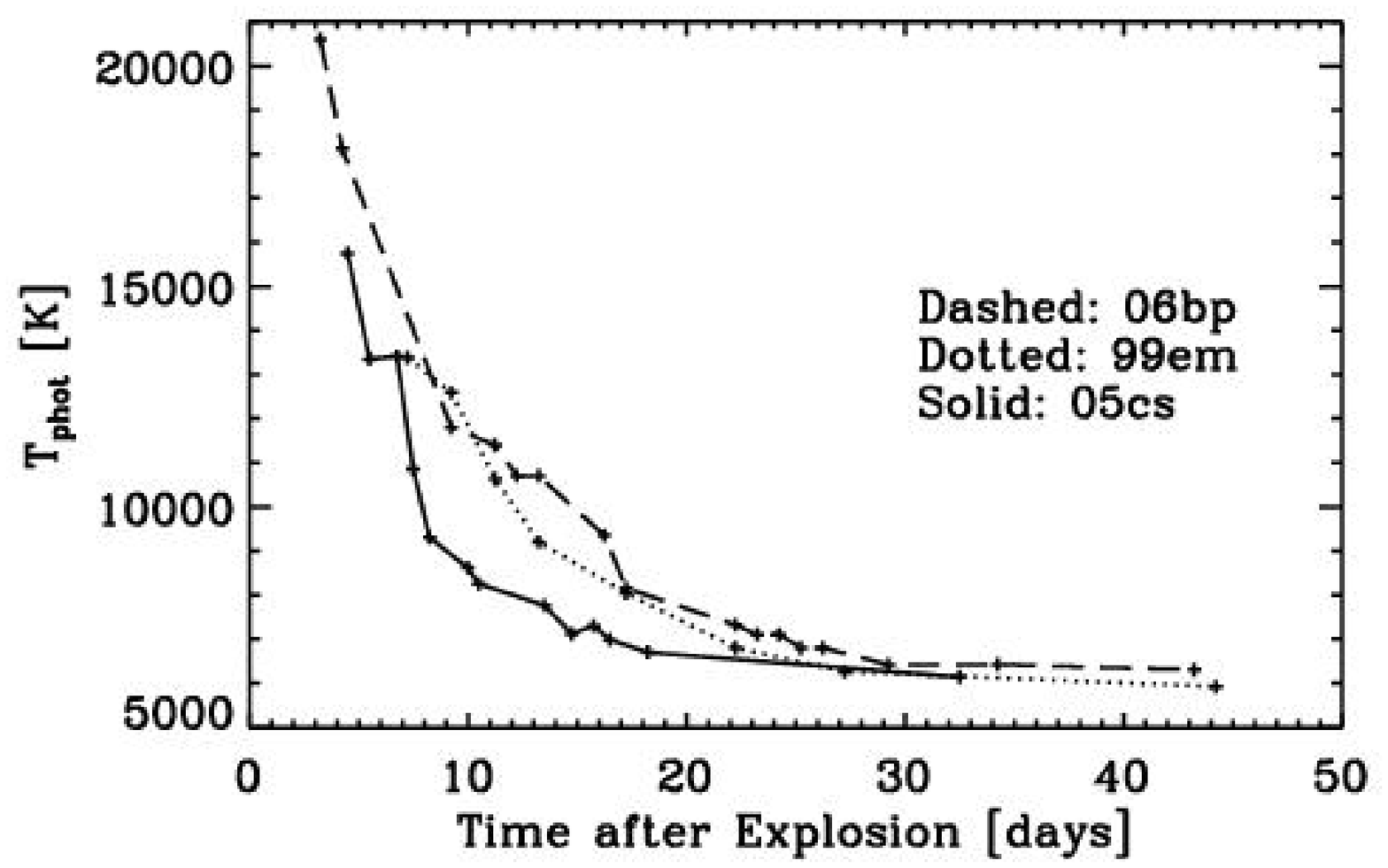}
% \plottwo{abs_mag_scaled.ps}{vphot.ps}
% \plottwo{rphot.ps}{tphot.ps}
\caption{{\it Top Left:} Evolution of the {\it absolute} magnitude in the $B$ (black), 
$V$ (blue), and $I$ (red) bands for 
SN 2005cs (solid), SN 2006bp (dashed; $i'$ is shown instead of $I$), and SN 1999em (dotted), 
with respect to the inferred time of explosion (JD 2453547.6 for SN 2005cs; 
JD 2453833.4 for SN 2006bp; JD 2451474.8 for SN 1999em), and
the inferred distance of 8.9\,Mpc and  reddening 
of 0.04 for SN 2005cs, 11.5\,Mpc and 0.1 for SN 1999em, and 17.5\,Mpc and 0.4 for SN 2006bp.
Note how the SN 2005cs is {\it intrinsically} underluminous compared to SN 1999em, by about 1.5\,mag
equivalent to a factor of four in luminosity (a valid comparison since their SED is comparable 
at a given time after explosion).
SN 2006bp is somewhat more luminous than SN 1999em, by about 0.5\,mag in each bandpass.
Crosses refer to the times of the observations that we use here, and given with respect to the inferred
time of explosion obtained in this work and in DH06 for SN 1999em.
{\it Top Right:} Same as left, but this time for the photospheric velocity. Notice how the
relationship in absolute magnitude carries over for the photospheric velocity (uncertain at the
10\% level), at {\it the same time} after explosion (note that the three ejecta have a similar 
density structure beyond one week after explosion). The larger the kinetic 
energy, the larger the bolometric luminosity of the ejecta.
{\it Bottom:} Same as the top right panel, but for the photospheric radius (left) and temperature (right).
[See the electronic edition of the Journal for a color version of this figure, 
and see \S\ref{sect_ener} for discussion.]
}
\label{fig_mag_vphot}
\end{figure*}

  Having inferred the reddening and the distance to SNe 2005cs and 2006bp, we can 
deduce the absolute magnitude for each event at all epochs studied.
Together with results presented for SN 1999em in DH06,
we show in the top-left panel of Fig.~\ref{fig_mag_vphot} the evolution with 
time since explosion of the absolute magnitude in $B$ (black), $V$ (blue), and $I$ (red), 
for SN 2005cs (solid line), SN 2006bp (dashed line; $i'$ is shown instead of $I$), and
SN 1999em (dotted line).
Note that the time of explosion is computed together with the distance, rather than estimated
based on prior non-detection of the SN on pre-explosion images of the corresponding 
host-galaxy\footnote{Our inferences are, however, compatible with these alternate estimates, 
in particular for SN 2006bp for which the explosion time is very tightly constrained by a non-detection
on April 9.15 and detection on April 10.15.}.
While the spectroscopic and photometric evolution of three events are comparable in all spectral
regions at a given time after explosion, they display a sizable range of intrinsic brightness.
SN 2005cs is about two magnitudes fainter than SN 2006bp, about a factor of six in (bolometric) 
luminosity, with SN 1999em somewhat less luminous than SN 2006bp. 
SN 2006bp was detected earliest of all three and is in fact the first Type II-P to be caught 
so soon after explosion, with a $\sim$2-day delay at most (note that 1987A and 1993J are
not Type II-P SNe).

In the top-right panel of Fig.~\ref{fig_mag_vphot}, we show the corresponding evolution
of the model photospheric velocity, for SNe 2005cs, 1999em, and 2006bp. The above trend is maintained,
i.e., larger intrinsic brightness is associated with larger ejecta 
velocity. Note that if the bolometric luminosity scaled with the square of the ejecta velocity,
a brightening by 1.5\,mag would result from a factor of 2 enhancement in velocity, which is compatible
with our inferences for SNe 2005cs and 2006bp (top row of Fig.~\ref{fig_mag_vphot}). Since we 
adopt similar ejecta structures (after one week past explosion), this suggests that 
the bolometric luminosity correlates with the ejecta kinetic energy in the hydrogen shell, 
as pointed out in the past by, for example, Falk \& Arnett (1977), Litvinova \& Nadezhin (1983),
Hamuy (2003), Zampieri et al. (2003), Pastorello et al. (2004).
The late time brightness is not controlled by the energy deposited at shock passage, but rather 
stems from the amount and location of the unstable nickel at the start of homologous expansion,
so the above trend may not carry over until much later times, although it might, but for different
reasons (higher energy explosions may systematically produce more $^{56}$Ni).

Similarly, and for completeness, we include in the bottom panels of Fig.~\ref{fig_mag_vphot} the 
temporal evolution of the photospheric radius (left) and temperature (right). Note the
smaller extent of the slowly-expanding ejecta of SN2005cs, the converging photospheric
temperatures, and the near-constant photospheric radii at the recombination epoch for all three SNe, tightly
linked to the Plateau appearance of the visual light-curves at such times.

% \clearpage
% \clearpage

\subsection{A common origin for the UV and the optical flux?}

   Our modeling approach has been first to fit the optical range, using 
both the overall shape of the SED and spectral lines formed in the SN ejecta.
As a consistency check, as well as an additional constraint on the 
resulting model parameters and the choice of reddening, we compared the
synthetic spectrum to the observed {\sl Swift} UVOT values. For both SN 2005cs 
and SN 2006bp, we obtain very satisfactory fits and we thus conclude that the observed 
UV and optical light emerges from the photospheric layers of the SN ejecta, 
over the first 1-2 months after explosion (starting 3-5 days afterwards).
This result is unambiguous, and, thus, no sizable contribution to the UV and optical fluxes
stems from an interaction with the CSM.

   A further evidence for the photospheric origin of the UV flux is 
the identification of the Mg{\,\sc ii}\,2800\AA\
line as a broad P-Cygni profile in {\sl Swift} UVOT grism spectra (Brown et al. 2007), 
with a width compatible with the photospheric velocity at the corresponding epoch.  
So, while circumstellar interaction may be important in some Type II SNe (the
IIn events),
it seems to contribute negligibly to the UV and optical light
we received from SN 2005cs and SN 2006bp, and more generally in agreement with the 
weak radio and X-ray emission from Type II-P SNe (see, e.g., Pooley et al. 2002; Chevalier et al. 2006). 

\subsection{Comparison with previous work}
\label{sect_comp_baron}

   The recent spectroscopic analysis by Baron et al. (2007) of SN 2005cs
at two epochs in the photospheric phase agrees in many respects with the findings 
of the present study for that object and, thus, gives some credence to the Type II SN spectroscopic 
analyzes performed with PHOENIX (Hauschildt \& Baron 1999) and CMFGEN 
(Hillier \& Miller 1998; Dessart \& Hillier 2005a). This is particularly satisfying
since PHOENIX and CMFGEN are two very distinct codes, and use different assumptions in the modeling.
A potential important difference in assumptions is our neglect of energy contribution from radioactive
decay throughout this work. By contrast with Baron et al. (2007), we find no strong evidence for nickel mixing and 
non-thermal energy deposition at the photosphere over the first month after explosion (in this work, we obtain
good fits to observations prior to the hydrogen-recombination epoch without invoking this additional
source of energy). The ejecta ionization structure computed by the two codes may thus differ.

   In this work, we infer a reddening $E(B-V) \sim 0.04$, in agreement with Baron et al. (2007) 
who obtain a value of 0.035.
We also concur with the findings of Baron et al. (2007) on the necessary CNO chemistry
to reproduce the observations of O{\,\sc ii} and N{\,\sc ii} lines in the early spectra of 
SN 2005cs (as well as in SN 2006bp), with a depletion of carbon and oxygen compared to
solar that is modest rather than severe. As discussed in \S\ref{sect_ion}, this constraint 
can be placed for SN 2005cs, by contrast with, e.g., SN 1999em, mostly because it was observed
at an earlier time when the ionization conditions at the photosphere are ripe for O{\sc ii} line emission.
The identification of O{\sc ii} and N{\sc ii} lines can explain most, if not all, the mysterious 
kinks seen in the blue wing of H$\beta$ or He{\sc i}\,5875\AA\ (see, e.g., Leonard et al. 2002a,b who
reported such kinks in the spectra of SNe 1999em and 1999gi).
Here, the abundances adopted at early times for both SNe 2005cs and 2006bp are, given as mass fractions, 
$X_{\rm H}$=0.55, $X_{\rm He}$=0.44, $X_{\rm C}$=0.0005, $X_{\rm N}$=0.002, $X_{\rm O}$=0.003,
and $X_{\rm Fe}$=0.0013. 
At later times, when in particular O{\,\sc i}\,7770\AA\ gains in strength, we enhance both the
carbon and the oxygen abundances, with C/He=0.001, O/He = 0.01, and reduce the nitrogen 
abundance, N/He=0.001. Again, observations support the same ejecta chemistry at the photosphere 
at corresponding epochs after explosion for both SNe.
We are however reserved concerning the sodium abundance, which we predicted, based on steady-state CMFGEN models, to be 
enhanced by a factor of four over its primordial value in SN 1999em (DH06). Comparable enhancements would be 
needed here for 2005cs and 2006bp, and, indeed, Baron et al. (2007) predict a sodium enhancement of a factor of 10.
{\it Time-dependent} CMFGEN models performed with a {\it primordial} sodium abundance, however, suggest that Na{\sc i}\,D may remain as 
strong and broad as observed (Dessart \& Hillier 2007a,b); further work is needed to quantify the effect of
time-dependence on abundance determinations. 

  For the observations of the 14th of July 2005, Baron et al. (2007) find at the photosphere a temperature
of 6000\,K, a velocity of 4000\,km\,s$^{-1}$ and a density exponent of 8. In the same order and for that date, 
we find 6700\,K, 3930\,km\,s$^{-1}$, and 10, in agreement with their values. 
For the observations of the 31st of July 2005, Baron et al. 
find at the photosphere a temperature of 5500\,K, a velocity of 6000\,km\,s$^{-1}$ and a density exponent of 12
(see their \S6).
In this work, for the observations of the 28th of July 2005, just three days before (note that observations 
do not change visibly over just a few days at such late times), we find in the same order
6140\,K, 2240\,km\,s$^{-1}$ and a density exponent of 10. While we fit the CfA observations reasonably well
with this parameter set on that date (Fig.\ref{05cs_0728}), 
Baron et al. require a photospheric velocity nearly three times as large to do so, and also larger
than the value on Day 17 (in fact nearly as large as on the first spectrum taken, for which we use 6880\,km\,s$^{-1}$). 
Upon inspection of their Figs.~12-13, it appears that H$\alpha$ peaks at $\sim$6480\AA,
while the observations they use, as originally published in P06, and our (restframe) CfA observations show a peak at 6550\AA.
This difference of 70\AA\ corresponds to a velocity shift of 3200\,km\,s$^{-1}$ at this wavelength 
and suggests that their extravagant proposition of an outward moving photosphere between Day 17 and Day 34 (see their \S6)
may stem solely from a problem with the wavelength of their flux datapoints.

Baron et al. (2007) obtained a SEAM distance to SN 2005cs of 7.9$^{+0.7}_{-0.6}$\,Mpc 
using all $UBVRI$ magnitudes, but at only two epochs, which is in agreement with our estimate 
of 8.9$\pm$0.5\,Mpc using $BVI$ and 12 or 13 epochs.

  Although the observed spectra of SNe 2005cs and 2006bp presented in this work are different in many ways,
these reflect mostly the difference in the ejecta expansion rate (small for 2005cs), the reddening in 
the line of sight (large for 2006bp), and the elapsed time between explosion and discovery (very short
for 2006bp). As shown in Fig.~\ref{fig_mag_vphot}, both SNe evolve in a similar fashion, 2005cs being merely
underluminous and boasting a lower ejecta kinetic energy.

\section{Conclusion}
\label{sect_conclusion}

  We have presented a multi-epoch multi-wavelength quantitative spectroscopic analysis
of the two Type II-P SN events 2005cs and 2006bp, over the first two months after explosion,
and have reached similar conclusions on the ejecta properties to those for SN 1999em (DH06).
The progenitor stars have a composition compatible with a BSG/RSG progenitor.
The spectral evolution shows a reddening of
the SED with time, with a flux peaking in the UV at very early times, but progressively
shifting to longer wavelengths. The photosphere recedes to deeper layers with time,
reaching mass shells moving at about 2000-3000\,\kms about 6 weeks after explosion
and having enhanced carbon and oxygen, and reduced nitrogen, abundances, compared 
to the surface layers. We find a density distribution that is very steep at early times,
and we surmise that this could be an imprint of the shock breakout epoch.
From about a week until six weeks after explosion, we find a more standard and constant 
density exponent of ten for the corresponding
power law distribution, compared to a value of 50 (for SN 2006bp) and 20 (for SN 2005cs)
used to model the first observation. {\sl Swift} UVOT photometry and ground-based
optical spectroscopy and photometry support a common origin for the UV and the optical 
light, so that we anticipate that if CSM interaction does occur, it does not contribute
significantly to the observed UV and optical fluxes.

  Using our analysis, we infer the distance and the explosion date 
of each SN. We find, in that order, JD 2453547.6$\pm$0.5 days (2.8 days 
prior to discovery) and 8.9$\pm$0.5\,Mpc for SN 2005cs, and JD 2453833.4$\pm$0.4\,days (1.7 days prior
to discovery) and 17.5$\pm$0.8\,Mpc for SN 2006bp. 
The large number of observations, the quality of our fits, the agreement
between the two methods followed suggests a very high level of internal consistency of the distances
determined here with Type II SNe. The actual level of accuracy on such determined distances, 
which we estimate to be on the order of 10\% (related to the spectral fitting accuracy), 
would be best assessed by confrontation
with alternate methods (Cepheids for example, although all alternate methods have their 
own inaccuracies, claimed and otherwise), the determination of distances to a galaxy 
host with well-recorded multiple Type II SN events (the typical Type II SN rate of 
2 per century per galaxy makes this proposition somewhat unrealistic), 
a Galactic SN that could also be resolved with an interferometer or a large-aperture telescope.
Our perspective on the uncertainty of our distance measurements is that,
were we to increase the number of Type II SNe suitable for analysis in M 51 
to 10 or 100, the scatter in the distance determined with our technique for
that sample would not be larger than 10\%. But this would still be no proof that the
resulting average distance is indeed the right one since enlarging the sample would 
only allow for an internal consistency check on our method.
In the short term, a Cepheid distance to the galaxy host of 2005cs and 2006bp is 
highly desirable, and would provide an independent check on the distance computed with 
Type II-P SNe, as was done with success in the past with SN 1999em (Leonard et al. 2003; 
Baron et al. 2004; DH06), giving further support to the potential use of such objects for distance 
determinations in the Universe.

 This work fits within an ambitious project to build a general understanding of 
SN ejecta, to infer properties of the progenitor star (with ramifications for 
massive star evolution) and the explosion mechanisms at the origin of the event.
A major ongoing theoretical development is the treatment of time-dependent
effects both in the radiation field and in the level populations to allow for
a direct modeling of the evolution over months after shock breakout, through the
photospheric and into the nebular phases, and based on a range of hydrodynamical 
models of the explosion.
In parallel, high-quality multi-epoch multi-wavelength observations 
for a substantial number of nearby SNe, irrespective of the type, galaxy host properties, 
covering from as soon as possible after explosion until the nebular phase,
is eagerly sought.

\acknowledgments

We thank Stan Woosley for providing the ejecta structure of a 15\,\mo model.
We also thank Andrea Pastorello, Robert Quimby, and Doug Leonard for providing their 
optical spectra of SN 2005cs, SN 2006bp, and SN 1999gi, respectively.
L.D. acknowledges support for this work from the Scientific Discovery through Advanced Computing
(SciDAC) program of the DOE, under grant numbers DE-FC02-01ER41184 and DE-FC02-06ER41452, 
and from the NSF under grant number AST-0504947. This research was supported in part by the 
National Science Foundation under Grant No. PHY05-51164 to the Kavli Institute for Theoretical
Physics at UC Santa Barbara and AST-0606772 to Harvard University.
This work is sponsored at PSU by NASA contract NAS5-00136.

\end{document}